\documentclass[fleqn,10pt]{article}
\usepackage{latexsym, graphicx, epsfig, amsmath, amssymb,amsfonts}
\usepackage{natbib,amsthm,version}
\usepackage{amsbsy,bm,multirow,enumerate}
\usepackage[titletoc,page]{appendix}
\usepackage[mathscr]{eucal}
\usepackage{mathtools}
\usepackage{color}
\usepackage{subfigure}
\usepackage[utf8]{inputenc}
\usepackage[english]{babel}
\usepackage{float}
\usepackage{caption}
\usepackage{systeme}

\newcommand{\bs}[1]{\boldsymbol{#1}}

\newtheorem{theorem}{Theorem}[section]

\newtheorem{remark}{Remark}

\graphicspath{{./Figures/stdtest/} {./Figures/capwave/}  {./Figures/risingbubble/rho100/}   {./Figures/risingbubble/rho1000/}   }

\title{ An Unconditionally Energy-Stable Scheme Based on an
  Implicit Auxiliary Energy Variable
  for Incompressible Two-Phase Flows with Different Densities
  Involving Only Precomputable Coefficient Matrices 
} 
\author{Zhiguo~Yang,\,
  Suchuan~Dong\thanks{Author of correspondence, Email: sdong@purdue.edu} \\
  Center for Computational and Applied Mathematics \\
  Department of Mathematics \\
  Purdue University 
 } 

\date{(November 18, 2018)}
\begin{document}
\maketitle



\begin{abstract}

  We present an energy-stable scheme for numerically approximating the governing
  equations for incompressible two-phase flows with different densities and
  dynamic viscosities for the two fluids. The proposed
  scheme employs a scalar-valued auxiliary
  energy variable in its formulation, and it satisfies a discrete energy stability
  property. More importantly, the scheme is computationally efficient.
  Within each time step, it computes two copies of the flow variables (velocity,
  pressure, phase field function) by solving individually a linear algebraic
  system involving a constant and time-independent coefficient matrix
  for each of these field variables. The coefficient matrices involved in these
  linear systems only need to be computed once and can be pre-computed.
  Additionally, within each time step
  the scheme requires the solution of a nonlinear algebraic
  equation about a {\em scalar-valued number} using the Newton's method.
  The cost for this nonlinear solver is very low, accounting
  for only a few percent of the total computation time per time step,
  because this nonlinear equation is about a scalar number, not a field function.
  Extensive numerical experiments have been presented for several two-phase
  flow problems involving large density ratios and large viscosity ratios.
  Comparisons with theory show that the proposed
  method produces physically accurate results. 
  Simulations with large time step sizes demonstrate the stability of computations
  and verify the robustness of the proposed method.
  An implication of this work is that energy-stable schemes
  for two-phase problems can also become computationally efficient
  and competitive, eliminating
  the need for expensive re-computations of coefficient matrices,
  even at large density ratios and  viscosity ratios.

\end{abstract}


\vspace{0.05cm}

\section{Introduction}
\label{sec:intro}


This work concerns the simulation of the dynamics of
a mixture of two immiscible
incompressible fluids with possibly very different densities and dynamic
viscosities based on the phase field approach.
The presence of fluid interfaces, the associated
surface tension, the density contrast
and viscosity contrast play an important role in the dynamics
of such systems, and these factors also make such numerical simulations
very challenging. Phase field (or diffuse interface)~\cite{Rayleigh1892,Waals1893,AndersonMW1998,LowengrubT1998,Jacqmin1999,LiuS2003}
is one of the main approaches for dealing with fluid interfaces
in the modeling of two-phase systems, along with other
related methods~\cite{OsherS1988,ScardovelliZ1999,Tryggvasonetal2001,SethianS2003}.
It has attracted an increasing interest from the community in the past years,
in part because of its physics-based nature.
With this approach the fluid interface is treated to be diffuse, as
a thin smooth transition layer~\cite{AndersonMW1998}.
The state of the system is characterized
by, apart from the hydrodynamic variables
such as velocity and pressure,
an order parameter (or phase field variable), which varies smoothly
within the transition layer and is mostly uniform in
the bulk phases. The evolution of the system is
characterized by, apart from the kinetic energy,
a free energy density function, which contains component terms
that promote the mixing of the two fluids and also component terms
that tend to separate these fluids. The interplay of these two opposing
tendencies determines the dynamic profile of the interface.
The adoption of the free energy in the formulation
makes it possible to relate to
other thermodynamic variables. Indeed, with the phase field approach
the governing equations of
the system can be rigorously derived based on the conservation laws
and thermodynamic principles. Several thermodynamically
consistent phase field models are already available in the literature
for two-phase and multiphase
flows, see e.g.~\cite{LowengrubT1998,KimL2005,AbelsGG2012,ShenYW2013,AkiDG2014,Dong2014,LiuSY2015,GongZW2017,Dong2018,RoudbariSBZ2018},
with various degrees of sophistication or the observance/violation
of other physical principles such as Galilean invariance and
reduction consistency.
The mass conservation of the individual fluid components in the system,
and the choice of an appropriate form for the free energy density function,
naturally give rise to the Cahn-Hilliard equation in two phases
or a system of coupled Cahn-Hilliard type equations (see e.g.~\cite{AbelsGG2012,Dong2018}, among others).


We focus on the numerical approximation and simulation of the
governing equations for incompressible two-phase flows with
different densities and viscosities in this work.
A computational scientist/engineer interested in such problems
confronts a compromise and needs to balance two seemingly incompatible
aspects: the desire to be able to use a larger time step
size (permissible by accuracy), and the computational cost.
On the one hand, semi-implicit splitting type
schemes (see e.g.~\cite{DongS2012,BadalassiCB2003,DingSS2007,Dong2012,Dong2014obc}, among others)
induces a
very low computational cost per time step, because among other
things only de-coupled linear algebraic systems need to be solved after discretization
and these linear systems only involve constant
and time-independent coefficient matrices
that can be pre-computed (see~\cite{DongS2012}), even with large density
ratios and viscosity ratios. 
The downside of these schemes lies in that they are only conditionally stable
and the time step size is restricted by CFL or related conditions.
On the other hand, energy-stable schemes (see e.g.~\cite{ShenY2010,Salgado2013,GuoLL2014,GrunK2014,ShenY2015,GuoLLW2017,YuY2017,RoudbariSBZ2018}, among others)
can potentially allow the use of much larger time step sizes
in dynamic simulations. The downside lies in that,
the computational cost per time step
of these schemes can be very high. Energy-stable schemes often require the solution
of coupled nonlinear algebraic field equations or coupled linear algebraic
equations. The linear algebraic systems associated with these schemes
involve time-dependent coefficient matrices, which require frequent
re-computations (or at every time step).

Mindful of the strengths and weaknesses of both types of
schemes, we would like to consider the following question.
Can we devise an algorithm
to combine the strengths of both types of schemes
that can achieve unconditional energy stability and simultaneously
require a relatively low computational cost?
Summarized in this paper is our attempt to tackle this question
and a numerical scheme that largely achieves this goal.


The current work has drawn  inspirations from several
previous studies in the literature. 
In the following we restrict our review of literature to
the energy-stable schemes for the hydrodynamic interactions
of two-phase flows with different densities and viscosities
based on the phase field framework.
This will leave out those algorithms that are devoted purely
to the phase field such as the Cahn-Hilliard or Allen-Cahn
equations~\cite{CahnH1958,AllenC1979} without hydrodynamic
interactions, for which a large volume of literature exists.
Those studies considering only matched densities for the two fluids
(see e.g.~\cite{AlandC2016,HanBYT2017,YangY2018},
among others) will also be largely left out.
In \cite{ShenY2010} a phase field model based on
the combined Navier-Stokes/Allen-Cahn equations with different
densities/viscosities for the two fluids
is considered, and several discretely energy-stable schemes
of first order in time are introduced based on projection,
Gauge-Uzawa, and pressure stabilization formulations. 
What is interesting lies in that all these schemes
are linear in nature. They only require the solution of linear,
albeit coupled,
algebraic equations for different flow variables
after discretization. To accommodate different densities
for the two fluids, the authors of \cite{ShenY2010} have
adopted a reformulation of the inertial term~\cite{GuermondQ2000}
in the Navier-Stokes equation, which as pointed out by~\cite{GrunK2014}
might not be consistent with the phase field equation employed therein.
Corresponding schemes for a Navier-Stokes/Cahn-Hilliard
model are presented in \cite{ShenY2010cam}.
A fractional-step scheme based on a pressure correction-type
strategy is developed in \cite{Salgado2013} for a phase-field model
in which the chemical potential contains a velocity term (see also \cite{LiuSY2015}),
together with a generalized Navier type boundary condition for contact
lines~\cite{QianWS2006}. This scheme is linear, and
the discrete equations about the phase field function and
the velocity are coupled together.
Improved algorithms for the Navier-Stokes/Cahn-Hilliard model
are later developed by~\cite{ShenY2015},
in which the discrete phase field equation and the momentum equations
are de-coupled thanks to an extra stabilization term~\cite{BoyerM2011}
employed for approximating the convection velocity in the Cahn-Hilliard
equation. The scheme is first order in time, and it is unclear
whether an analogous second-order stabilization term exists for the
approximation of the convection velocity.
In \cite{GrunK2014}
a discretely energy stable scheme for the phase field model
of \cite{AbelsGG2012} is developed, which gives rise to
a system of nonlinear algebraic equations that couple together all
the flow variables. It is interesting to note that
the algorithm developed in \cite{YuY2017}
for the same phase field model only requires the solution of linear
equations and that the phase-field and momentum equations are
de-coupled owing to the same treatment of the discrete convection
velocity of the Cahn-Hilliard equation as in \cite{ShenY2015}.
Two discretely energy stable schemes are described in
\cite{GuoLL2014,GuoLLW2017} for the quasi-incompressible
Navier-Stokes/Cahn-Hilliard model of \cite{LowengrubT1998}.
The schemes can preserve the mass conservation on the discrete level,
and they lead to coupled highly-nonlinear algebraic systems
after discretization. Numerical
schemes for related quasi-incompressible hydrodynamic phase field
models have also been proposed in~\cite{GongZW2017,RoudbariSBZ2018,GongZYW2018}.
In particular, in \cite{GongZYW2018}
the so-called invariant energy quadratization method has been
used to reformulate the phase field equation, and
the resultant numerical schemes are second-order in time,
and involve the solution of linear algebraic systems that
couple together the different flow variables.


Apart from the many contributions discussed above,
an interesting strategy for formulating
energy-stable schemes for gradient-type dynamical systems
(gradient flows) based on certain auxiliary variables
has emerged recently~\cite{Yang2016,ShenXY2018}.
The invariant energy quadratization (IEQ) method~\cite{Yang2016}
introduces an auxiliary field function related to the square root of
the potential free energy density function together with
a dynamic equation for this auxiliary variable, and allows one
to reformulate the gradient-flow evolution equation
to facilitate schemes for ensuring the energy stability relatively easily.
The scalar auxiliary variable (SAV) method~\cite{ShenXY2018}
introduces an auxiliary variable, which is a scalar-valued
number rather than a field function, related to the square
root of the total potential energy integral, and a dynamic
equation about this scalar variable. Both types of auxiliary
variables can simplify the formulation 
of schemes to achieve energy stability for gradient flows.

Another recent development that inspires the current work is~\cite{LinD2018},
in which an energy-stable scheme for the incompressible Navier-Stokes
equations based on a scalar auxiliary variable related to
the total kinetic energy has been developed. Because
the Navier-Stokes equation is not a gradient-type system,
the scalar auxiliary variable formulation as developed
in \cite{ShenXY2018} for gradient flows
cannot be directly used. Indeed, it is observed that if one reformulates the viscous
term in the Navier-Stokes equation using
the auxiliary variable, in a way analogous
to the treatment of the dissipation term in
the evolution equation for gradient flows, the simulation results
turn out to be very poor. Instead, a viable strategy for Navier-Stokes equations
seems to be to control
the convection term with the auxiliary variable.
Such a  strategy is presented 
in \cite{LinD2018}, which hinges on reformulating the convection-term
contribution into a boundary integral in the dynamic equation
for the auxiliary total kinetic energy.


In this paper we build upon
the scalar auxiliary variable idea and present an energy-stable
scheme for the numerical approximation of the two-phase
governing equations with different densities and viscosities for
the two fluids based on the phase field model
of~\cite{AbelsGG2012}. 
By introducing a scalar-valued variable related to
the total of the kinetic energy and the potential free energy of
the two-phase system, we reformulate the two-phase governing equations
into an equivalent form. By carefully treating
the variable-density and variable-viscosity terms,
we show that the proposed scheme
honors a discrete energy stability relation.
We  present an efficient solution
algorithm and a procedure for
dealing with the integrals of the unknown
field functions. Within each time step, our algorithm
only requires the solution of several de-coupled individual
linear algebraic systems, each for an individual field function,
together with the solution of a nonlinear
algebraic equation about a {\em scalar-valued number}.
More importantly, those linear algebraic systems to be solved within
a time step each involves
a {\em constant and time-independent} coefficient matrix,
which only needs to be computed once and can be pre-computed.
The nonlinear algebraic equation involved therein
requires Newton iterations in its solution.
But its computational cost is very low,
because this equation is about a
single {\em scalar number}, not a field function.
Numerical experiments show that
the solution for the nonlinear equation accounts for around $8\%$ of the
total solver time within a time step, 
and essentially all this time is spent on computing the coefficients
involved in the nonlinear equation in preparation for the Newton method,
rather than in the actual Newton iterations.
Because of these attractive properties, our algorithm is computationally
very efficient.
We will present a number of numerical experiments to demonstrate the stability
of our method at large time step sizes for two-phase problems.


The new aspects of this work include the following:
(i) the energy-stable scheme for the two-phase governing equations
with different densities and viscosities for the two fluids, and
(ii) the efficient solution algorithm for implementing the proposed scheme.
The property of our solution algorithm that it involves only linear
algebraic systems with constant and time-independent coefficient matrices,
even at large density ratios and viscosity ratios,
is particular attractive, because this makes the cost low
and the method computationally
very efficient.
To the best of the authors' knowledge, this is the first energy-stable
scheme which involves only constant and time-independent coefficient matrices
for incompressible two-phase flows with different densities and viscosities for
the two fluids.


The rest of this paper is organized as follows.
In Section \ref{sec:method}, we introduce a scalar-valued auxiliary
variable and reformulate the two-phase governing equations into
an equivalent system employing
this variable. Then we present a scheme for temporal discretization
for the reformulated equivalent system, and prove that the scheme
satisfies a discrete energy law. An efficient solution algorithm
for implementing the scheme will be presented, and
the spatial discretization based on $C^0$ spectral elements will be
discussed.
In Section \ref{sec:tests} we present a few numerical examples
of two-phase flows involving large density ratios and viscosity
ratios to demonstrate the accuracy of our method
and also its stability at large time step sizes.
Section \ref{sec:summary} then concludes the discussions
with some closing remarks.

\section{Energy-Stable Scheme for Incompressible Two-Phase Flows}
\label{sec:method}

\subsection{Governing Equations}

Consider a mixture of two immiscible incompressible fluids contained in
some domain $\Omega$ in two or three dimensions, whose boundary is denoted by $\partial \Omega$. Let $\rho_1$ and $\rho_2$ respectively denote the constant
densities of the two fluids, and $\mu_1$ and $\mu_2$ denote their
constant dynamic viscosities. The conservations of mass/momentum and
the second law of thermodynamics for this two-phase system
lead to the following coupled
system of equations (see e.g.~\cite{AbelsGG2012,Dong2014} for details):
\begin{subequations}\label{eq:NphaseModel}
\begin{align}
&\rho \frac{D\bs u}{Dt}+\tilde{\bs J} \cdot \nabla \bs u=-\nabla p + \nabla \cdot[\mu  \mathcal{\bs D}( \bs u)]-\lambda \nabla \cdot (\nabla \phi \nabla \phi) +\bs f(\bs x,t), \label{eq:NphaseEq1orig}\\
& \nabla \cdot \bs u=0,\label{eq:NphaseEq2}\\
&\frac{D\phi}{Dt}=m\nabla^2  \mathcal{C}+g(\bs x,t),  \label{eq:NphaseEq3} \\
&  \mathcal{C}=-\lambda\nabla^2 \phi +h(\phi)     , \label{eq:NphaseEqC}
\end{align}
\end{subequations}
where $\frac{D}{Dt}=\frac{\partial}{\partial t} +\bs u\cdot \nabla  $ is the material derivative, and $\mathcal{\bs D}( \bs u)=\nabla \bs u+\nabla \bs u^T.$
In the above equations, $\bs u(\bs x,t)$ is the velocity, $p(\bs x, t)$ is the
pressure, and $\bs f(\bs x,t)$ is an external body force,
where $t$ is time and $\bs x$ is the spatial coordinate and $(\cdot)^T$ denotes the transpose of $(\cdot).$ $\phi(\bs x,t)$ denotes the phase field function, $-1 \leq \phi \leq 1.$ The flow regions with $\phi=1$ and $\phi=-1$ respectively represent the first and the second fluids. The iso-surface $\phi(\bs x,t)=0$ marks the interface between the two fluids at time $t.$ The function $h(\phi)$ in equation \eqref{eq:NphaseEqC} is given by
\begin{equation}\label{eq:hphi}
h(\phi)=F'(\phi)= \frac{\lambda}{\eta^2}\phi (\phi^2-1)\;\;\text{with}\;\; F(\phi)=\frac{\lambda}{4\eta^2}(1-\phi^2)^2,
\end{equation}
where $\eta$ is a characteristic length scale of the interface thickness,
and $F(\phi)$ is the potential free energy density (double-well function) of the
system. $\mathcal{C}$ in the above equations denotes the chemical potential,
satisfying the relation
$
\mathcal{C} = \frac{\delta W}{\delta \phi},
$
where $W$ is the free energy of the system given by
$
W =\int_{\Omega}\left[ \frac{\lambda}{2}\nabla\phi\cdot\nabla\phi + F(\phi) \right].
$
$\lambda$ is the mixing energy density coefficient
and is related to the surface
tension by~\cite{YueFLS2004}
\begin{equation}\label{eq:lambda}
\lambda=\frac{3}{2\sqrt{2}}\sigma \eta,
\end{equation}
where $\sigma$ is the interface surface tension and is assumed to be a constant.
$m>0$ is the mobility of the interface, and is assumed to be a constant in the current paper.  The density $\rho$ and the dynamic viscosity $\mu$ of the two-phase
mixture are related to the phase field function by
\begin{equation}\label{eq:rhomu}
\rho(\phi)=\frac{ \rho_1+ \rho_2}{2}+ \frac{ \rho_1-  \rho_2}{2}\phi,\quad \mu(\phi)=\frac{\mu_1+ \mu_2}{2}+\frac{ \mu_1- \mu_2}{2}\phi.
\end{equation}
In equation \eqref{eq:NphaseEq1orig}, $\tilde {\bs J}$ is given by
\begin{equation}\label{eq:tildeJ}
\tilde {\bs J}(\phi)=J_0 \nabla \mathcal{C}=J_0 \nabla \big(-\lambda\nabla^2 \phi +h(\phi) \big),\quad {\rm with}\;\;J_0=-\frac{1}{2}( \rho_1- \rho_2)m.
\end{equation}
Note that $\tilde {\bs J}$ and $\rho$ given above satisfy the following relation
\begin{equation}\label{eq:rhoJrelation}
\frac{D\rho}{Dt}=-\nabla \cdot \tilde{\bs J}.
\end{equation}
$g(\bs x,t)$ in equation \eqref{eq:NphaseEq3} is a prescribed source term
for the purpose of numerical testing only, and will be set to
$g=0$ in actual simulations.

To facilitate subsequent development of the unconditionally
energy-stable scheme, we transform equation \eqref{eq:NphaseEq1orig}  into an
equivalent form:
\begin{equation}\label{eq:NphaseEq1}
\rho \frac{D\bs u}{Dt}+\tilde{\bs J} \cdot \nabla \bs u=-\nabla P + \nabla \cdot[\mu  \mathcal{\bs D}( \bs u)]+\mathcal{C}\nabla \phi+\bs f(\bs x,t), 
\end{equation}
where $P=p+\frac{\lambda}{2}\nabla \phi \cdot \nabla \phi +F(\phi)$ is an effective pressure and we have used the following identity
\begin{equation*}
\nabla \cdot (\nabla \phi \nabla \phi)=\frac{1}{2}\nabla(\nabla \phi \cdot \nabla \phi)+(\nabla^2 \phi) \nabla \phi.
\end{equation*}
It can be shown that the system consisting of equations
\eqref{eq:NphaseEq1orig}--\eqref{eq:NphaseEqC}, with $\bs f=0$ and $g=0$,
satisfies the following energy law (assuming that all boundary flux terms vanish),
\begin{equation}
  \frac{d}{dt}\int_{\Omega}\left[
  \frac{1}{2}\rho|\bs u|^2 + \frac{\lambda}{2}\nabla\phi\cdot\nabla\phi
  + F(\phi)
  \right]
  = -\int_{\Omega}\frac{\mu}{2}\left\|\nabla\bs u \right\|^2
  - \int_{\Omega} m\nabla \mathcal{C}\cdot\nabla\mathcal{C}.
\end{equation}

Equations \eqref{eq:NphaseEq1}, \eqref{eq:NphaseEq2}-\eqref{eq:NphaseEqC} are to be supplemented by appropriate boundary and initial conditions for the velocity and the phase field function. We assume the following boundary conditions:
\begin{align}
&\bs u|_{\partial \Omega}= \bs w(\bs x,t), \label{eq:bc1}\\
&\bs n \cdot \nabla \phi|_{\partial \Omega}=d_a(\bs x,t),\label{eq:bc2}\\
& \bs n \cdot \nabla (\nabla^2 \phi)|_{\partial \Omega}=d_b(\bs x,t), \label{eq:bc3}
\end{align}
and the initial conditions:
\begin{align}
& \bs u(\bs x,0)=\bs u_{in}(\bs x),\label{eq:ic1}\\
& \phi(\bs x,0)=\phi_{in}(\bs x).\label{eq:ic2}
\end{align}
In the above equations $d_a$ and $d_b$ are prescribed source
terms for the purpose of numerical testing only, and
will be set to $d_a=0$ and $d_b=0$ in actual simulations.
The boundary conditions \eqref{eq:bc2} and \eqref{eq:bc3}
with $d_a=0$ and $d_b=0$
correspond to a solid-wall boundary of neutral wettability, i.e.~contact
angle is $90^0$ when the fluid interface intersects the wall.
$\bs u_{in}(\bs x)$ and $\phi_{in}(\bs x)$ are the initial velocity and
phase field distributions.


\subsection{Reformulated Equivalent System of Equations}

To derive the energy-stable scheme for the system  \eqref{eq:NphaseEq1}, \eqref{eq:NphaseEq2}-\eqref{eq:NphaseEqC}, we introduce a shifted energy consisting of
the kinetic energy and the potential free energy component as follows,
\begin{equation}\label{eq:energyE}
E(t)=\int_{\Omega}\Big[ \frac{1}{2}\rho|\bs u|^2+F(\phi) \Big]d\Omega +C_0
\end{equation}
where $C_0$ is a chosen constant such that $E(t)>0$ for all $t\geq 0.$
Note that $E(t)$ is a scalar-valued number, not a field function.
Define an auxiliary variable
\begin{equation}\label{eq:energyfunc}
R(t)=\sqrt{E(t)}.
\end{equation}
With the help of the identities
\begin{equation}
\left\{
\begin{split}
& \rho\frac{D\bs u}{Dt}\cdot \bs u= \frac{D}{Dt}\Big( \frac{1}{2}\rho |\bs u|^2 \Big)-\frac{D\rho}{Dt} \Big( \frac{1}{2}|\bs u|^2  \Big) \\
& \Big(  \tilde {\bs J} \cdot \nabla \bs u \Big) \cdot \bs u= \nabla \cdot \Big(  \frac{1}{2}|\bs u|^2 \tilde {\bs J} \Big) - \nabla \cdot \tilde {\bs J} \Big( \frac{1}{2}|\bs u|^2 \Big)\\
&\bs u \cdot \nabla\Big(\frac{1}{2}\rho |\bs u|^2  \Big)=\nabla \cdot \Big(\frac{1}{2} \rho |\bs u|^2 \bs u  \Big)-(\nabla \cdot \bs u)  \frac{1}{2}\rho |\bs u|^2 
\end{split}
\right.
\end{equation}
and the equations \eqref{eq:NphaseEq2} and \eqref{eq:rhoJrelation}, we  can obtain that
\begin{equation}\label{eq:energyderivative}
\frac{\partial }{\partial t}\Big( \frac{1}{2}\rho |\bs u|^2 \Big)=\rho \frac{D\bs u}{D t}\cdot \bs u+\Big( \tilde {\bs J}\cdot \nabla \bs u \Big)\cdot \bs u-\nabla \cdot \Big(\frac{1}{2}\tilde {\bs J} |\bs u|^2   \Big) -\nabla \cdot \Big( \frac{1}{2}\rho |\bs u|^2 \bs u  \Big).
\end{equation}
Thus, taking the derivative of $R(t)$ leads to
\begin{equation}
\begin{split}
2R\frac{dR}{dt}&=\frac{dE}{dt}=\int_{\Omega} \frac{\partial}{\partial t} \Big[ \frac{1}{2}\rho |\bs u|^2+F(\phi)  \Big]\\
&= \int_{\Omega} \Big[\rho \Big( \frac{\partial \bs u}{\partial t} +\bs u \cdot \nabla \bs u  \Big)+\tilde {\bs J}\cdot \nabla \bs u  \Big]\cdot \bs u+ \int_{\Omega} h(\phi) \frac{\partial \phi}{\partial t} 
-  \int_{\partial \Omega} (\bs n \cdot \bs u) \frac12\rho |\bs u|^2 -  \int_{\partial \Omega} \Big(\bs n \cdot \tilde {\bs J} \Big)\frac12 |\bs u|^2,
\end{split}
\end{equation}
where $\bs n$ is the outward-pointing unit vector normal to the boundary $\partial \Omega,$ and we have used equation \eqref{eq:energyderivative}, integration by part, and the divergence theorem.  It is crucial to note that both $E(t)$ and $R(t)$
are scalar-valued variables depending only on $t$, and the fact that
\begin{equation}\label{eq:REidentyty}
\frac{R(t)}{\sqrt{E(t)}}=1.
\end{equation}

In light of equation \eqref{eq:REidentyty}, we can
re-write the system consisting of equations \eqref{eq:NphaseEq1}, \eqref{eq:NphaseEq2}-\eqref{eq:NphaseEqC} into the following equivalent form
\begin{equation}\label{eq:NphaseEqnew1}
\begin{split}
  \rho  \frac{\partial \bs u}{\partial t} +\frac{R(t)}{\sqrt{E(t)}} \Big \{ \rho \bs u \cdot \nabla \bs u + & \tilde{\bs J} \cdot \nabla \bs u -\nabla \mu \cdot \mathcal{\bs D}(\bs u)  +\Big(1-\frac{\rho}{\rho_0}   \Big) \nabla P
  +(\mu-\nu_m \rho)\nabla\times\nabla\times \bs u  \Big \} \\
&=-\frac{\rho}{\rho_0}\nabla P +\nu_m \rho  \nabla^2 \bs u  +\mathcal{C}\nabla \phi+\bs f,
\end{split}
\end{equation}
\begin{equation}\label{eq:NphaseEqnew2}
\nabla \cdot \bs u=0,
\end{equation}
\begin{equation}\label{eq:NphaseEqnew3}
\frac{\partial \phi}{\partial t} +\frac{R(t)}{\sqrt{E(t)}} \bs u \cdot \nabla \phi= m\nabla^2  \mathcal{C}+g,
\end{equation}
\begin{equation}\label{eq:NphaseEqnew4}
  \mathcal{C}=-\lambda \nabla^2 \phi
  + S\Big( \phi- \phi   \Big)
  +   \frac{R(t)}{\sqrt{E(t)}}h(\phi),
\end{equation}
\begin{equation}\label{eq:NphaseEqnew5}
\begin{split}
2 R \frac{dR}{dt}&= \int_{\Omega}\rho  \frac{\partial \bs u}{\partial t} \cdot \bs u - \int_{\Omega}  \mathcal{C} \nabla \phi \cdot \bs u +   \frac{R(t)}{\sqrt{E(t)}} \int_{\Omega}  \mathcal{C} \nabla \phi \cdot \bs u +  \frac{R(t)}{\sqrt{E(t)}}  \int_{\Omega} h(\phi) \frac{\partial \phi}{\partial t}    \\
&+ \frac{R(t)}{\sqrt{E(t)}} \int_{\Omega}  \Big \{ \rho \bs u \cdot \nabla \bs u
+  \tilde{\bs J} \cdot \nabla \bs u -\nabla \mu \cdot \mathcal{\bs D}(\bs u)  +\Big(1-\frac{\rho}{\rho_0}   \Big) \nabla P
+(\mu-\nu_m \rho)\nabla\times\nabla\times \bs u  \Big \}\cdot \bs u\\
& + \int_{\Omega} \Big \{ \nabla\cdot\left[ \mu \mathcal{\bs D}(\bs u)\right] +\Big(\frac{\rho}{\rho_0}  -1 \Big) \nabla P -\nu_m \rho\nabla^2 \bs u    \Big\} \cdot \bs u\\
& -\frac{1}{2}  \int_{\partial \Omega} (\bs n \cdot \bs u) \rho |\bs u|^2 -\frac{1}{2}  \int_{\partial \Omega} \Big(\bs n \cdot \tilde {\bs J} \Big) |\bs u|^2.
\end{split}
\end{equation}
To obtain equations \eqref{eq:NphaseEqnew1} and \eqref{eq:NphaseEqnew5},
we have used equation \eqref{eq:NphaseEq2} and the identities
\begin{equation*}
  \left\{
  \begin{split}
    &
    \nabla \cdot [\mu \mathcal{D}(\bs u)]=\mu [\nabla^2 \bs u+ \nabla(\nabla \cdot \bs u)]+\nabla \mu \cdot \mathcal{D}(\bs u), \\
    & \nabla^2\bs u = \nabla(\nabla\cdot\bs u) - \nabla\times\nabla\times\bs u,
    \end{split}
  \right.
\end{equation*}
and have added/subtracted appropriate terms such as
$\frac{\rho}{\rho_0}\nabla P$, $\nu_m\rho\nabla^2\bs u$, $S\phi$,
and $\mathcal{C}\nabla\phi\cdot\bs u$.
In these equations, $\rho_0$ is a constant given
by $\rho_0={\rm min}(\rho_1, \rho_2),$ $\nu_m$ is a chosen constant
satisfying
$\nu_m\geq \frac{1}{2}\max\left(\frac{\mu_1}{\rho_1},\frac{\mu_2}{\rho_2} \right),
$
and $S$ is a chosen constant satisfying
a condition to be specified later in equation \eqref{eq:alpha}.

The original system consisting of equations  \eqref{eq:NphaseEq1}, \eqref{eq:NphaseEq2}-\eqref{eq:NphaseEqC}, \eqref{eq:bc1}-\eqref{eq:ic2} is equivalent to the reformulated system consisting of equations \eqref{eq:NphaseEqnew1}-\eqref{eq:NphaseEqnew5}, together with the boundary conditions \eqref{eq:bc1}-\eqref{eq:bc3}, and the initial conditions \eqref{eq:ic1}-\eqref{eq:ic2}, supplemented by an extra initial condition for $R(t),$ i.e.
\begin{equation}\label{eq:Rin}
R(0)=\Big(\int_{\Omega}\Big[ \frac{1}{2}\rho(\phi_{in}) |\bs u_{in}|^2+F(\phi_{in}) \Big]d\Omega +C_0 \Big)^{1/2}.
\end{equation}
In the reformulated system, the dynamic variables are $\bs u(\bs x,t)$, $P(\bs x,t)$,
$\phi(\bs x,t)$ and $R(t)$. Note that $E(t)$ is computed by equation \eqref{eq:energyE}.
We next focus on this reformulated equivalent system of equations, and present an unconditionally energy-stable scheme for approximating this system.

\subsection{Formulation of Numerical Scheme}

Let $n\geq 0$ denote the time step index, and $(\cdot)^n$ denote the variable $(\cdot)$ at time step $n.$ Let $J \;(J=1 \;\text{or}\; 2)$ denote the temporal order of accuracy of the scheme. We set
\begin{equation}
\phi^0=\phi_{in},\quad \bs u^0=\bs u_{in}, \quad R^0=R(0),
\end{equation}
where $R(0)$ is given in equation \eqref{eq:Rin}. 
Define a scalar-valued variable $\xi^{n+1}$ 
\begin{equation}\label{eq:En1}
  \xi^{n+1}=\frac{R^{n+1}}{\sqrt{E^{n+1}}}.
\end{equation}
Given $(\phi, \bs u,P, R)$ at time step $n$ and previous time steps,
we compute $(\phi^{n+1}, \bs u^{n+1},P^{n+1}, R^{n+1})$ through the following scheme
\begin{equation}\label{eq:scheme1eq}
\begin{split}
\bar{\rho}^{n+1}\Big(  \frac{\gamma_0 \bs u^{n+1} -\hat {\bs u}}{\Delta t}  \Big) &+\xi^{n+1} \bar{\rho}^{n+1} \bs{\mathcal{N}} =\\
&-\frac{\bar{\rho}^{n+1}}{\rho_0}\nabla P^{n+1}+\nu_m \bar{\rho}^{n+1} \nabla^2 \bs u  ^{n+1}+\mathcal{C}^{n+1}\nabla \bar{\phi}^{n+1}+\bs f^{n+1},
\end{split}
\end{equation}
\begin{equation}\label{eq:scheme2eq}
\nabla \cdot \bs u^{n+1}=0,
\end{equation}
\begin{equation}\label{eq:scheme3eq}
 \frac{\gamma_0 \phi^{n+1} -\hat \phi}{\Delta t}+\xi^{n+1} \big(\bs u^{*,n+1} \cdot \nabla \phi^{*,n+1}\big)=m \nabla^2 \mathcal{C}^{n+1} + g^{n+1},
\end{equation}
\begin{equation}\label{eq:scheme5eq}
\mathcal{C}^{n+1}=-\lambda \nabla^2 \phi^{n+1}+S\big( \phi^{n+1}- \phi^{*,n+1} \big)      +\xi^{n+1} h(\phi^{*,n+1}),
\end{equation}
\begin{equation}\label{eq:scheme6eq}
\begin{split}
&2R^{n+1}  \frac{\gamma_0 R^{n+1} -\hat R}{\Delta t}=\int_{\Omega} \bar{\rho}^{n+1} \Big( \frac{\gamma_0 \bs u^{n+1} -\hat {\bs u}}{\Delta t}  \Big)\cdot \bs u^{n+1} -\int_{\Omega} \mathcal{C}^{n+1} \big(  \bs u^{n+1} \cdot \nabla \bar{\phi}^{n+1} \big) \\
&+\xi^{n+1} \int_{\Omega}   \mathcal{C}^{n+1} \big(  \bs u^{*,n+1} \cdot \nabla \phi^{*,n+1} \big) +\xi^{n+1} \int_{\Omega} h(\phi^{*,n+1}) \frac{\gamma_0 \phi^{n+1} -\hat \phi}{\Delta t}+\xi^{n+1}\int_{\Omega}  \bar{\rho}^{n+1}\bs{\mathcal{N}}\cdot \bs u^{n+1} \\
&+\int_{\Omega}\Big\{   \nabla \bar \mu^{n+1} \cdot \mathcal{\bs D}(\bs u^{n+1}) +\Big(\frac{\bar{\rho}^{n+1}}{\rho_0}  -1 \Big) \nabla P^{n+1} +(\bar \mu^{n+1}-\nu_m \bar{\rho}^{n+1})\nabla^2 \bs u^{n+1}    \Big\}   \cdot \bs u^{n+1}\\
& -\frac{1}{2}  \int_{\partial \Omega} (\bs n \cdot \bs u^{n+1}) \bar \rho^{n+1} \big|\bs u^{n+1}\big|^2 -\frac{1}{2}  \int_{\partial \Omega} \Big(\bs n \cdot \tilde {\bs J}^{n+1} \Big) \big|\bs u^{n+1}\big|^2,
\end{split}
\end{equation}
\begin{align}
&\bs u^{n+1} =\bs w^{n+1}\;\;\;{\rm on}\;\; \partial \Omega,\label{eq:udiscrete} \\
&\bs n \cdot \nabla \phi^{n+1}=d_a^{n+1}\;\;\;{\rm on}\;\; \partial \Omega \label{eq:phidiscrete1}\\
& \bs n \cdot \nabla (\nabla^2 \phi^{n+1})=d_b^{n+1}\;\;\;{\rm on}\;\; \partial \Omega. \label{eq:phidiscrete2}
\end{align}

The symbols in the above equations are defined as follows.
In equations \eqref{eq:scheme1eq} and \eqref{eq:scheme6eq}
\begin{equation}
\bs{\mathcal{N}}= {\bs{{Q}}}+\Big(\frac{ \bar{\mu} ^{n+1}}{\bar {\rho} ^{n+1}}-\nu_m \Big)\nabla \times \nabla \times \bs u^{*,n+1},
\end{equation}
and
\begin{equation}
\begin{split}
{\bs{{Q}}}={\bs{{Q}}}(\bar \phi^{n+1}, \bs u^{*,n+1})&= \bs u^{*,n+1} \cdot \nabla \bs u^{*,n+1} + \frac{1}{\bar \rho ^{n+1}} \bar{\tilde{\bs J}}^{n+1} \cdot \nabla \bs u^{*,n+1} -\frac{1}{\bar \rho ^{n+1}}\nabla \bar \mu^{n+1} \cdot \mathcal{\bs D}(\bs u^{*,n+1})\\
 &  +\Big(\frac{1}{\bar \rho^{n+1} }-\frac{1}{\rho_0}   \Big) \nabla P^{*,n+1} .
\end{split}
\end{equation}
In equation \eqref{eq:En1}, $E^{n+1}$ is given by (see equation \eqref{eq:energyE})
\begin{equation}
  E^{n+1} = \int_{\Omega} \left[
    \frac{1}{2}\bar{\rho}^{n+1} |\bs u^{n+1}|^2
    + F(\phi^{n+1})
    \right]d\Omega + C_0.
  \label{eq:Enp1}
\end{equation}
In these equations, $\bar\phi^{n+1}$ is a $J$-order approximation
of $\phi$ to be specified later in equation \eqref{eq:phibar},
and $\bar \rho^{n+1},$ $\bar \mu^{n+1}$ and $\bar {\tilde {\bs J}}^{n+1}$ are given by
\begin{equation}\label{eq:rhomuJ}
\bar \rho^{n+1}=\rho(\bar {\phi}^{n+1}),\quad \bar \mu^{n+1}=\mu(\bar {\phi}^{n+1}),\quad \bar {\tilde {\bs J}}^{n+1}=\tilde {\bs J}(\bar \phi^{n+1});
\end{equation}
see equations \eqref{eq:rhomu}-\eqref{eq:tildeJ}.
Let $\chi$ denote a generic variable.
Then in equations \eqref{eq:scheme1eq}--\eqref{eq:scheme6eq},
$\frac{1}{\Delta t}(\gamma_0\chi^{n+1}-\hat{\chi})$
represents an approximation
of $\left.\frac{\partial \chi}{\partial t} \right|^{n+1}$
with the $J$-th order backward differentiation formula (BDF),
with $\gamma_0$ and $\hat{\chi}$ given by
\begin{equation}
  \hat{\chi} = \left\{
  \begin{array}{ll}
    \chi^n, & J=1, \\
    2\chi^n-\frac{1}{2}\chi^{n-1},& J=2;
  \end{array}
  \right.
  \qquad
  \gamma_0 = \left\{
  \begin{array}{ll}
    1, & J=1, \\
    3/2, & J=2.
  \end{array}
  \right.
  \label{equ:def_var_hat}
\end{equation}
$\chi^{*,n+1}$ denotes a $J$-th order explicit approximation of $\chi^{n+1}$
given by
\begin{equation}
  \chi^{*,n+1} = \left\{
  \begin{array}{ll}
  \chi^n, & J=1, \\
  2\chi^n - \chi^{n-1}, & J=2.
  \end{array}
  \right.
  \label{equ:def_var_star}
\end{equation}


It should be noted that the different approximations for various terms
in equations \eqref{eq:scheme6eq}, \eqref{eq:scheme1eq} and
\eqref{eq:scheme3eq} are crucial.
They are to ensure a discrete energy
stability property (see Section \ref{sec:energy_law}),
and simultaneously allow an implementation
in which the resultant linear algebraic systems involve only
constant and time-independent coefficient matrices after discretization.

\subsection{Discrete Energy Law}
\label{sec:energy_law}

Our numerical scheme satisfies a discrete energy stability property, and
thus can be potentially favorable for long-time simulations. More specifically,
the following discrete energy law holds:
\begin{theorem}\label{thm:stability}
  In the absence of the external force $\bs f^{n+1}$ and source term $g^{n+1},$
  and with zero boundary conditions $\bs w^{n+1}=\bs 0,$ $d_a^{n+1}=d_b^{n+1}=0,$ the scheme consisting of equations \eqref{eq:En1}-\eqref{eq:phidiscrete2} satisfies the following property:
\begin{equation}\label{eq:energylaw}
\mathcal{E}^{n+1}-\mathcal{E}^n=-D^{n+1}-\frac{1}{2}\int_{\Omega}\bar  \mu^{n+1}\|\mathcal{D}(\bs u^{n+1}) \|^2-\int_{\Omega}m |\nabla \mathcal{C}^{n+1}|^2,
\end{equation}
where $\mathcal{E}^n$ is a discrete energy 
\begin{equation}\label{eq:disenergy}
\mathcal{E}^n=
\begin{cases}
 \frac{\lambda }{2 \Delta t} \|\nabla \phi^n \|^2 +\frac{1}{\Delta t}|R^n|^2,& J=1,\\
 \frac{S}{2\Delta t}\| \phi^n -\phi^{n-1} \|^2 +\frac{\lambda}{4 \Delta t} \Big( \|\nabla \phi^n  \|^2 +\| \nabla \phi^{*,n} \|^2\Big)+\frac{1}{2\Delta t}\Big( |R^n|^2+|R^{*,n}|^2 \Big),& J=2,
\end{cases}
\end{equation}
and $D^{n+1}$ is the discrete dissipation 
\begin{equation}\label{eq:dissipation}
D^{n+1}=\begin{cases}
\frac{S}{\Delta t}\| \phi^{n+1}-\phi^n \|^2+\frac{\lambda}{2\Delta t} \| \nabla \phi^{n+1}-\nabla \phi^n \|^2+\frac{1}{\Delta t}|R^{n+1}-R^{n}|^2,& J=1,\\
\frac{S}{\Delta t}\| \phi^{n+1}-\phi^{*,n+1}  \|^2+\frac{\lambda}{4 \Delta t}\| \nabla \phi^{n+1} -\nabla \phi^{*,n+1}  \|^2+\frac{1}{2\Delta t}|R^{n+1}-R^{*,n+1}|^2,& J=2.
\end{cases}
\end{equation}
\end{theorem}
\begin{proof}
  Multiplying equation \eqref{eq:scheme1eq} by $\bs u^{n+1},$
  equation \eqref{eq:scheme3eq} by $\mathcal{C}^{n+1},$ and equation \eqref{eq:scheme5eq}
  by $-\frac{\gamma_0 \phi^{n+1}-\hat \phi}{\Delta t},$ and taking the $L^2$ inner product lead to
\begin{equation}\label{eq:testv}
\begin{split}
&\int_{\Omega} \bar \rho^{n+1} \Big( \frac{\gamma_0 \bs u^{n+1} -\hat {\bs u}}{\Delta t}  \Big)\cdot \bs u^{n+1} +\xi^{n+1} \int_{\Omega}\bar \rho^{n+1} \bs{\mathcal{N}}\cdot \bs u^{n+1}\\
&=\int_{\Omega} \Big\{-\frac{\bar \rho^{n+1}}{\rho_0}\nabla P^{n+1}+\nu_m \bar \rho^{n+1}  \nabla^2 \bs u  ^{n+1} \Big\}\cdot \bs u^{n+1} + \int_{\Omega} \mathcal{C}^{n+1}\big( \bs u^{n+1} \cdot \nabla \bar{\phi}^{n+1} )+\int_{\Omega}\bs f^{n+1}\cdot \bs u^{n+1},\\
&\int_{\Omega}  \frac{\gamma_0 \phi^{n+1}-\hat \phi}{\Delta t} \mathcal{C}^{n+1}+\xi^{n+1} \int_{\Omega}  \mathcal{C}^{n+1} \big( \bs u^{*,n+1}\cdot \nabla \phi^{*,n+1}\big)=m\int_{\Omega} \mathcal{C}^{n+1}  \nabla^2 \mathcal{C}^{n+1}+ \int_{\Omega}g^{n+1}\mathcal{C}^{n+1},\\
&\int_{\Omega} \frac{\hat \phi-\gamma_0 \phi^{n+1}}{\Delta t} \mathcal{C}^{n+1}=\frac{\lambda}{\Delta t} \int_{\Omega} \nabla^2 \phi^{n+1} \big(\gamma_0 \phi^{n+1}-\hat \phi \big)- \frac{S}{\Delta t} \int_{\Omega} \big(\phi^{n+1}-\phi^{*,n+1}\big)\big(\gamma_0 \phi^{n+1}-\hat \phi \big)\\
&\qquad \qquad \qquad \qquad \qquad   -\xi^{n+1} \int_{\Omega} h(\phi^{*,n+1}) \frac{\gamma_0 \phi^{n+1}-\hat \phi}{\Delta t}.
\end{split}
\end{equation}
Summing up the three equations in \eqref{eq:testv}
and the equation \eqref{eq:scheme6eq}, we arrive at
\begin{equation}\label{eq:Rstablity1}
\begin{split}
&\frac{2}{\Delta t}  R^{n+1}\big(\gamma_0 R^{n+1}-\hat R\big)=-\int_{\Omega} \nabla P^{n+1}\cdot \bs u^{n+1} +\int_{\Omega} \nabla \cdot \Big[ \bar \mu^{n+1}\mathcal{D}(\bs u^{n+1})  \Big]\cdot \bs u^{n+1}+m\int_{\Omega}\mathcal{C}^{n+1} \nabla^2 \mathcal{C}^{n+1}  \\
&+\frac{\lambda}{\Delta t} \int_{\Omega} \nabla^2 \phi^{n+1} (\gamma_0 \phi^{n+1}-\hat \phi)- \frac{S}{\Delta t} \int_{\Omega} (\phi^{n+1}-\phi^{*,n+1})(\gamma_0 \phi^{n+1}-\hat \phi)  +\int_{\Omega}\bs f^{n+1}\cdot \bs u^{n+1} +\int_{\Omega}g^{n+1}\mathcal{C}^{n+1} \\
& -\frac{1}{2}  \int_{\partial \Omega} (\bs n \cdot \bs u^{n+1}) \bar \rho^{n+1} \big|\bs u^{n+1}\big|^2 -\frac{1}{2}  \int_{\partial \Omega} \Big(\bs n \cdot \tilde {\bs J}^{n+1} \Big) \big|\bs u^{n+1}\big|^2.
\end{split}
\end{equation}
By equation \eqref{eq:scheme2eq}, integration by part and the divergence theorem, we obtain that
\begin{equation}\label{eq:divp}
\int_{\Omega}\nabla P^{n+1} \cdot \bs u^{n+1}=\int_{\partial \Omega} P^{n+1}\big( \bs n \cdot \bs u^{n+1}  \big).
\end{equation}
Using equation \eqref{eq:scheme5eq} and the boundary conditions \eqref{eq:phidiscrete1} and \eqref{eq:phidiscrete2}, we have 
\begin{equation}\label{eq:ndotC}
\bs n \cdot \nabla \mathcal{C}^{n+1}=-\lambda d_b^{n+1}+S(d_a^{n+1}-d_a^{*,n+1})+\xi^{n+1}h'(\phi^{*,n+1})d_a^{*,n+1}\;\;\text{on}\;\;\partial \Omega.
\end{equation}

With the help of equation \eqref{eq:divp} and the relations
\begin{equation}\label{eq:idstability1}
\left\{
\begin{split}
&
\int_{\Omega} \nabla \cdot \Big[ \bar \mu^{n+1}\mathcal{D}(\bs u^{n+1})  \Big]\cdot \bs u^{n+1}=\int_{\partial \Omega} \bar \mu^{n+1}\bs n \cdot \mathcal{D}(\bs u^{n+1})\cdot \bs u^{n+1}-\frac{1}{2}\int_{\Omega} \bar \mu^{n+1}\|\mathcal{D}(\bs u^{n+1}) \|^2, \\
&
\int_{\Omega}\mathcal{C}^{n+1} \nabla^2 \mathcal{C}^{n+1}=\int_{\partial \Omega} \big( \bs n \cdot \nabla \mathcal{C}^{n+1} \big)\mathcal{C}^{n+1}-\int_{\Omega} \big | \nabla \mathcal{C}^{n+1}\big |^2 , 
 \\
&
 \int_{\Omega} \nabla^2 \phi^{n+1} \big(\gamma_0 \phi^{n+1}-\hat \phi \big)=\int_{\partial \Omega}\big( \bs n \cdot \nabla \phi^{n+1} \big)\big( \gamma_0 \phi^{n+1}-\hat \phi  \big)-\int_{\Omega} \nabla \phi^{n+1}\cdot \big( \gamma_0 \nabla \phi^{n+1}-\nabla \hat \phi  \big),
\end{split} 
\right.
\end{equation}
equation \eqref{eq:Rstablity1} can be transformed into
\begin{equation}\label{eq:Rid2}
\begin{split}
&\frac{2}{\Delta t}  R^{n+1} \big(\gamma_0 R^{n+1}-\hat R\big)=-\frac{1}{2}\int_{\Omega} \bar \mu^{n+1}\|\mathcal{D}(\bs u^{n+1}) \|^2 -\int_{\Omega}m |\nabla \mathcal{C}^{n+1}|^2 +\int_{\Omega}\bs f^{n+1}\cdot \bs u^{n+1} +\int_{\Omega}g^{n+1}\mathcal{C}^{n+1}  \\
&  -\frac{\lambda}{\Delta t}\int_{\Omega} \nabla \phi^{n+1}\cdot \big( \gamma_0 \nabla \phi^{n+1}-\nabla \hat \phi  \big)- \frac{S}{\Delta t} \int_{\Omega} \big(\phi^{n+1}-\phi^{*,n+1}\big)\big(\gamma_0 \phi^{n+1}-\hat \phi \big)\\
&+\int_{\partial \Omega} m\big( -\lambda d_b^{n+1}+S(d_a^{n+1}-d_a^{*,n+1})+\xi^{n+1}h'(\phi^{*,n+1})d_a^{*,n+1} \big)\mathcal{C}^{n+1} + \frac{\lambda}{\Delta t} \int_{\partial \Omega} d_a^{n+1} \big( \gamma_0 \phi^{n+1}-\hat \phi \big)\\
&+\int_{\partial \Omega} \Big[-P^{n+1}\bs n +\bar \mu^{n+1} \bs n \cdot \mathcal{D}(\bs u^{n+1})-\frac{1}{2}\bar \rho^{n+1}\big(\bs n \cdot \bs w^{n+1}  \big)\bs w^{n+1}-\frac{1}{2}\big(\bs n \cdot \tilde {\bs J}^{n+1} \big) \bs w^{n+1}   \Big]\cdot \bs w^{n+1},
\end{split}
\end{equation}
where we have used the boundary conditions
\eqref{eq:udiscrete}-\eqref{eq:phidiscrete2} and \eqref{eq:divp}-\eqref{eq:ndotC}.

Note the following relations for a generic variable $\chi$,
\begin{equation*}
\left\{
\begin{split}
&\chi^{n+1}(\gamma_0 \chi^{n+1}-\hat \chi)=\frac{1}{2}\Big( |\chi^{n+1}|^2-|\chi^n|^2+|\chi^{n+1}-\chi^n|^2  \Big) \\
& (\chi^{n+1}-\chi^{*,n+1})(\gamma_0 \chi^{n+1}-\hat \chi)=|\chi^{n+1}-\chi^n|^2,\;\;\text{for}\;\;J=1,
\end{split}
\right.
\end{equation*}
and
\begin{equation*}
\left\{
\begin{split}
& \chi^{n+1} \big(\gamma_0 \chi^{n+1}-\hat \chi\big)=\frac{1}{4}\Big(|\chi^{n+1}|^2-|\chi^n|^2   \Big)+\frac{1}{4} |\chi^{n+1}-2 \chi^n +\chi^{n-1}  |^2+\frac{1}{4}\Big( |2\chi^{n+1}-\chi^n|^2 -|2\chi^n-\chi^{n-1}|^2  \Big) \\
& (\chi^{n+1}-\chi^{*,n+1})(\gamma_0 \chi^{n+1}-\hat \chi)=\frac{1}{2}\Big(| \chi^{n+1}-\chi^n |^2-| \chi^{n}-\chi^{n-1} |^2   \Big)+|\chi^{n+1}-2 \chi^n+\chi^{n-1} |^2,\;\;\text{for}\;\;J=2.
\end{split}
\right.
\end{equation*}
Combining the above relations with equation \eqref{eq:Rid2} and letting  $\bs f^{n+1}=\bs 0,$ $g^{n+1}=0$ in $\Omega,$ and $\bs w^{n+1}=\bs 0,$ $d_a^{n+1}=d_b^{n+1}=0$
on $\partial \Omega,$ all the boundary terms vanish, and equation \eqref{eq:Rid2} is transformed into 
\begin{equation}\label{eq:stabJ1}
\begin{split}
& \frac{S}{\Delta t} \| \phi^{n+1} -\phi^n\|^2 +\frac{\lambda}{2 \Delta t} \Big( \| \nabla \phi^{n+1} \|^2 +\|\nabla \phi^{n+1}-\nabla \phi^n  \|^2  \Big)+\frac{1}{\Delta t}\Big( |R^{n+1}|^2 +|R^{n+1}-R^n|^2 \Big)\\
&=\frac{\lambda }{2\Delta t}\| \nabla \phi^n\|^2+\frac{1}{\Delta t}|R^n|^2-\frac{1}{2}\int_{\Omega} \bar \mu^{n+1}\|\mathcal{D}(\bs u^{n+1}) \|^2-\int_{\Omega}m |\nabla \mathcal{C}^{n+1}|^2\;\;\text{for}\;\;J=1,
\end{split}
\end{equation}
and
\begin{equation}\label{eq:stabJ2}
\begin{split}
&\frac{S}{\Delta t}\Big( \frac{1}{2}\big\| \phi^{n+1}-\phi^n \big\|^2+\big\| \phi^{n+1}-2\phi^n+\phi^{n-1}\big\|^2  \Big) \\
&+\frac{\lambda}{4 \Delta t}\Big(  \big\|\nabla \phi^{n+1} \big\|^2 + \big\| \nabla \phi^{n+1}-2\nabla \phi^n +\nabla \phi^{n-1}\big\|^2 +\big\|2\nabla \phi^{n+1}-\nabla \phi^n \big\|^2  \Big)\\
&+\frac{1}{2\Delta t}\Big(\big|R^{n+1} \big|^2+ \big|R^{n+1}-2 R^n +R^{n-1}  \big|^2  + \big|2R^{n+1}-R^n\big|^2 \Big)\\
&=\frac{\lambda}{4\Delta t} \Big( \big\|\nabla \phi^n \big\|^2 +\big\|2 \nabla \phi^n-\nabla \phi^{n-1} \big\|^2  \Big)+\frac{S}{2\Delta t}\big\|\phi^n- \phi^{n-1} \big\|^2 \\
&\frac{1}{2\Delta t}\Big(\big| R^n \big|^2+\big| 2R^n-R^{n-1} \big|^2   \Big)-\frac{1}{2}\int_{\Omega} \bar \mu^{n+1}\|\mathcal{D}(\bs u^{n+1}) \|^2-\int_{\Omega}m |\nabla \mathcal{C}^{n+1}|^2\;\;\text{for}\;\;J=2.
\end{split}
\end{equation}
The energy stability result in Theorem \ref{thm:stability} can be obtained directly from the above equations \eqref{eq:stabJ1}-\eqref{eq:stabJ2} by defining the discrete energy and dissipation in equations \eqref{eq:disenergy}-\eqref{eq:dissipation}.
\end{proof}

\subsection{Efficient Solution Algorithm}

We next consider how to implement the algorithm represented by the equations \eqref{eq:En1}-\eqref{eq:phidiscrete2}.  Although $E^{n+1},$ $R^{n+1}$ as well as $\xi^{n+1}$ are implicit and $E^{n+1}$ involves the integral of the unknown field functions
$\bs u^{n+1}$ and $\phi^{n+1}$  over the domain,
the scheme can be implemented in an efficient way. 

Combining equations \eqref{eq:scheme5eq} and \eqref{eq:scheme3eq} leads to
\begin{equation}\label{eq:implem1}
\frac{\gamma_0 \phi^{n+1}-\hat \phi}{\Delta t}+\xi^{n+1}   \big(\bs u^{*,n+1}\cdot \nabla \phi^{*,n+1} \big)=m \nabla^2\Big[ -\lambda \nabla^2 \phi^{n+1}+S\big( \phi^{n+1}- \phi^{*,n+1} \big)      +\xi^{n+1} h(\phi^{*,n+1}) \Big]+ g^{n+1}.
\end{equation}
This equation can be re-written as
\begin{equation}\label{eq:phi4order}
\frac{\gamma_0}{\lambda m \Delta t }\phi^{n+1}+\nabla^2 \big( \nabla^2 \phi^{n+1}\big)-\frac{S}{\lambda}\nabla^2 \phi^{n+1}=Z_1^n+\xi^{n+1} Z_2^n,
\end{equation}
where
\begin{equation}\label{eq:Z1Z2}
Z_1^n=\frac{1}{\lambda m}\big(  \frac{ \hat \phi}{\Delta t}     -mS \nabla^2  \phi^{*,n+1}  + g^{n+1} \big),\quad Z_2^n=\frac{1}{\lambda m}\Big(  m \nabla^2 h(\phi^{*,n+1} )-   \bs u^{*,n+1}\cdot \nabla \phi^{*,n+1}   \Big).
\end{equation}

Barring the unknown scalar-valued variable $\xi^{n+1},$
equation \eqref{eq:phi4order} is a fourth-order equation about $\phi^{n+1},$ which can be  transformed into two decoupled Helmholtz-type equations (see e.g.~\cite{YueFLS2004,DongS2012}).
Basically, by adding/subtracting a term $\alpha \nabla^2 \phi^{n+1}$ ($\alpha$ denoting a constant to be determined) on the left hand side (LHS), we can transform equation \eqref{eq:phi4order} into
\begin{equation}\label{eq:phidec}
 \nabla^2 \Big(\nabla^2 \phi^{n+1}+\alpha \phi^{n+1}\Big)-\Big( \alpha+\frac{S}{\lambda}  \Big)\Big[ \nabla^2 \phi^{n+1}-\frac{\gamma_0}{m \Delta t (S+\alpha \lambda)} \phi^{n+1} \Big]=Z_1^n+\xi^{n+1} Z_2^n.
\end{equation}
By requiring that
$ 
\alpha=-\frac{\gamma_0}{m \Delta t (S+\alpha \lambda)},
$ 
we obtain 
\begin{equation}\label{eq:alpha}
\alpha= \frac{1}{2\lambda}\bigg[  -S +\sqrt{S^2-4\frac{\gamma_0 \lambda}{m \Delta t}} \bigg], \quad \text{and the condition}\;\; S\geq \sqrt{\frac{4\gamma_0 \lambda}{m \Delta t}}.
\end{equation}
The chosen constant $S$ must satisfy the above condition.
Therefore, equation \eqref{eq:phidec} can be written equivalently as
\begin{subequations}\label{eq:phidecsys}
\begin{align}
&\nabla^2 \psi^{n+1}-\Big( \alpha+\frac{S}{\lambda} \Big)\psi^{n+1}=Z_1^n+\xi^{n+1}Z_2^n,  \label{eq:psieq}  \\
&\nabla^2 \phi^{n+1} + \alpha \phi^{n+1}=\psi^{n+1},\label{eq:phieq}
\end{align}
\end{subequations}
where $\psi^{n+1}$ is an auxiliary variable defined by equation \eqref{eq:phieq}. Note that if the scalar variable $\xi^{n+1}$ is given, equations \eqref{eq:psieq} and \eqref{eq:phieq} are decoupled. One can solve equation \eqref{eq:psieq} for $\psi^{n+1},$ and then solve equation \eqref{eq:phieq} for $\phi^{n+1}.$

Correspondingly, the boundary condition \eqref{eq:phidiscrete2} can be transformed into
\begin{equation}\label{eq:bcpsi}
\bs n \cdot \nabla \psi^{n+1}=\alpha d_a^{n+1}+d_b^{n+1},\;\;\text{on}\;\;\partial \Omega,
\end{equation}
where we have used equation \eqref{eq:phieq} and \eqref{eq:phidiscrete1}.

Taking advantage of the fact that $\xi^{n+1}$ is a scalar number,
instead of a field function, we introduce two sets of phase field functions $(\psi_i^{n+1},\phi_i^{n+1}),\;i=1,2 $  as solutions to the
following two Helmholtz-type problems:
\begin{subequations}
\begin{align}
&\nabla^2 \psi_i^{n+1}-\Big( \alpha+\frac{S}{\lambda} \Big)\psi_i^{n+1}=Z_i^n,\;\;i=1,2,\quad  \bs n \cdot \nabla \psi_1^{n+1}=\alpha d_a^{n+1}+d_b^{n+1}, \quad \bs n \cdot \nabla \psi_2^{n+1}=0,\;\; \text{on}\;\;\partial \Omega;\label{eq:psieqdecomp}\\
&\nabla^2 \phi_i^{n+1}+\alpha \phi_i^{n+1}=\psi_i^{n+1},\;\; i=1,2,\quad  \bs n \cdot \nabla \phi_1^{n+1}=d_a^{n+1},\quad \bs n\cdot \nabla \phi_2^{n+1}=0,\;\; \text{on}\;\;\partial \Omega.\label{eq:phieqdecomp}
\end{align}
\end{subequations}
Then we have the following result.
\begin{theorem}
\label{thm:thm_2}

Given scalar value $\xi^{n+1}$,
the following field functions solve the system consisting of equations
\eqref{eq:psieq}-\eqref{eq:bcpsi} and \eqref{eq:phidiscrete1}:
\begin{subequations}
  \begin{equation}
    \psi^{n+1} = \psi_1^{n+1} + \xi^{n+1}\psi_2^{n+1}, 
    \label{equ:psi_expr}
  \end{equation}
  \begin{equation}
    \phi^{n+1} = \phi_1^{n+1} + \xi^{n+1}\phi_2^{n+1}, 
    \label{equ:phi_expr}
  \end{equation}
\end{subequations}
where $(\psi_i^{n+1},\phi_i^{n+1})$ ($i=1,2$) are given by
equations \eqref{eq:psieqdecomp}--\eqref{eq:phieqdecomp}.
\end{theorem}

It should be noted that $\psi_1^{n+1}$ and $\psi_2^{n+1}$
defined in \eqref{eq:psieqdecomp} are not coupled,
and $\phi_1^{n+1}$ and $\phi_2^{n+1}$ defined in
\eqref{eq:phieqdecomp} are not coupled either.
One can first compute $\psi_1^{n+1}$ and
$\psi_2^{n+1}$ from \eqref{eq:psieqdecomp}, and then
compute $\phi_1^{n+1}$ and $\phi_2^{n+1}$ from \eqref{eq:phieqdecomp}.

The corresponding weak formulations for $(\psi_i^{n+1},\phi_i^{n+1})$ ($i=1,2$)
can be obtained by taking the $L^2$ inner product between equations
\eqref{eq:psieqdecomp}--\eqref{eq:phieqdecomp} and a test function, and they
are as follows:
\\
\noindent\underline{Find $\psi_1^{n+1} \in H^1(\Omega)$, such that}
\begin{equation}\label{eq:weakpsi1}
\begin{split}
\int_{\Omega} \nabla \psi_1^{n+1} \cdot \nabla \varphi +&\Big( \alpha+ \frac{S}{\lambda}\Big)\int_{\Omega}\psi_1^{n+1} \varphi=-\frac{S}{\lambda}\int_{\Omega}  \nabla \phi^{*,n+1}\cdot \nabla \varphi  -\frac{1}{\lambda m }\int_{\Omega} \Big(  \frac{\hat \phi}{\Delta t}+  g^{n+1}    \Big)\varphi  \\
&+\int_{\partial \Omega}\Big[\frac{S}{\lambda} d_a^{*,n+1}+\alpha d_a^{n+1}+d_b^{n+1} \Big]\varphi,\quad \forall \varphi \in H^1(\Omega).
\end{split}
\end{equation}
\noindent\underline{Find $\phi_1^{n+1} \in H^1(\Omega)$, such that}
\begin{equation}\label{eq:weakphi1}
\int_{\Omega} \nabla \phi_1^{n+1}\cdot \nabla \varphi -\alpha \int_{\Omega} \phi_1^{n+1} \varphi=-\int_{\Omega} \psi_1^{n+1} \varphi +\int_{\partial \Omega} d_a^{n+1} \varphi,\quad \forall \varphi \in H^1(\Omega).
\end{equation}
\noindent\underline{Find $\psi_2^{n+1} \in H^1(\Omega)$, such that}
\begin{equation}\label{eq:weakpsi2}
\begin{split}
\int_{\Omega} \nabla \psi_2^{n+1}& \cdot \nabla \varphi +\Big( \alpha+ \frac{S}{\lambda}\Big)\int_{\Omega}\psi_2^{n+1} \varphi= 
 \frac{1}{\lambda}\int_{\Omega} \nabla  h(\phi^{*,n+1}  )\cdot \nabla \varphi \\
& + \frac{1}{ \lambda m} \int_{\Omega} \big(\bs u^{*,n+1}  \cdot \nabla \phi^{*,n+1} \big)\varphi -\frac{1}{\lambda}\int_{\partial \Omega} h'(\phi^{*,n+1})d_a^{*,n+1}   ,\quad \forall \varphi \in H^1(\Omega).
\end{split}
\end{equation}
\noindent\underline{Find $\phi_2^{n+1} \in H^1(\Omega)$, such that}
\begin{equation}\label{eq:weakphi2}
\int_{\Omega} \nabla \phi_2^{n+1}\cdot \nabla \varphi -\alpha \int_{\Omega} \phi_2^{n+1} \varphi=-\int_{\Omega} \psi_2^{n+1} \varphi ,\quad \forall \varphi \in H^1({\Omega}).
\end{equation}

We define $\bar \phi^{n+1}$
as a $J$-th order approximation of $\phi^{n+1}$ given by
\begin{equation}\label{eq:phibar}
\bar \phi^{n+1}=\phi_1+ \phi_2.
\end{equation}
Then $\bar \rho^{n+1},$ $\bar \mu^{n+1}$ and $\bar {\tilde {\bs J}}^{n+1}$ can be evaluated accordingly by equation \eqref{eq:rhomuJ}.

Given $\{ \psi_i^{n+1}, \phi_i^{n+1} \}\;(i=1,2),$ $\mathcal{C}^{n+1}$ in \eqref{eq:scheme5eq} can be decomposed into
\begin{equation}\label{eq:Ctwoparts}
\mathcal{C}^{n+1}=\mathcal{C}^{n+1}_1+\xi^{n+1} \mathcal{C}^{n+1}_2,
\end{equation}
where by equations \eqref{eq:phieqdecomp} and
\eqref{equ:psi_expr}-\eqref{equ:phi_expr} we have
\begin{subequations}\label{eq:C1C2}
\begin{align}
& \mathcal{C}^{n+1}_1=-\lambda \psi_1^{n+1}+(\alpha \lambda +S)\phi_1^{n+1}-S \phi^{*,n+1}, \label{eq:C1}\\
&\mathcal{C}^{n+1}_2=-\lambda \psi_2^{n+1} +(\alpha \lambda +S)\phi_2^{n+1}+h(\phi^{*,n+1}),
\end{align}
\end{subequations}

Substituting \eqref{eq:Ctwoparts} into \eqref{eq:scheme1eq} leads to
\begin{equation}\label{eq:pu}
\begin{split}
 \nabla P^{n+1}=&-\rho_0\nu_m   \nabla \times \nabla \times \bs u  ^{n+1}- \frac{\gamma_0\rho_0}{\Delta t} \bs u  ^{n+1}+\rho_0\Big\{ \frac{\hat {\bs u}}{\Delta t}+\frac{\mathcal{C}^{n+1}_1 }{\bar \rho^{n+1}} \nabla \bar{\phi}^{n+1} +\frac{\bs f^{n+1}}{\bar \rho^{n+1}} \Big \}\\
&+\xi^{n+1}\rho_0\Big\{ \frac{\mathcal{C}^{n+1}_2 }{\bar \rho^{n+1}} \nabla \bar{\phi}^{n+1}-{\bs{{Q}}}- \Big(\frac{\bar \mu ^{n+1}}{\bar \rho ^{n+1}}-\nu_m \Big)\nabla \times \nabla \times \bs u^{*,n+1}  \Big \},
\end{split}
\end{equation}
where we have used equation \eqref{eq:scheme2eq}.
Let $q(\bs x)$ denote an arbitrary test function in the continuous space. 
Take the $L^2$-inner product between $\nabla q$ and equation \eqref{eq:pu}, and we obtain
\begin{equation}\label{eq:weakpold}
\begin{split}
\int_{\Omega}& \nabla P^{n+1}\cdot \nabla q=-\frac{\rho_0  \gamma_0}{\Delta t} \int_{\partial \Omega} \bs n \cdot \bs w^{n+1} q-\rho_0 \nu_m \int_{\partial \Omega} \bs n \times {\bs \omega}^{n+1} \cdot \nabla q\\
&+ \rho_0 \int_{\Omega} \Big( \frac{\hat {\bs u}}{\Delta t}+\frac{\mathcal{C}^{n+1}_1 }{\bar \rho^{n+1}} \nabla \bar{\phi}^{n+1} +\frac{\bs f^{n+1}}{\bar \rho^{n+1}} \Big)\cdot \nabla q - \xi ^{n+1} \rho_0  \int_{\partial \Omega} \Big( \frac{\bar \mu^{n+1}}{\bar \rho^{n+1}}-\nu_m   \Big)\bs n \times {\bs \omega}^{*,n+1}\cdot \nabla q \\
&+\xi^{n+1} \rho_0\int_{\Omega}  \Big( \frac{\mathcal{C}^{n+1}_2 }{ \bar \rho^{n+1}} \nabla \bar{\phi}^{n+1}-\bs Q +\nabla \Big( \frac{\bar \mu^{n+1}}{\bar \rho^{n+1}} \Big)\times {\bs \omega}^{*,n+1}   \Big) \cdot \nabla q ,\;\;\; \forall q
\end{split}
\end{equation}
where  $\bs \omega=\nabla \times \bs u$ is the vorticity, and we have used integration by part, the divergence theorem, equations \eqref{eq:scheme2eq} and \eqref{eq:udiscrete}, and the identity
\begin{equation}\label{eq:vorticityeq}
\frac{\mu}{\rho}\nabla \times  {\bs \omega} \cdot \nabla q=\nabla \cdot \Big( \frac{\mu}{\rho} {\bs \omega} \times \nabla q  \Big)-\nabla\Big( \frac{\mu}{\rho}  \Big)\times {\bs \omega} \cdot \nabla q.
\end{equation}

Note that in equation \eqref{eq:weakpold},
$P^{n+1}$ is weakly coupled with $\bs u^{n+1}$ as the boundary integral
involves $\bs \omega^{n+1}=\nabla \times \bs u  ^{n+1}.$
We have implemented two methods to solve equation \eqref{eq:weakpold}:
\begin{itemize}

\item
  We solve equation \eqref{eq:weakpold}, and the velocity equation
  \eqref{eq:weaku1} below, via a sub-iteration procedure (fixed-point iteration).
  We start by setting
  $\bs n \times \bs\omega^{n+1}|_{\partial\Omega}
  =\bs n \times\bs\omega^{*,n+1}|_{\partial\Omega}$ on the right hand side (RHS)
  of \eqref{eq:weakpold},
  and solve  \eqref{eq:weakpold} and \eqref{eq:weaku1} alternately
  until convergence. See Remark \ref{rem:coupled} below for details.

\item
  We make the following approximation on the boundary,
  \begin{equation}
  \bs n\times \bs \omega^{n+1}|_{\partial {\Omega}}\approx
\frac{R^{n+1}}{\sqrt{E^{n+1}}} \bs n \times \bs \omega^{*,n+1}|_{\partial \Omega} =
\xi^{n+1} \bs n \times \bs \omega^{*,n+1}|_{\partial \Omega}.
\label{eq:vort_approx}
\end{equation}
Then, except for the scalar number $\xi^{n+1}$, the computation for
$P^{n+1}$ from equation \eqref{eq:weakpold}
is not coupled with the computation for the velocity $\bs u^{n+1}$.
This implementation is similar to the first step in
the sub-iteration procedure (but without sub-iteration).
How then to deal with $\xi^{n+1}$ in equations \eqref{eq:weakpold}
and \eqref{eq:weaku1} with the approximation \eqref{eq:vort_approx}
is described below.

\end{itemize}
We observe that there is 
essentially no difference in the simulation results obtained
with these two implementations, and that the approximation
\eqref{eq:vort_approx} on the boundary does not
weaken the stability of the method, e.g.~the ability to use large
$\Delta t$ in simulations. 
In Section \ref{sec:tests} we will provide numerical results
comparing these two methods,
and show that there is very little and basically no difference in their
simulation results.
Because the method with \eqref{eq:vort_approx}
is much simpler and produces the same results as the sub-iteration
procedure, the majority of simulations
in Section \ref{sec:tests} and the simulations there by default
(unless otherwise specified)
are performed using this method.

Let us now consider how to deal with $\xi^{n+1}$
in \eqref{eq:weakpold}, with the approximation \eqref{eq:vort_approx}.
The following discussions also pertain to the
implementation with the sub-iteration procedure.
With the approximation \eqref{eq:vort_approx},
we can transform \eqref{eq:weakpold} into
\begin{equation}\label{eq:Psolver}
\begin{split}
&\int_{\Omega}\nabla P^{n+1}\cdot \nabla q= \rho_0 \int_{\Omega} \Big( \frac{\hat {\bs u}}{\Delta t}+\frac{\mathcal{C}^{n+1}_1 }{\bar \rho^{n+1}} \nabla \bar{\phi}^{n+1} +\frac{\bs f^{n+1}}{\bar \rho^{n+1}} \Big)\cdot \nabla q -\frac{\rho_0  \gamma_0}{\Delta t} \int_{\partial \Omega} \bs n \cdot \bs w^{n+1} q\\
&+\xi^{n+1}\Big\{ \rho_0\int_{\Omega}  \Big( \frac{\mathcal{C}^{n+1}_2 }{ \bar \rho^{n+1}} \nabla \bar{\phi}^{n+1}-\bs Q +\nabla \Big( \frac{\bar \mu^{n+1}}{\bar \rho^{n+1}} \Big)\times {\bs \omega}^{*,n+1}   \Big) \cdot \nabla q  - \rho_0  \int_{\partial \Omega} \frac{\bar \mu^{n+1}}{\bar \rho^{n+1}}\bs n \times {\bs \omega}^{*,n+1}\cdot \nabla q  \Big \},\;\;\; \forall q.
\end{split}
\end{equation}
Thanks to the fact that $\xi^{n+1}$ is a scalar number (not a field function),
we can solve equation \eqref{eq:Psolver} as follows. Define $(P_1^{n+1},P_2^{n+1})$  as solutions to the following two Poisson-type problems:
\\
\noindent\underline{Find $P_1^{n+1} \in \mathbb{V}(\Omega)=\big \{  v\in H^1(\Omega),\int_{\Omega}v=0  \big \}$ such that}
  \begin{equation}
  \int_{\Omega}\nabla P_1^{n+1}\cdot \nabla q= \rho_0 \int_{\Omega} \Big( \frac{\hat {\bs u}}{\Delta t}+\frac{\mathcal{C}^{n+1}_1 }{\bar \rho^{n+1}} \nabla \bar{\phi}^{n+1} +\frac{\bs f^{n+1}}{\bar \rho^{n+1}} \Big)\cdot \nabla q -\frac{\rho_0  \gamma_0}{\Delta t} \int_{\partial \Omega} \bs n \cdot \bs w^{n+1} q,\;\;\; \forall q \in H^1(\Omega)
    \label{equ:P_1_equ}
  \end{equation}
\noindent\underline{Find $P_2^{n+1} \in \mathbb{V}(\Omega)$ such that}
  \begin{equation}
  \begin{split}
   \int_{\Omega}\nabla P_2^{n+1}\cdot \nabla q=& \rho_0\int_{\Omega}  \Big( \frac{\mathcal{C}^{n+1}_2 }{ \bar \rho^{n+1}} \nabla \bar{\phi}^{n+1}-\bs Q +\nabla \Big( \frac{\bar \mu^{n+1}}{\bar \rho^{n+1}} \Big)\times {\bs \omega}^{*,n+1}   \Big) \cdot \nabla q \\
   &  - \rho_0  \int_{\partial \Omega} \frac{\bar \mu^{n+1}}{\bar \rho^{n+1}}\bs n \times {\bs \omega}^{*,n+1}\cdot \nabla q,\quad \forall q \in H^1(\Omega). \label{equ:P_2_equ}
   \end{split}
  \end{equation}
Then it is straightforward to verify that the solution to equation \eqref{eq:Psolver} is given by
\begin{equation}\label{eq:p1p2}
P^{n+1}=P_1^{n+1}+\xi^{n+1} P_2^{n+1},
\end{equation}
where $\xi^{n+1}$ is to be determined. 

In light of the equations \eqref{eq:p1p2} and \eqref{eq:Ctwoparts},
we can transform equation \eqref{eq:scheme1eq} into 
\begin{equation}\label{eq: ueqnew}
\begin{split}
\frac{\gamma_0}{\nu_m \Delta t} \bs u  ^{n+1}&-\nabla^2 \bs u  ^{n+1}=\frac{1}{\nu_m}\Big [ -\frac{1}{\rho_0} \nabla P_1^{n+1}+ \frac{\hat {\bs u}}{\Delta t}+\frac{\mathcal{C}^{n+1}_1 }{\bar \rho^{n+1}} \nabla \bar{\phi}^{n+1} +\frac{\bs f^{n+1}}{\bar \rho^{n+1}}   \Big]\\
&+\xi^{n+1} \frac{1}{\nu_m}\Big[ -\frac{1}{\rho_0} \nabla P_2^{n+1} + \frac{\mathcal{C}^{n+1}_2 }{\bar \rho^{n+1}} \nabla \bar{\phi}^{n+1}-\bs Q  -\Big(\frac{\bar \mu ^{n+1}}{\bar \rho ^{n+1}}-\nu_m \Big)\nabla \times  \bs \omega^{*,n+1}  \Big) \Big].
\end{split}
\end{equation}
Let $\varphi(\bs x)$ denote an arbitrary test function that vanishes on $\partial \Omega,$ i.e. $\varphi|_{\partial \Omega}=0.$ Taking the $L^2$ inner product between $\varphi$ and the equation \eqref{eq: ueqnew}, we obtain the weak formulation for $\bs u^{n+1}:$
\begin{equation}\label{eq:weaku1}
\begin{split}
\frac{\gamma_0}{\nu_m \Delta t} \int_{\Omega} \bs u^{n+1} \varphi  &+\int_{\Omega}\nabla \bs u^{n+1}\cdot \nabla \varphi  =\frac{1}{\nu_m} \int_{\Omega}\Big ( -\frac{1}{\rho_0} \nabla P_1^{n+1}+ \frac{\hat {\bs u}}{\Delta t}+\frac{\mathcal{C}^{n+1}_1 }{\bar \rho^{n+1}} \nabla \bar{\phi}^{n+1} +\frac{\bs f^{n+1}}{\bar \rho^{n+1}}   \Big)  \varphi\\
&+\xi^{n+1} \frac{1}{\nu_m}\Big\{\int_{\Omega} \Big(-\frac{1}{\rho_0} \nabla P_2^{n+1} + \frac{\mathcal{C}^{n+1}_2 }{\bar \rho^{n+1}} \nabla \bar{\phi}^{n+1}-\bs Q+\nabla \Big(\frac{\bar \mu ^{n+1}}{\bar \rho ^{n+1}}\Big)\times \bs \omega^{*,n+1}\Big)\varphi  \\
&\qquad\qquad \qquad  - \int_{\Omega}\Big(\frac{\bar \mu ^{n+1}}{\bar \rho ^{n+1}}-\nu_m\Big)\bs \omega^{*,n+1}\times \nabla \varphi  \Big\},\;\;\; \forall \varphi \in H_0^1(\Omega),
\end{split}
\end{equation}
where we have used equation \eqref{eq:vorticityeq}.

Again we exploit the fact that $\xi^{n+1}$ is a scalar number, and equation \eqref{eq:weaku1} can be solved as follows.  Introducing two auxiliary velocity field functions $(\bs u_1^{n+1},\bs u_2^{n+1})$ as the solutions of the following Helmholtz-type problems:
\\
\noindent\underline{Find $\bs u_1^{n+1} \in [H^1(\Omega)]^d$ such that}
\begin{subequations}\label{eq:u1solve}
  \begin{equation}
\frac{\gamma_0}{\nu_m \Delta t} \int_{\Omega} \bs u_1^{n+1} \varphi  +\int_{\Omega}\nabla \bs u_1^{n+1}\cdot \nabla \varphi =\frac{1}{\nu_m} \int_{\Omega}\Big ( -\frac{1}{\rho_0} \nabla P_1^{n+1}+ \frac{\hat {\bs u}}{\Delta t}+\frac{\mathcal{C}^{n+1}_1 }{\bar \rho^{n+1}} \nabla \bar{\phi}^{n+1} +\frac{\bs f^{n+1}}{\bar \rho^{n+1}}   \Big)  \varphi,\; \forall \varphi\in H_0^1(\Omega),
    \label{equ:u_1_equ}
  \end{equation}
  \begin{equation}
 \bs u_1^{n+1}=\bs w^{n+1},\;\;\text{on}\;\;\partial \Omega.    \label{equ:u_1_cons}
  \end{equation}
\end{subequations}
\noindent\underline{Find $\bs u_2^{n+1}\in [H_0^1(\Omega)]^d$ such that}
  \begin{equation}\label{eq:u2solve}
  \begin{split}
 \frac{\gamma_0}{\nu_m \Delta t} \int_{\Omega} \bs u_2^{n+1}& \varphi   +\int_{\Omega}\nabla \bs u_2^{n+1}\cdot \nabla \varphi = -  \frac{1}{\nu_m} \int_{\Omega}\Big(\frac{\bar \mu ^{n+1}}{\bar \rho ^{n+1}}-\nu_m\Big)\bs \omega^{*,n+1}\times \nabla \varphi \\
& +  \frac{1}{\nu_m}\int_{\Omega} \Big(-\frac{1}{\rho_0} \nabla P_2^{n+1} + \frac{\mathcal{C}^{n+1}_2 }{\bar \rho^{n+1}} \nabla \bar{\phi}^{n+1}-\bs Q+\nabla \Big(\frac{\bar \mu ^{n+1}}{\bar \rho ^{n+1}}\Big)\times \bs \omega^{*,n+1}\Big)\varphi,  \;\; \forall \varphi\in H_0^1(\Omega),
  \end{split}
  \end{equation}
where $H_0^1(\Omega) =\{ v\in H^1(\Omega), v|_{\partial\Omega}=0  \}$.
It is straightforward  to verify that the solution
$\bs u^{n+1}$ to equations \eqref{eq:weaku1} and \eqref{eq:udiscrete} is computed by
\begin{equation}\label{eq:u1u2}
\bs u^{n+1}=\bs u_1^{n+1}+\xi^{n+1} \bs u_2^{n+1},
\end{equation}
provided that the scalar variable $\xi^{n+1}$ is given.

The scalar value $\xi^{n+1}$ still needs to be determined.
Note that equation \eqref{eq:scheme6eq}
can be transformed into equation \eqref{eq:Rid2}, and this equation
  can be written as
\begin{equation}\label{eq:Rid3}
\begin{split}
&\frac{2\xi^{n+1} }{\Delta t} R^{n+1} \big(\gamma_0 R^{n+1}-\hat R\big)+\frac{\xi^{n+1}}{2}\int_{\Omega} \bar \mu^{n+1}\|\mathcal{D}(\bs u^{n+1}) \|^2 +m\xi^{n+1}\int_{\Omega} |\nabla \mathcal{C}^{n+1}|^2 -\xi^{n+1}\int_{\Omega}\bs f^{n+1}\cdot \bs u^{n+1}  \\
&-\xi^{n+1}\int_{\Omega}g^{n+1}\mathcal{C}^{n+1}   +\frac{\lambda \xi^{n+1}}{\Delta t}\int_{\Omega} \nabla \phi^{n+1}\cdot \nabla\big( \gamma_0  \phi^{n+1}- \hat \phi  \big)+ \frac{S \xi^{n+1}}{\Delta t}  \int_{\Omega} \big(\phi^{n+1}-\phi^{*,n+1}\big)\big(\gamma_0 \phi^{n+1}-\hat \phi \big)\\
&+\xi^{n+1}\int_{\partial \Omega}m \big( \lambda d_b^{n+1}-S(d_a^{n+1}-d_a^{*,n+1})-\xi^{n+1}h'(\phi^{*,n+1})d_a^{*,n+1}      \big)\mathcal{C}^{n+1} - \frac{\lambda \xi^{n+1}}{\Delta t} \int_{\partial \Omega}d_a^{n+1} \big( \gamma_0 \phi^{n+1}-\hat \phi \big)\\
&-\xi^{n+1}\int_{\partial \Omega} \Big[-P^{n+1}\bs n +\bar \mu^{n+1} \bs n \cdot \mathcal{D}(\bs u^{n+1})-\frac{1}{2}\bar \rho^{n+1}\big(\bs n \cdot \bs w^{n+1}  \big)\bs w^{n+1}-\frac{1}{2}\big(\bs n \cdot \tilde {\bs J}^{n+1} \big) \bs w^{n+1}   \Big]\cdot \bs w^{n+1}=0,
\end{split}
\end{equation}
where we have multiplied both sides of the equation by
$\frac{R^{n+1}}{\sqrt{E^{n+1}}}$, which we observe can
improve the robustness of the scheme.
This is a nonlinear equation with the unknown scalar variable $\xi^{n+1}$, and
it will be solved for $\xi^{n+1}.$
In light of equation \eqref{eq:Enp1}, we have
\begin{equation}\label{eq:Edecomp}
  \begin{split}
  E^{n+1}&=E(\xi^{n+1})=\int_{\Omega}\Big[ \frac{\bar \rho^{n+1}}{2} |\bs u^{n+1}|^2+F(\phi^{n+1}) \Big]d\Omega +C_0 \\
  &=A_0+A_1 \xi^{n+1}+A_2(\xi^{n+1})^2+A_3(\xi^{n+1})^3+ A_4 (\xi^{n+1})^4,
  \end{split}
\end{equation}
where 
\begin{equation}\label{eq:coeffA}
\begin{split}
&A_0=\int_{\Omega}\Big\{  \frac{1}{2}\bar \rho^{n+1} \big|\bs u_1^{n+1}\big|^2 +\frac{\lambda}{4\eta^2} \Big( 1-\big(\phi_1^{n+1}\big)^2  \Big)^2  \Big \}d\Omega+C_0,\\
&A_1=\int_{\Omega} \Big\{ \bar \rho^{n+1} \bs u_1^{n+1}\cdot \bs u_2^{n+1} -\frac{\lambda}{\eta^2}\Big( 1-\big(\phi_1^{n+1} \big)^2  \Big) \phi_1^{n+1} \phi_2^{n+1}   \Big \}d\Omega,\\
&A_2=\int_{\Omega} \Big\{ \frac{1}{2}\bar \rho^{n+1} \big| \bs u_2^{n+1} \big|^2  + \frac{\lambda}{4\eta^2} \Big(6\big(\phi_1^{n+1}\big)^2-2  \Big)\big( \phi_2^{n+1} \big)^2 \Big \}d\Omega,\\
  &A_3= \frac{\lambda}{\eta^2} \int_{\Omega} \phi_1^{n+1} \big(\phi_2^{n+1} \big)^3   d\Omega,\\
  &A_4=\frac{\lambda}{4\eta^2} \int_{\Omega} \big(\phi_2^{n+1} \big)^4d\Omega.
\end{split}
\end{equation}
In light of equations \eqref{eq:En1}, \eqref{eq:phidiscrete1}, \eqref{eq:Ctwoparts}, \eqref{eq:p1p2} and \eqref{eq:u1u2},  we can transform equation \eqref{eq:Rid3} into
\begin{equation}\label{eq:nonlinear}
F(\xi^{n+1})=\frac{2\gamma_0}{\Delta t}(\xi^{n+1})^3E(\xi^{n+1})-\frac{2\hat R}{\Delta t}(\xi^{n+1})^2\sqrt{E(\xi^{n+1})}+B_0 \xi^{n+1}+B_1(\xi^{n+1})^2+B_2(\xi^{n+1})^3=0,
\end{equation}
where $\hat R$ is defined by \eqref{equ:def_var_hat} and
\begin{equation}\label{eq:B0}
\begin{split}
B_0=&\frac{1}{2}\int_{\Omega} \bar \mu^{n+1}\|\mathcal{D}(\bs u_1^{n+1}) \|^2 +m\int_{\Omega} |\nabla \mathcal{C}_1^{n+1}|^2+\frac{\lambda}{\Delta t}\int_{\Omega} \nabla \phi_1^{n+1}\cdot \big( \gamma_0 \nabla \phi_1^{n+1}-\nabla \hat \phi  \big)\\
&+ \frac{S}{\Delta t} \int_{\Omega} \big(\phi_1^{n+1}-\phi^{*,n+1}\big)\big(\gamma_0 \phi_1^{n+1}-\hat \phi \big) -\int_{\Omega }\bs f^{n+1} \cdot \bs u_1^{n+1}-\int_{\Omega}g^{n+1}\mathcal{C}_1^{n+1}\\
&-m \int_{\partial \Omega} \big( Sd_a^{n+1}-\lambda d_b^{n+1}-S d_a^{*,n+1} \big)\mathcal{C}_1^{n+1}  -\frac{\lambda}{\Delta t} \int_{\partial \Omega} d_a^{n+1}\big(\gamma_0 \phi_1^{n+1}-\hat \phi \big) \\
&-\int_{\partial \Omega} \Big[-P_1^{n+1}\bs n +\bar \mu^{n+1} \bs n \cdot \mathcal{D}(\bs u_1^{n+1})-\frac{1}{2}\bar \rho^{n+1}|\bs w^{n+1}|^2\bs n\Big]\cdot  \bs w^{n+1}\\
&+\frac{J_0}{2}\int_{\partial \Omega} \Big\{  \frac{\lambda}{\eta^2}d_a^{n+1}\big(3(\phi_1^{n+1})^2-1\big)-\lambda d_b^{n+1}     \Big\}|\bs w^{n+1}|^2,
\end{split}
\end{equation}
\begin{equation}\label{eq:B1}
\begin{split}
B_1=&\int_{\Omega} \bar \mu^{n+1}\mathcal{D}(\bs u_1^{n+1}): \mathcal{D}(\bs u_2^{n+1})   +2m\int_{\Omega} \nabla \mathcal{C}_1^{n+1}\cdot \nabla \mathcal{C}_2^{n+1}-\int_{\Omega }\bs f^{n+1} \cdot \bs u_2^{n+1}-\int_{\Omega}g^{n+1}\mathcal{C}_2^{n+1}\\
& +\frac{\lambda}{\Delta t}\int_{\Omega}  \big( 2\gamma_0 \nabla \phi_1^{n+1}-\nabla \hat \phi  \big) \cdot \nabla \phi_2^{n+1}+ \frac{S}{\Delta t} \int_{\Omega} \big(2\gamma_0 \phi_1^{n+1}-\gamma_0 \phi^{*,n+1}-\hat \phi    \big)\phi_2^{n+1}  \\
&- m \int_{\partial \Omega}\Big\{ \big( Sd_a^{n+1}-\lambda d_b^{n+1}-S d_a^{*,n+1} \big)\mathcal{C}_2^{n+1} +h'( \phi^{*,n+1}) d_a^{*,n+1} \mathcal{C}_1^{n+1} \Big\}-\frac{\lambda \gamma_0}{\Delta t} \int_{\partial \Omega} d_a^{n+1}\phi_2^{n+1}\\
& -\int_{\partial \Omega} \Big[-P_2^{n+1}\bs n +\bar \mu^{n+1} \bs n \cdot \mathcal{D}(\bs u_2^{n+1})\Big]\cdot  \bs w^{n+1}
+  \frac{6\lambda J_0}{2\eta^2}\int_{\partial \Omega}  d_a^{n+1}\phi_1^{n+1} \phi_2^{n+1}   |\bs w^{n+1}|^2,
\end{split}
\end{equation}
\begin{equation}\label{eq:B2}
\begin{split}
B_2=&\frac{1}{2}\int_{\Omega} \bar \mu^{n+1}\|\mathcal{D}(\bs u_2^{n+1}) \|^2      +m\int_{\Omega} |\nabla \mathcal{C}_2^{n+1}|^2 +\frac{\lambda\gamma_0}{\Delta t}\int_{\Omega} |\nabla \phi_2^{n+1}|^2+ \frac{S\gamma_0}{\Delta t} \int_{\Omega} |\phi_2^{n+1}|^2\\
&  -m \int_{\partial \Omega}h'( \phi^{*,n+1}) d_a^{*,n+1}  \mathcal{C}_2^{n+1}+ \frac{3\lambda J_0}{2\eta^2} \int_{\partial \Omega} d_a^{n+1} (\phi_2^{n+1})^2 | \bs w^{n+1}|^2.
\end{split}
\end{equation}

To determine $\xi^{n+1}$, we solve the {\em scalar} nonlinear
equation \eqref{eq:nonlinear}  about $\xi^{n+1}$ using the Newton's method
with an initial guess $\xi^{n+1}=1$.
The cost of this computation is very small 
compared to the total cost  within a time step, because this equation is about
a scalar number (not a field function).
Once $\xi^{n+1}$ is known, $E^{n+1}$ can be computed based on equation \eqref{eq:Edecomp}, and $R^{n+1}$ can be computed using equation \eqref{eq:En1}. The phase field $\phi^{n+1},$ pressure $P^{n+1}$ and velocity $\bs u^{n+1}$
are then given by equations \eqref{equ:phi_expr},
\eqref{eq:p1p2} and \eqref{eq:u1u2}, respectively. 

\begin{remark}\label{rem:coupled}
  The weakly coupled equation \eqref{eq:weakpold} can be solved by a
  sub-iteration method.  In the first iteration step, we use $\bs \omega_{(0)}^{n+1}=\bs \omega^{*,n+1}$ to replace $\bs w^{n+1}$. As a result, we can solve $P^{n+1}_{(1)}$ and $\bs u_{(1)}^{n+1},$ as well as the vorticity $\bs \omega_{(1)}^{n+1}=\nabla \times \bs u_{(1)}^{n+1}$ by the algorithm discussed above. Then we replace $\bs \omega^{n+1}$ with $\bs \omega_{(1)}^{n+1}$  in equation \eqref{eq:weakpold} and repeat the
  iterations until $\|\bs \omega_{(k+1)}^{n+1}-\bs \omega_{(k)}^{n+1}  \|$ becomes
  sufficiently small. See Section \ref{sec:tests}
  (e.g.~Figure \ref{fig:subiter}) for numerical results with this
  method.
\end{remark}

\subsection{Spectral Element Implementation}

Let us now consider the spatial discretization of equations \eqref{eq:weakpsi1}-\eqref{eq:weakphi2}, \eqref{equ:P_1_equ}-\eqref{equ:P_2_equ}
and \eqref{eq:u1solve}-\eqref{eq:u2solve}. We discretize the domain $\Omega$ with a mesh consisting of $N_{el}$
conforming $C^0$ spectral elements~\cite{KarniadakisS2005}.
Let the positive integer $K$ denote the element order, which represents a measure of the highest polynomial degree in the polynomial expansions of the field variables within an element. Let $\Omega_h$ denote the discretized domain, $\partial \Omega_h$ denote the boundary of $\Omega_h,$ and $\Omega_h^e (1\leq e \leq N_{el})$ denote the element $e.$ Define three function spaces
\begin{equation}\label{Approxspace}
\begin{split}
&\mathbb{X}=\big\{ v\in H^1(\Omega_h): v\; \text {is a polynomial characterized by}  \;K\;\text{on}\;\Omega_h^e,\; \text{for}\;1\leq e \leq N_{el} \big \},\\
  &\mathbb{V}_0=\big \{ v \in \mathbb{X}: \int_{\Omega_h}\!\!vd\Omega_h=0 \big \}, \\
  & \mathbb{V}_1=\big \{  v \in \mathbb{X}: v|_{\partial \Omega_h}=0 \big \}.
\end{split}
\end{equation}
In the following, let $d\; (d=2\;\text{or}\;3)$ denote the spatial dimension, and the subscript in $(\cdot)_h$ denote the discretized version of the variable $(\cdot)$. The fully discretized equations corresponding to equations \eqref{eq:weakpsi1}-\eqref{eq:weakphi2} are:
\\
\noindent\underline{Find $\psi_{1h}^{n+1} \in \mathbb{X}$, such that}
\begin{equation}\label{eq:apppsi1}
\begin{split}
\int_{\Omega_h} \nabla \psi_{1h}^{n+1} \cdot \nabla \varphi_h +&\Big( \alpha+ \frac{S}{\lambda}\Big)\int_{\Omega_h}\psi_{1h}^{n+1} \varphi_h=-\frac{S}{\lambda}\int_{\Omega_h}  \nabla \phi^{*,n+1}_h \cdot \nabla \varphi_h  -\frac{1}{\lambda m }\int_{\Omega_h} \Big(  \frac{\hat \phi_h}{\Delta t}+  g_h^{n+1}    \Big)\varphi_h  \\
&+\int_{\partial \Omega_h}\Big[\frac{S}{\lambda} d_{ah}^{*,n+1}+\alpha d_{ah}^{n+1}+d_{bh}^{n+1} \Big]\varphi_h,\quad \forall \varphi_h \in \mathbb{X}.
\end{split}
\end{equation}
\noindent\underline{Find $\phi_{1h}^{n+1} \in \mathbb{X}$, such that}
\begin{equation}\label{eq:appphi1}
\int_{\Omega_h} \nabla \phi_{1h}^{n+1}\cdot \nabla \varphi_h -\alpha \int_{\Omega_h} \phi_{1h}^{n+1} \varphi_h=-\int_{\Omega_h} \psi_{1h}^{n+1} \varphi_h +\int_{\partial \Omega_h} d_{ah}^{n+1} \varphi_h,\quad \forall \varphi_h \in \mathbb{X},
\end{equation}
\noindent\underline{Find $\psi_{2h}^{n+1} \in \mathbb{X}$, such that}
\begin{equation}\label{eq:apppsi2}
\begin{split}
\int_{\Omega_h} \nabla \psi_{2h}^{n+1}& \cdot \nabla \varphi_h +\Big( \alpha+ \frac{S}{\lambda}\Big)\int_{\Omega_h}\psi_{2h}^{n+1} \varphi_h= 
 \frac{1}{\lambda}\int_{\Omega_h} \nabla  h(\phi_h^{*,n+1}  )\cdot \nabla \varphi_h \\
& + \frac{1}{ \lambda m} \int_{\Omega_h} \big(\bs u_h^{*,n+1}  \cdot \nabla \phi_h^{*,n+1} \big)\varphi_h -\frac{1}{\lambda}\int_{\partial \Omega_h} h'(\phi_h^{*,n+1})d_{ah}^{*,n+1}   ,\quad \forall \varphi_h \in \mathbb{X},
\end{split}
\end{equation}
\noindent\underline{Find $\phi_{2h}^{n+1} \in \mathbb{X}$, such that}
\begin{equation}\label{eq:appphi2}
\int_{\Omega_h} \nabla \phi_{2h}^{n+1}\cdot \nabla \varphi_h -\alpha \int_{\Omega_h} \phi_{2h}^{n+1} \varphi_h=-\int_{\Omega_h} \psi_{2h}^{n+1} \varphi_h ,\quad \forall \varphi_h \in \mathbb{X},
\end{equation}
\noindent
The fully discretized equations corresponding to \eqref{equ:P_1_equ}-\eqref{equ:P_2_equ} are:
\\
\noindent\underline{Find $P_{1h}^{n+1} \in \mathbb{V}_0$ such that}
  \begin{multline}
  \int_{\Omega_h}\nabla P_{1h}^{n+1}\cdot \nabla q_h= \rho_0 \int_{\Omega_h} \Big( \frac{\hat {\bs u}_h}{\Delta t}+\frac{\mathcal{C}^{n+1}_{1h} }{\bar \rho_h^{n+1}} \nabla \bar{\phi}_h^{n+1} +\frac{\bs f_h^{n+1}}{\bar \rho_h^{n+1}} \Big)\cdot \nabla q_h -\frac{\rho_0  \gamma_0}{\Delta t} \int_{\partial \Omega_h} \bs n_h \cdot \bs w_h^{n+1} q_h, \\ \forall q_h \in \mathbb{X}
    \label{equ:appP_1_equ}
  \end{multline}
\noindent\underline{Find $P_{2h}^{n+1} \in \mathbb{V}_0$ such that}
  \begin{equation}
  \begin{split}
   \int_{\Omega_h}\nabla P_{2h}^{n+1}\cdot \nabla q_h=& \rho_0\int_{\Omega_h}  \Big( \frac{\mathcal{C}^{n+1}_{2h} }{ \bar \rho_h^{n+1}} \nabla \bar{\phi}_h^{n+1}-\bs Q_h +\nabla \Big( \frac{\bar \mu_h^{n+1}}{\bar \rho_h^{n+1}} \Big)\times {\bs \omega_h}^{*,n+1}   \Big) \cdot \nabla q_h \\
   &  - \rho_0  \int_{\partial \Omega_h} \frac{\bar \mu_h^{n+1}}{\bar \rho_h^{n+1}}\bs n_h \times {\bs \omega_h}^{*,n+1}\cdot \nabla q_h,\quad \forall q_h \in \mathbb{X}. \label{equ:appP_2_equ}
   \end{split}
  \end{equation}
  The fully discretized equations corresponding
  to \eqref{eq:u1solve}-\eqref{eq:u2solve} are: 
  \\
\noindent\underline{Find $\bs u_1^{n+1} \in [\mathbb{X}]^d$ such that}
\begin{subequations}\label{eq:appu1solve}
  \begin{multline}
\frac{\gamma_0}{\nu_m \Delta t} \int_{\Omega_h} \bs u_{1h}^{n+1} \varphi_h  +\int_{\Omega_h}\nabla \bs u_{1h}^{n+1}\cdot \nabla \varphi_h =\frac{1}{\nu_m} \int_{\Omega_h}\Big ( -\frac{1}{\rho_0} \nabla P_{1h}^{n+1}+ \frac{\hat {\bs u}_h}{\Delta t}+\frac{\mathcal{C}^{n+1}_{1h} }{\bar \rho_h^{n+1}} \nabla \bar{\phi}_h^{n+1} +\frac{\bs f_h^{n+1}}{\bar \rho_h^{n+1}}   \Big)  \varphi_h, \\ \forall \varphi_h\in \mathbb{V}_1,
    \label{equ:appu_1_equ}
  \end{multline}
  \begin{equation}
 \bs u_{1h}^{n+1}=\bs w_h^{n+1},\;\;\text{on}\;\;\partial \Omega.    \label{equ:appu_1_cons}
  \end{equation}
\end{subequations}
\noindent\underline{Find $\bs u_2^{n+1} \in [\mathbb{V}_1]^d$ such that}
  \begin{equation}\label{eq:appu2solve}
  \begin{split}
 \frac{\gamma_0}{\nu_m \Delta t} \int_{\Omega_h} \bs u_{2h}^{n+1}& \varphi_h   +\int_{\Omega_h}\nabla \bs u_{2h}^{n+1}\cdot \nabla \varphi_h = -  \frac{1}{\nu_m} \int_{\Omega_h}\Big(\frac{\bar \mu_h ^{n+1}}{\bar \rho_h ^{n+1}}-\nu_m\Big)\bs \omega_h^{*,n+1}\times \nabla \varphi_h \\
& +  \frac{1}{\nu_m}\int_{\Omega_h} \Big(-\frac{1}{\rho_0} \nabla P_{2h}^{n+1} + \frac{\mathcal{C}^{n+1}_{2h} }{\bar \rho_h^{n+1}} \nabla \bar{\phi}_h^{n+1}-\bs Q_h+\nabla \Big(\frac{\bar \mu_h ^{n+1}}{\bar \rho_h ^{n+1}}\Big)\times \bs \omega_h^{*,n+1}\Big)\varphi_h,  \;\; \forall \varphi_h \in \mathbb{V}_1,
  \end{split}
  \end{equation}

We summarize the final solution algorithm as the following steps:
\begin{itemize}
\item[(i)] Solve equations \eqref{eq:apppsi1}-\eqref{eq:appphi2} for $(\psi_1^{n+1},\psi_2^{n+1})$ and $(\phi_1^{n+1},\phi_2^{n+1})$ successively;\\
Compute $\bar \phi^{n+1}$, $\bar \rho^{n+1},$ $\bar \mu^{n+1}$ and $\bar {\tilde {\bs J}}^{n+1}$ based on equations \eqref{eq:phibar} and \eqref{eq:rhomuJ}, respectively;\\
Compute $\mathcal{C}_1^{n+1}$ and $\mathcal{C}_2^{n+1}$ based on equation \eqref{eq:C1C2}.
\item[(ii)] Solve equations \eqref{equ:appP_1_equ}-\eqref{equ:appP_2_equ} for $(P_1^{n+1}, P_2^{n+1}).$
\item[(iii)] Solve equations \eqref{eq:appu1solve}-\eqref{eq:appu2solve} for $(\bs u_1^{n+1},\bs u_2^{n+1})$.
\item[(iv)] Compute the coefficients $A_0$-$A_4$ using equation \eqref{eq:coeffA}.\\
                Compute the coefficients $B_0$-$B_2$ using equations \eqref{eq:B0}-\eqref{eq:B2}.
\item[(v)] Solve equation \eqref{eq:nonlinear} for $\xi^{n+1}$ using the Newton's method with an initial guess $\xi^{n+1}=1.$
\item[(vi)] Compute $\phi^{n+1}$ based on equation \eqref{equ:phi_expr};\\ 
Compute $P^{n+1}$ based on equation \eqref{eq:p1p2}; \\
Compute $\bs u^{n+1}$ based on equation \eqref{eq:u1u2};\\
                Compute $ E^{n+1}$ based on equation \eqref{eq:Edecomp};\\
                Compute $R^{n+1}$ based on equation \eqref{eq:En1}.

\end{itemize}
%
It can be noted that with the above algorithm two sets of field variables
for each of the velocity, pressure, and phase-field functions are
computed, together with a nonlinear algebraic equation about
a {\em scalar-valued} number. The computations for different field variables
are de-coupled.
When computing the field variables, all the resultant linear
algebraic systems after discretization only involve constant and
time-independent coefficient matrices, which only need to be computed
once and can be pre-computed.

\section{Representative Numerical Examples}
\label{sec:tests}

In this section, we provide numerical results in two dimensions
to illustrate the accuracy
and robustness of the energy-stable scheme described in the previous
section. Both spatial and temporal convergence rates are presented via a manufactured
analytic solution. We then use the capillary wave problem as
a benchmark to demonstrate that the proposed algorithm
produces physically accurate results. We also show numerical
simulations of a rising bubble problem with large density contrast
and viscosity contrast.
It should be noted that when the physical variables and parameters
are normalized consistently, the dimensional and non-dimensionalized
governing equations, boundary conditions and initial conditions
will have the same form~\cite{Dong2014,Dong2015}. Let
$L$, $U_0$ and $\varrho_d$ denote the characteristic length, velocity
and density scales, respectively.
Table \ref{tab:normalization} lists the constants for consistently
normalizing different physical variables and parameters.
In subsequent discussions all the physical variables and parameters are
assumed to have been normalized according to Table \ref{tab:normalization}.
The majority of simulation results reported in this section, and
unless otherwise specified, the simulations of this section
are computed using the method with the approximation~\eqref{eq:vort_approx}
when solving equation \eqref{eq:weakpold}.

\begin{table}[tbp]
  \centering
  \begin{tabular}{ll | ll}
    \hline
    variable & normalization constant & variable & normalization constant \\
    \hline
    $\bs x$, $\eta$ & $L$
    & $\bs u$, $\bs w$, $\bs u_{in}$ & $U_0$ \\
    $t$, $\Delta t$, $J_0$ & $L/U_0$
    & $\bs g_r$ (gravity) & $U_0^2/L$ \\
    $p, P, P_1, P_2, h(\phi), F(\phi)$, $\mathcal{C}$, $\mathcal{C}_1$, $\mathcal{C}_2$, $S$ & $\varrho_d U_0^2$
    & $\phi$, $\phi_1$, $\phi_2$, $\phi_{in}$, $\xi$ & $1$ \\
    $\rho_1, \rho_2, \rho$, $\rho_0$ & $\varrho_d$
    & $\mu_1, \mu_2, \mu$ & $\varrho_d U_0 L$ \\
    $m$ & $L/(\varrho_d U_0)$
    & $\sigma$ & $\varrho_d U_0^2L$ \\
    $\lambda$ & $\varrho_d U_0^2L^2$ 
    & $\nu_m$ & $U_0L$ \\
    $\bs f$ & $\varrho_d U_0^2/L$
    & $\tilde{\bs J}$ & $\varrho_d U_0$ \\
    $g$, $\bs\omega$ & $U_0/L$
    & $d_a$ & $1/L$ \\
    $E$, $C_0$, $A_0, A_1, A_2, A_3, A_4$ & $\varrho_dU_0^2L^d$ ($d=2$ or $3$)
    & $d_b$ & $1/L^3$ \\
    $R$ & $\sqrt{\varrho_dU_0^2L^d}$
    & $\psi$, $\psi_1$, $\psi_2$, $\alpha$ & $1/L^2$ \\
    $B_0, B_1, B_2$ & $\varrho_dU_0^3L^{d-1}$
    & & \\
    \hline
  \end{tabular}
  \caption{Normalization constants for various physical variables and parameters.}
  \label{tab:normalization}
\end{table}


\subsection{Convergence rates}

To begin with, we employ a manufactured analytic solution to demonstrate the convergence rates of the scheme developed herein against both the element order
and the time step size. Consider the computational domain $\Omega=\big\{ (x,y)|\;0\leq x \leq 2, -1 \leq y \leq 1  \big \}$ and a mixture of two immiscible fluids  contained in this domain. We assume the following analytic solution to the
Navier-Stokes/Cahn-Hilliard coupled system of equations \eqref{eq:NphaseEq1}, \eqref{eq:NphaseEq2}-\eqref{eq:NphaseEqC}:
\begin{equation}\label{eq:contrivsolu}
\begin{cases}
&u=\sin(\pi x) \cos(\pi y) \sin(t),\\
&v=-\cos(\pi x) \sin(\pi y) \sin(t),\\
&P=\sin(\pi x) \sin (\pi y) \cos(t),\\
& \phi=\cos(\pi x) \cos(\pi y) \sin(t),
\end{cases}
\end{equation} 
where $(u,v)$ are the two components of the velocity $\bs u.$ Accordingly, the source terms $\bs f$ and $g$ in equations \eqref{eq:NphaseEq1} and \eqref{eq:NphaseEq3} are chosen such that the analytic expressions \eqref{eq:contrivsolu} satisfy equations \eqref{eq:NphaseEq1} and \eqref{eq:NphaseEq3}, respectively. The source terms $\bs w$, $d_a$ and $d_b$ in equations \eqref{eq:bc1}-\eqref{eq:bc3} are chosen such that the contrived solution in \eqref{eq:contrivsolu} satisfies the boundary conditions. The initial conditions \eqref{eq:ic1}-\eqref{eq:ic2} are imposed for the velocity and phase field function,
respectively, where $\bs u_{in}$ and $\phi_{in}$ are chosen by letting $t=0$ in
the contrived solution \eqref{eq:contrivsolu}.

We discretize the domain using two equal-sized quadrilateral elements, with the element order and the time step size ${\Delta}t$ varied systematically in the spatial and temporal convergence tests.
The numerical algorithm from Section \ref{sec:method} is employed to numerically integrate the governing equations
in time from $t=0$ to $t=t_f.$ Then the numerical solution and the exact solution as given by equation \eqref{eq:contrivsolu} at $t=t_f$ are compared, and the errors in the  $L^2$ and $L^{\infty}$  norm for various flow variables are computed.
All the physical and numerical parameters involved in the simulation of this problem are tabulated in Table \ref{tab:simu_param}.

\begin{table}
  \centering
  \begin{tabular}{ll| ll}
    \hline
    parameter & value & parameter & value \\
    $C_0$ & $0$ & $\lambda$ & $0.001$  \\
    $m$ & $0.001$ & $\eta$ & $0.1$  \\
    $(\rho_1,\rho_2)$& $(1.0,3.0)$& $(\mu_1,\mu_2)$ &$(0.01,0.02)$\\
    $\rho_0$ & ${\rm min}( \rho_1,\rho_2) $   &$\nu_m$ & $2\times \frac{{\rm max}(\mu_1, \mu_2)}{{\rm min}(\rho_1, \rho_2)}$  \\
   $\Delta t_{\min}$ & $1e-4$ & $\Delta t$ & (varied) \\
   $t_f$ & $0.1$ (spatial tests) or $1.0$ (temporal tests) & $S$ & $\sqrt{\frac{4\gamma_0\lambda}{m\Delta t}}$
    or $\sqrt{\frac{4\gamma_0\lambda}{m\Delta t_{\min}}}$ \\
    Element order & (varied) & Number of elements & $2$ \\
    \hline
  \end{tabular}
  \caption{Simulation parameter values for convergence tests.}
  \label{tab:simu_param}
\end{table}

\begin{figure}[tbp]
  \centering
 \subfigure[Errors vs Element order]{ \includegraphics[scale=.39]{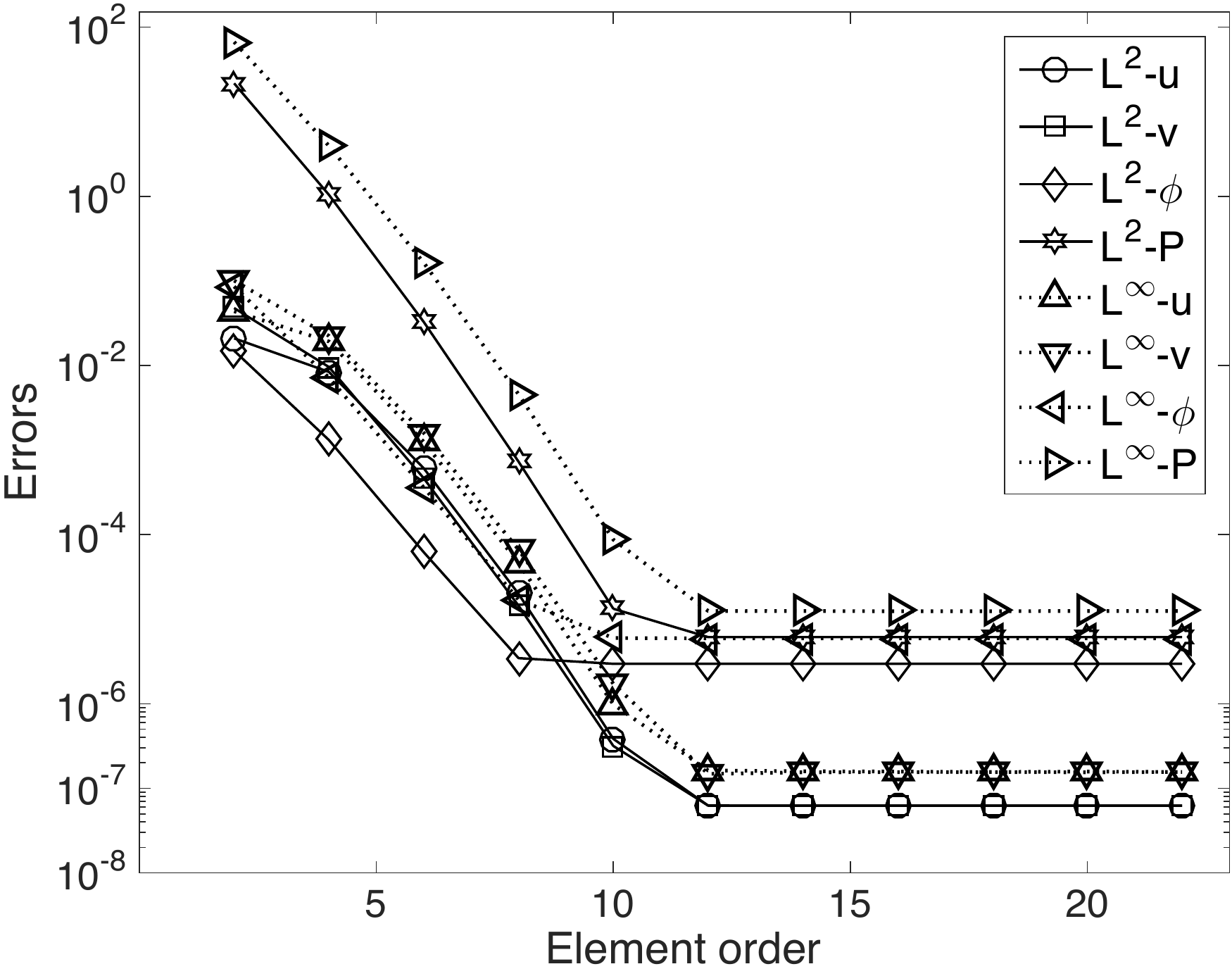}} \quad
 \subfigure[Errors vs ${\Delta}t$]{ \includegraphics[scale=.38]{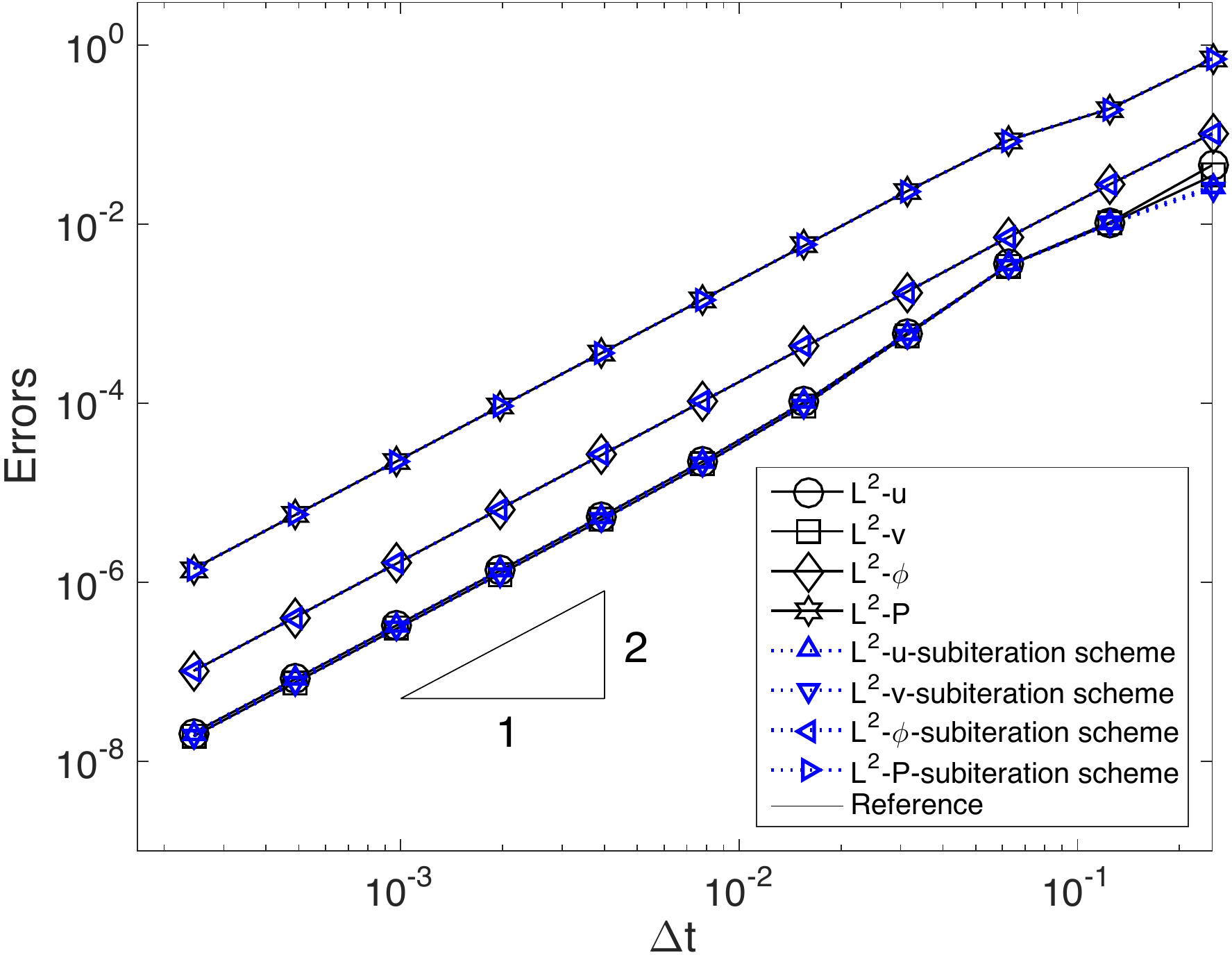}} \\
 \caption{Spatial/temporal convergence tests: (a) $L^{2}$ and $L^{\infty}$ errors of flow variables versus element order (fixed $\Delta t=0.001$ and $t_f=0.1$); (b) $L^{2}$ errors of flow variables versus $\Delta t$ (fixed element order $22$ and $t_f=1$) and comparison with the sub-iteration method discussed in Remark \ref{rem:coupled}.
 }
\label{fig:standtest}
\end{figure}

In the spatial convergence test, we fix the integration time at $t_f=0.1$ and the time step size at $\Delta t=0.001$ ($100$ time steps), and vary the element order systematically between $2$ and $22.$ The same element order has been used for these two spectral elements. Figure \ref{fig:standtest}(a) depicts the numerical errors at $t=t_f$ in $L^2$ and $L^{\infty}$ norms for different flow variables as a function of the element order. It is evident that within a specific range of the element order (below around $12$), the errors decrease exponentially with
increasing element order, displaying an exponential convergence rate in space. Beyond the element order of about $12,$ the error curves level off as the element order further increases, showing a saturation caused by the temporal truncation error.

In the temporal convergence test, we fix the integration time at $t_f=1$ and the element order at a large value $22,$ and vary the time step size systematically between $\Delta t=2.44141\times 10^{-4}$ and $\Delta t=0.25.$ Figure \ref{fig:standtest}(b) shows the numerical errors at $t=t_f$ in $L^2$ norm for different variables as a function of $\Delta t$ in logarithmic scales. It can be observed that the numerical errors exhibit a second order convergence rate in time.

\begin{figure}[tbp]
  \centering
 \subfigure[Comparison of methods with and without sub-iterations]{ \includegraphics[scale=.38]{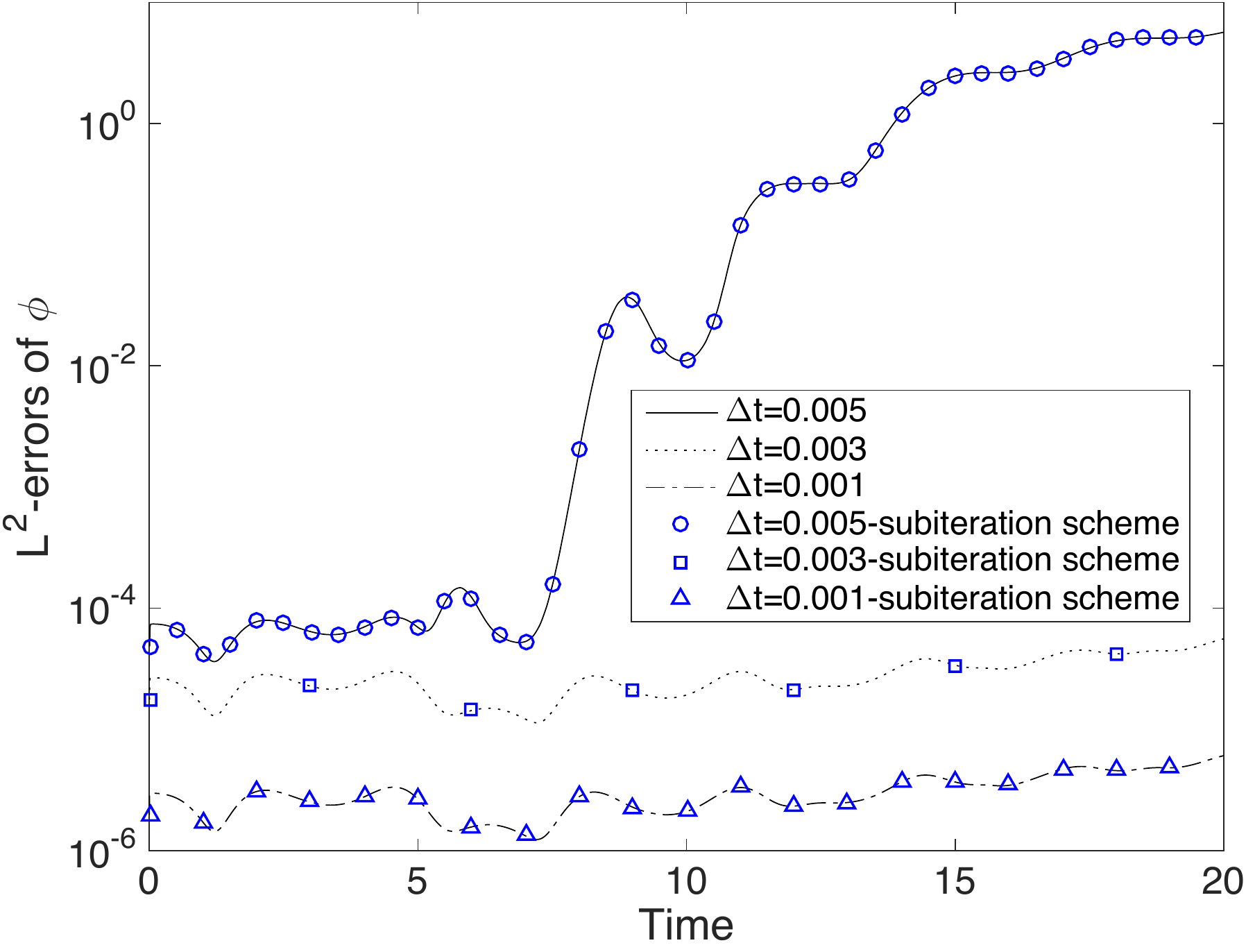}} \quad
\subfigure[Number of sub-iterations within a time step]{ \includegraphics[scale=.38]{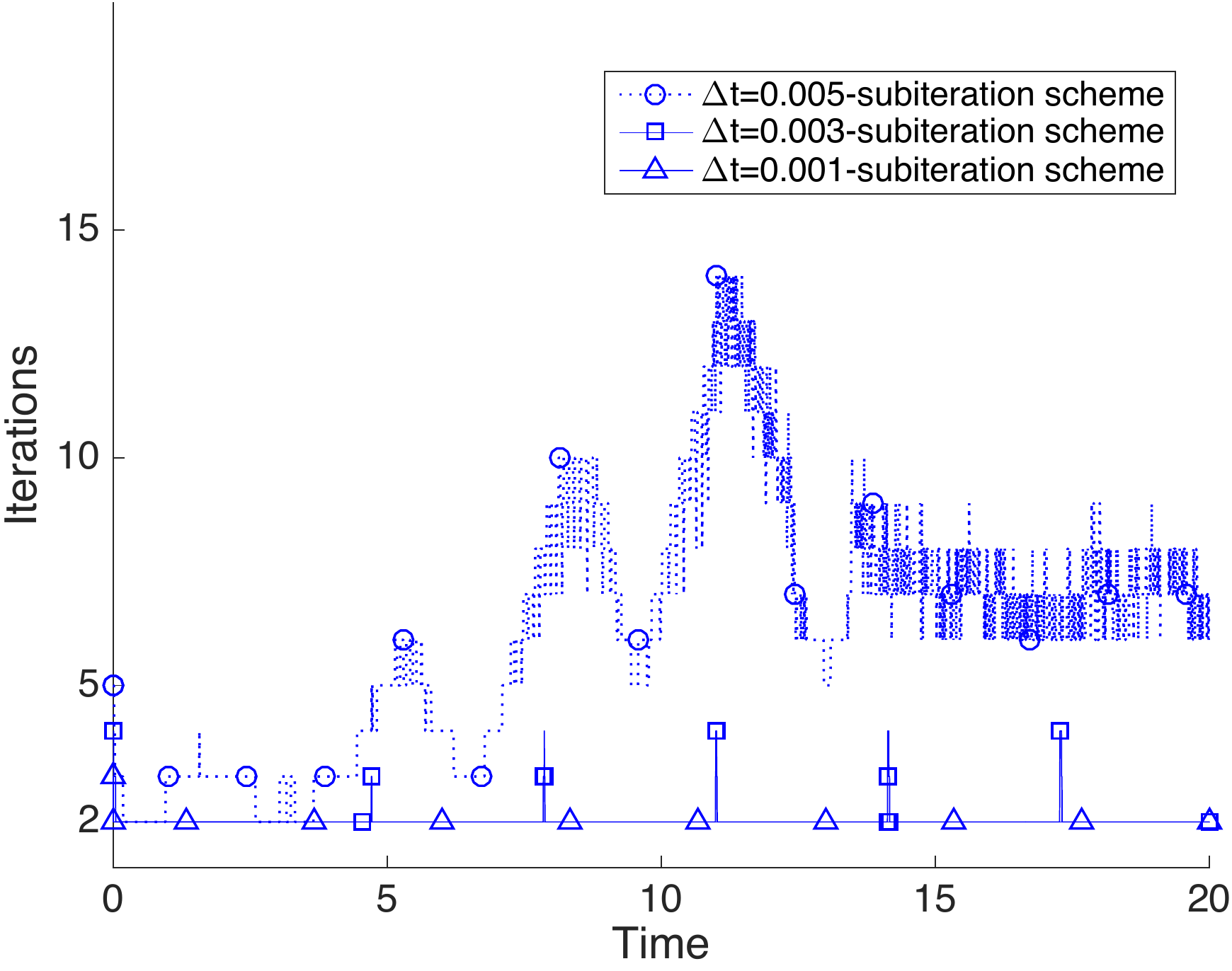}} 
\caption{Computations with sub-iterations: (a) Comparison of $L^2$ errors
  of $\phi$ between the method using approximation~\eqref{eq:vort_approx} and the
  sub-iteration method with several ${\Delta}t$ values;
  (b) Evolution of the number of sub-iterations needed within a time step
  over time for the sub-iteration method.
  Results are obtained with a fixed element order $22$ and $t_f=20.$
  Those curves in (a) with legends not marked by ``subiteration scheme'' are
  obtained with the method using the approximation ~\eqref{eq:vort_approx}.
}
\label{fig:subiter}
\end{figure}

As discussed in Section \ref{sec:method},
two methods have been implemented for solving equation \eqref{eq:weakpold},
based on a sub-iteration procedure as discussed in Remark \ref{rem:coupled}
and based on the approximation~\eqref{eq:vort_approx} on $\partial\Omega$.
The simulation results presented so far are obtained based on
the method with the approximation~\eqref{eq:vort_approx} by default.
To compare these two methods,
in Figure \ref{fig:standtest}(b) we have also included the numerical errors
obtained with the sub-iteration method (marked by ``subiteration scheme'' in
the legend).
It is observed that the difference in the numerical errors between
these two methods is negligible.
In Figure \ref{fig:subiter}(a) we compare  the time histories
of the $L^2$-errors for $\phi$ computed with several ${\Delta}t$ values
for longer-time simulations ($t_f=20$) with both methods.
We observe that the error histories of both methods essentially overlap with each other. In Figure \ref{fig:subiter}(b), we plot the time histories of the number of sub-iterations
needed for convergence within a time step for the sub-iteration method.
The results indicate that for the smaller $\Delta t$ values around $2 \sim 4$
sub-iterations are needed for convergence within a time step, while
for the larger $\Delta t=0.005$ it can take $8$ or more sub-iterations within
a time step. 
These results show that the method with the approximation~\eqref{eq:vort_approx}
is clearly advantageous. It is cheaper computationally, and produces essentially
the same simulation results. 

The above numerical results indicate that the numerical algorithm developed herein exhibits a spatial exponential convergence rate and a temporal second-order convergence rate.

\subsection{Capillary Wave Problem}

\begin{table}[tbp]
\centering 
\begin{tabular}{l c| l c} 
\hline 
Parameter & Value & Parameter & Value \\  
\hline 
 $H_0$         &0.01            &   $k_w$ (wave number)          &  $2\pi$             \\
  $\sigma$        &   1.0         &    $|\bs g_r|$ (gravity)          &     1.0          \\
 $ \rho_1$         & 1.0           &   $ \rho_2$           &  (varied)             \\
 $ \mu_1$          & 0.01           & $\mu_2$             & $ \mu_1 \frac{ \rho_2}{ \rho_1}$              \\
$\rho_0$ & ${\rm min}( \rho_1,\rho_2) $   &$\nu_m$ & $2\times \frac{{\rm max}(\mu_1, \mu_2)}{{\rm min}(\rho_1, \rho_2)}$  \\
$C_0$ & (varied) & $S$ & $\sqrt{\frac{4\gamma_0\lambda}{m\Delta t}}$\\
 $J$ (temporal order)   &  2& Number of elements  & 240  \\ 
$m$   & $5\times 10^{-4}$ or $10^{-5}$     &    $\eta$         & (varied)              \\
$\lambda$& $\frac{3}{2\sqrt{2}}\sigma \eta$   &       $\Delta t$    & (varied)   \\ 
\hline 
\end{tabular}
\caption{Simulation parameter values for the capillary wave problem.}
\label{table:capillary} 
\end{table}

\begin{figure}[tbp]
 \subfigure[$H$ vs element order]{ \includegraphics[scale=.35]{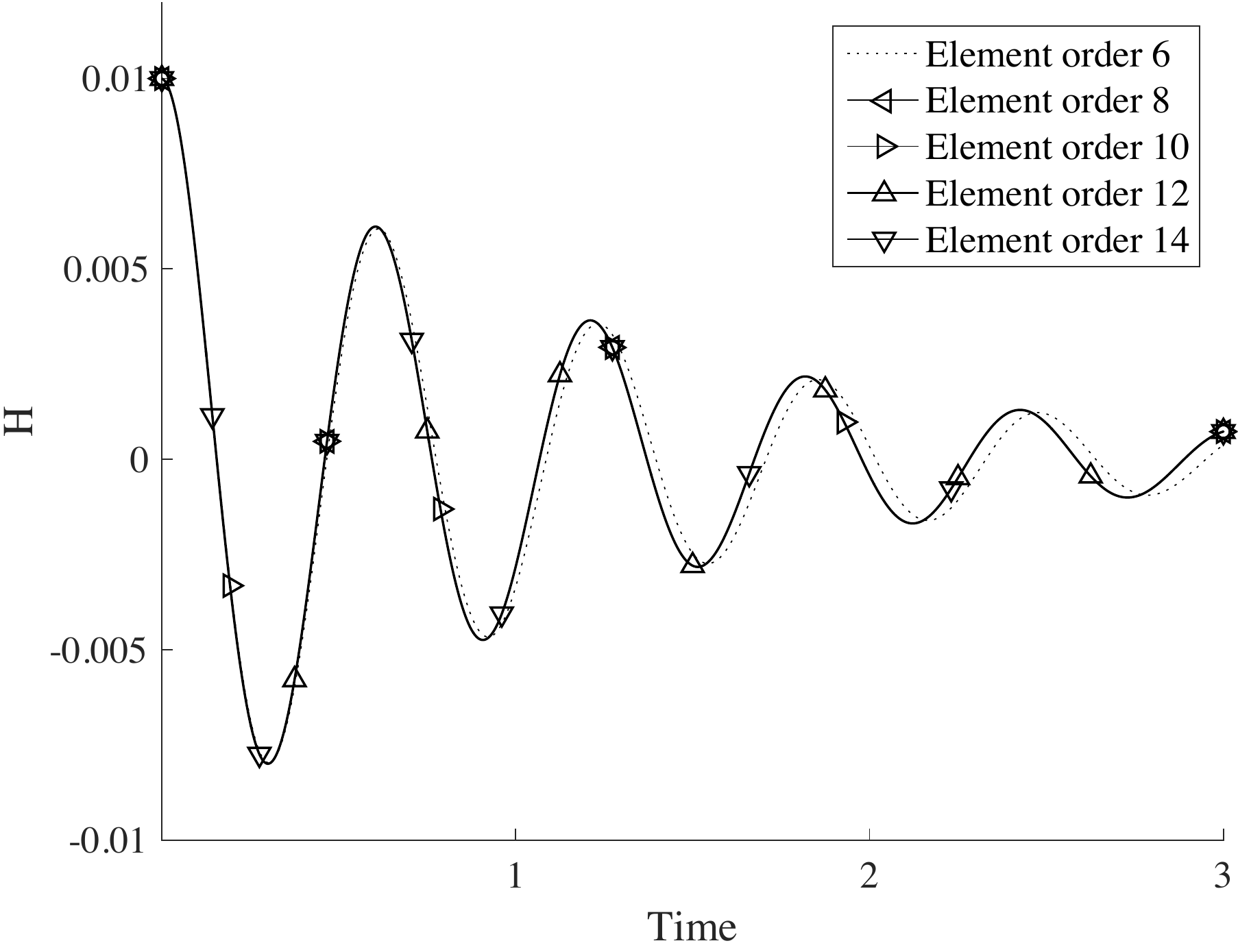}} 
\quad
\subfigure[$H$ vs interfacial thickness $\eta$]{ \includegraphics[scale=.35]{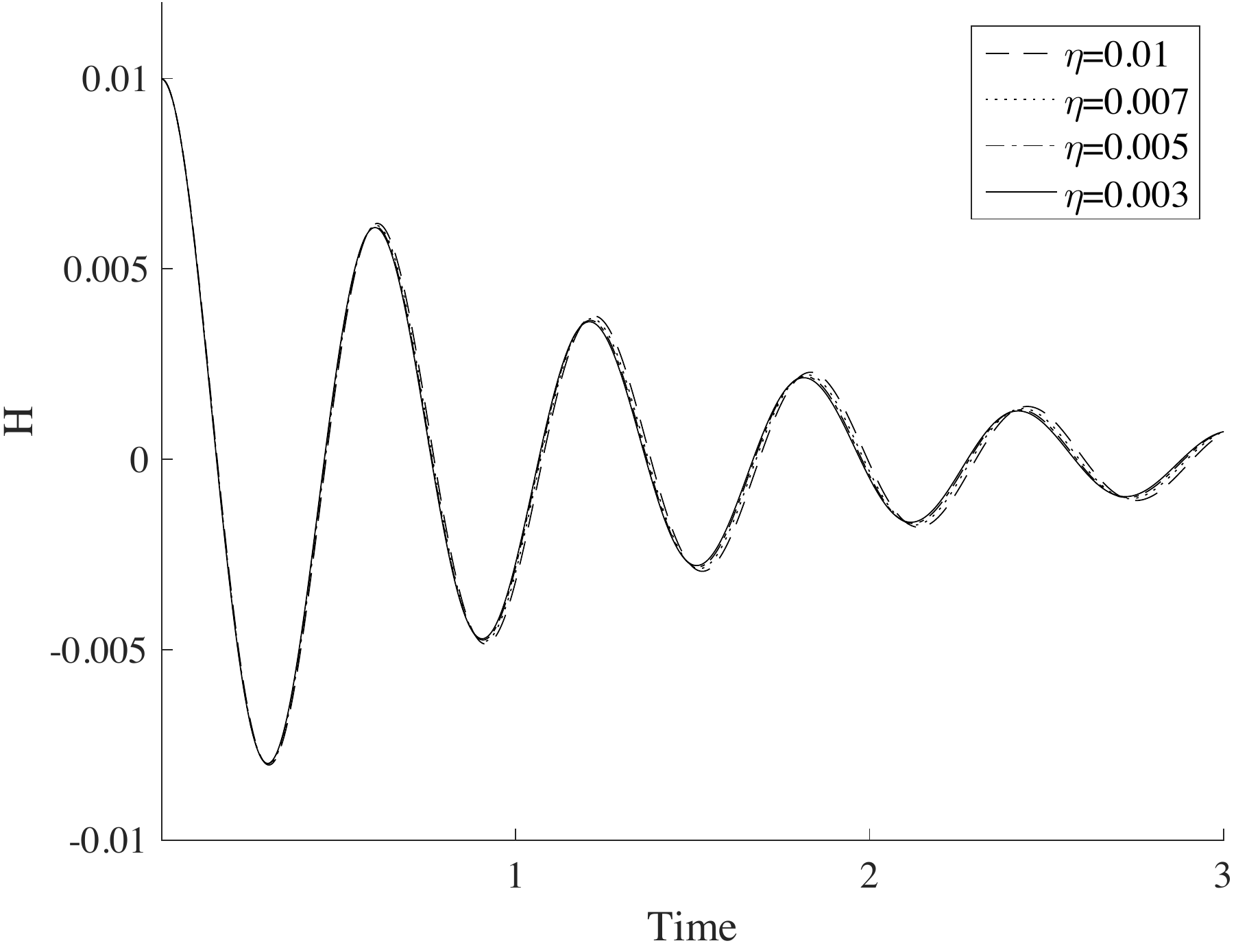}} \\
\subfigure[$H$ vs simulation parameter $C_0$]{ \includegraphics[scale=.35]{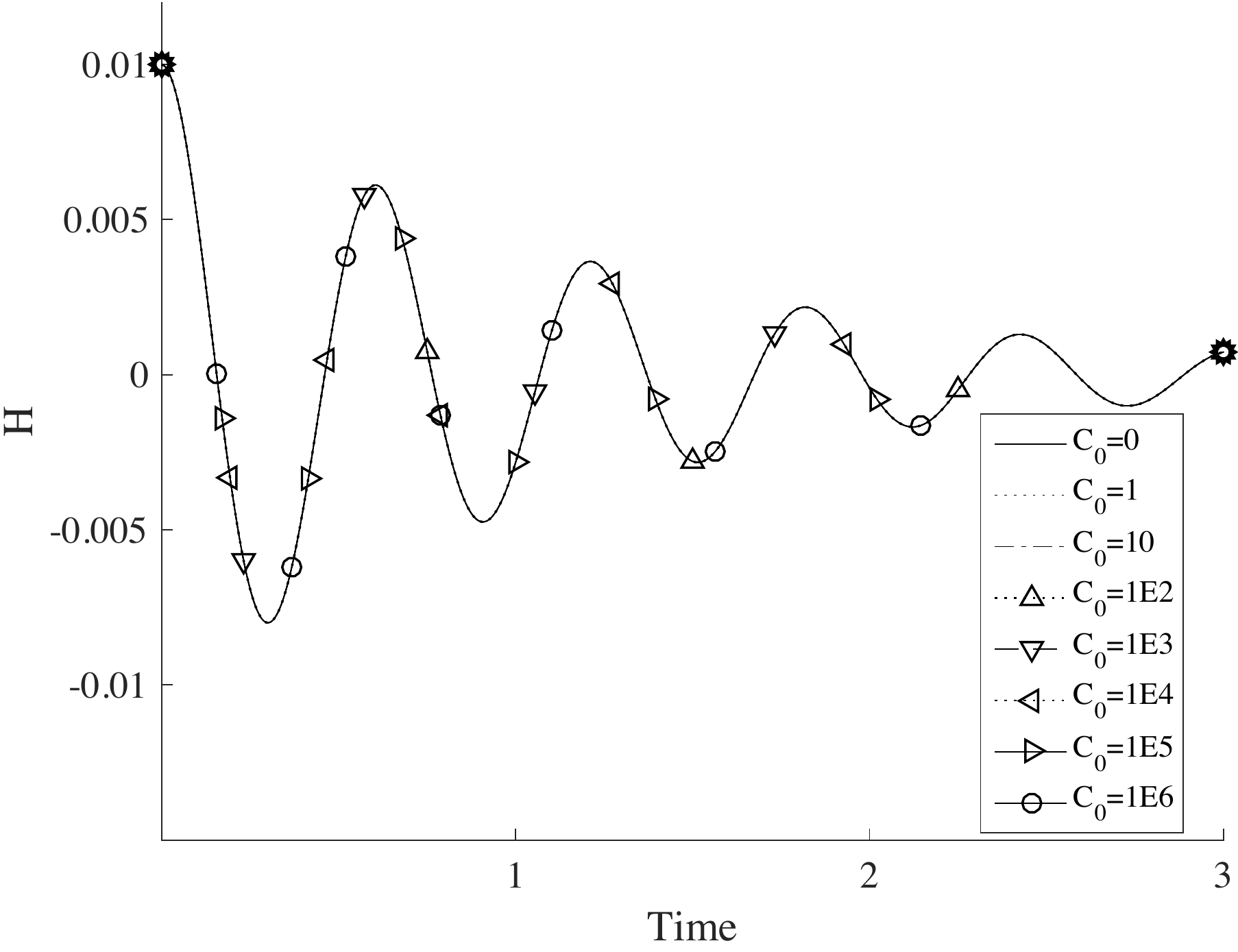}} \quad
\subfigure[$H$ vs Prosperetti's exact solution]{ \includegraphics[scale=.35]{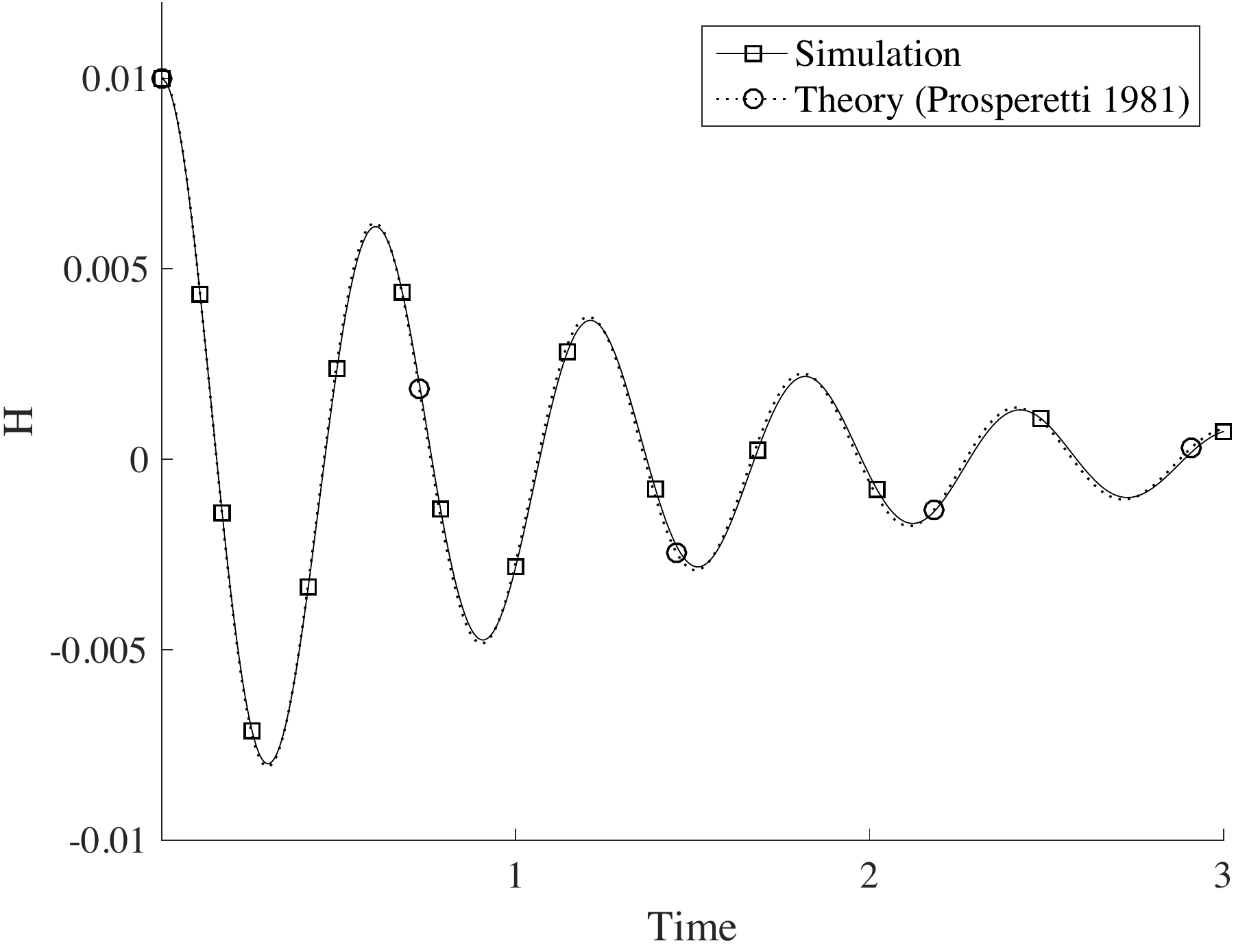}} 
\caption{
  Capillary wave problem (matched density $\rho_1=\rho_2=1$): Comparison of capillary amplitude histories with different element orders (a), with different interfacial thickness $\eta$ values (b), with different $C_0$ (c) and between simulation result and Prosperetti's exact solution. In (a), $\eta=0.005$, $C_0=0$; In (b), element order 10, $C_0=0$; In (c), $\eta=0.005,$ element order=10; In (d), element order 10, $\eta=0.005$, $C_0=0$.
}
\label{fig:capwavevarycoeff}
\end{figure}

In this subsection, we use the capillary wave problem, for which an exact solution is available, as a benchmark to
test the accuracy of the proposed algorithm.

The problem setting is as follows.
We consider two immiscible incompressible fluids contained in an infinite domain.
The upper half of the domain is occupied by the lighter fluid
(with density $\rho_1$), and the lower half of the domain is occupied by the heavier
fluid (with density $\rho_2$). The gravity is assumed to be in the downward direction.
The interface formed between these two fluids is perturbed from its
horizontal equilibrium position by a small-amplitude sinusoidal wave form, and starts
to oscillate at $t=0.$ The objective here is to study the motion of
the interface over time. 

In \cite{Prosperetti1981} an exact time-dependent standing-wave solution to this
capillary wave problem was derived, under the condition
that the two fluids must have  matched kinematic viscosities
(but their densities and dynamic viscosities can differ).
To compare with Prosperetti's exact physical solution (see \cite{Prosperetti1981}),
we adopt the following settings to simulate the capillary wave problem: 
(i) the capillary amplitude
is small compared with  the
vertical dimension of the domain; and (ii)
the kinematic viscosity $\nu$ satisfies $ \nu=\frac{ \mu_1}{ \rho_1}=\frac{ \mu_2}{ \rho_2}$.

\begin{figure}[tbp]
 \subfigure[$H$ vs various ${\Delta}t$]{ \includegraphics[scale=.35]{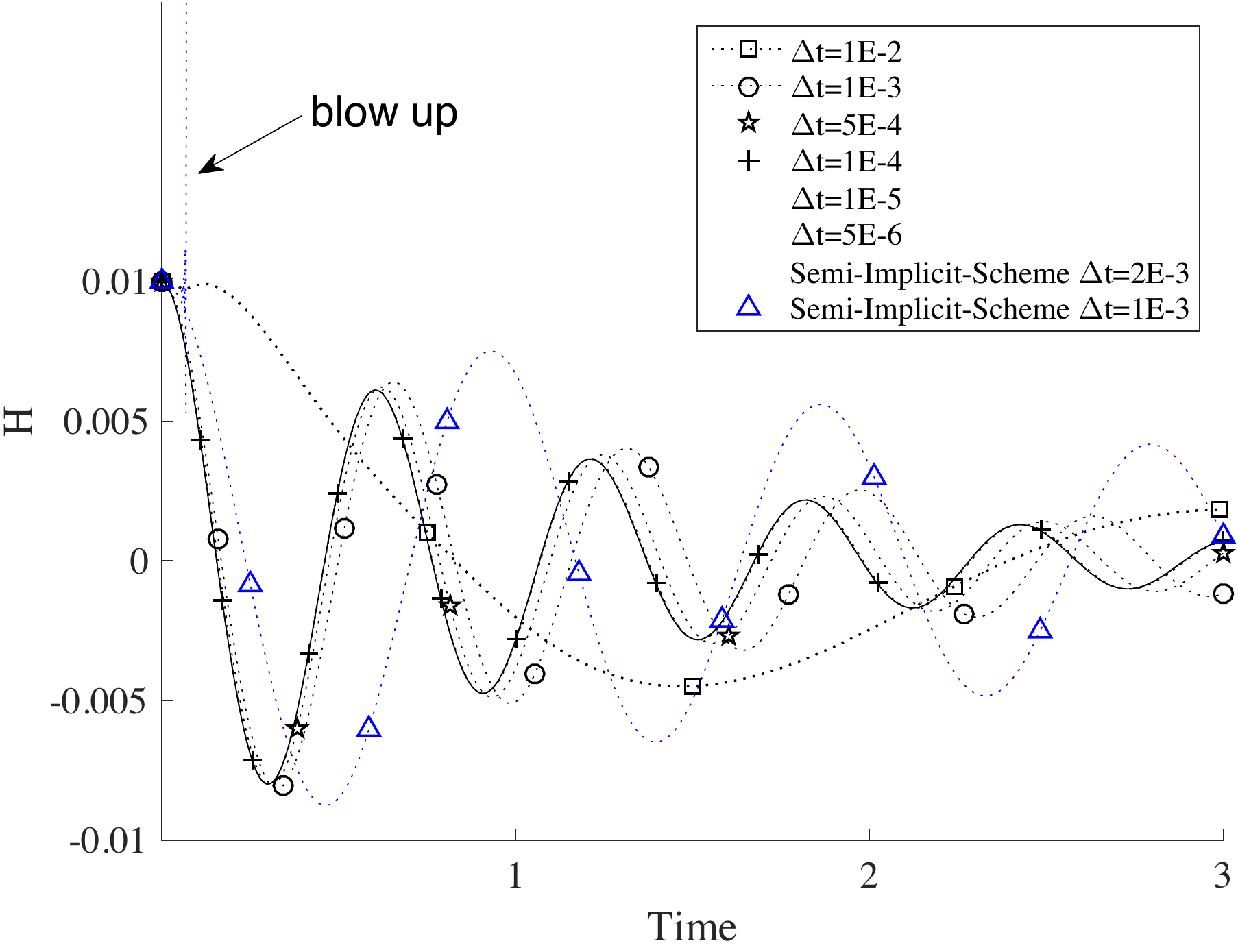}} \quad 
\subfigure[Time history of $\xi^{n+1}$]{ \includegraphics[scale=.35]{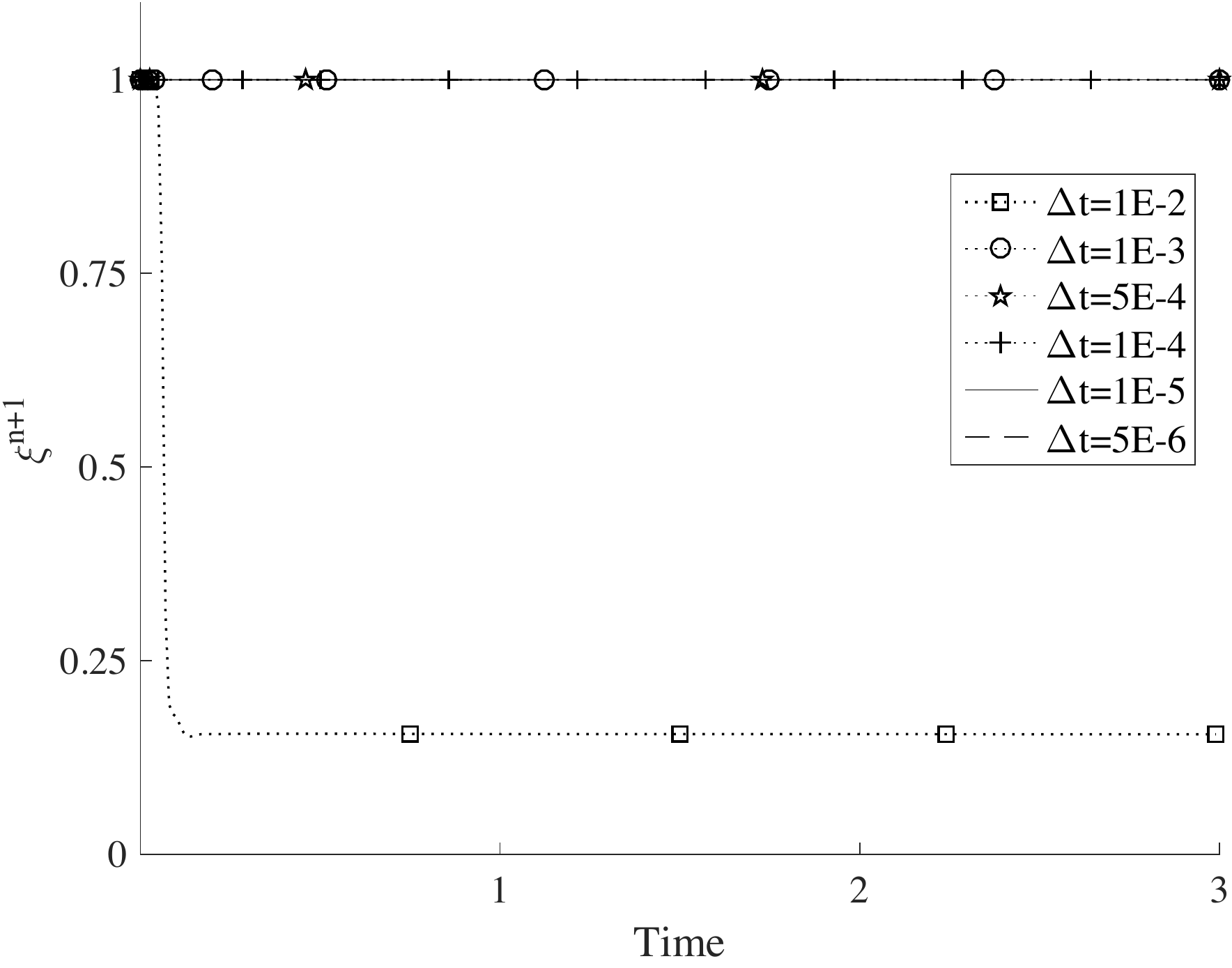}} \\
\subfigure[$H$ vs large ${\Delta}t$]{ \includegraphics[scale=.35]{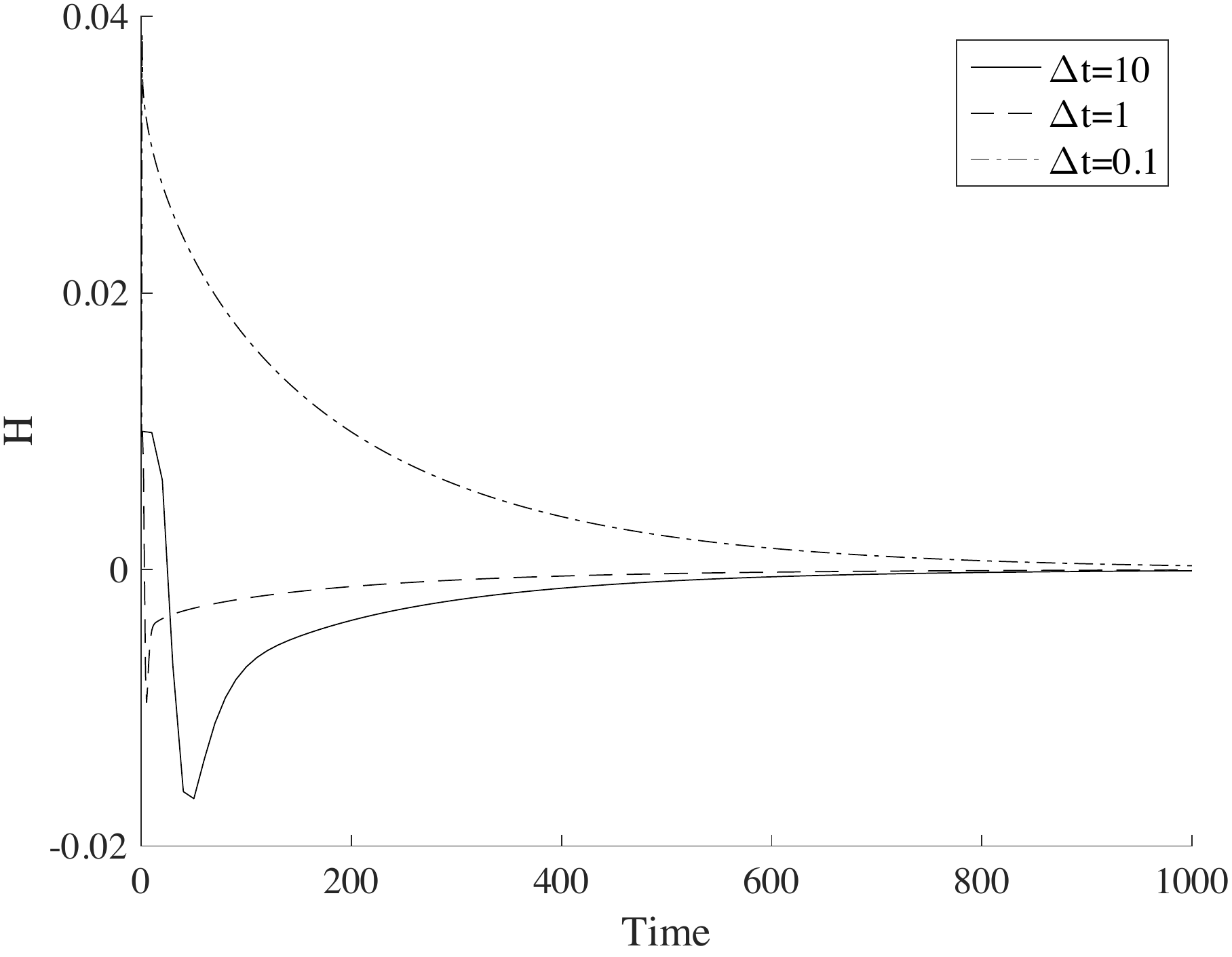}}\quad 
\subfigure[Time history of  $\xi^{n+1}$]{ \includegraphics[scale=.35]{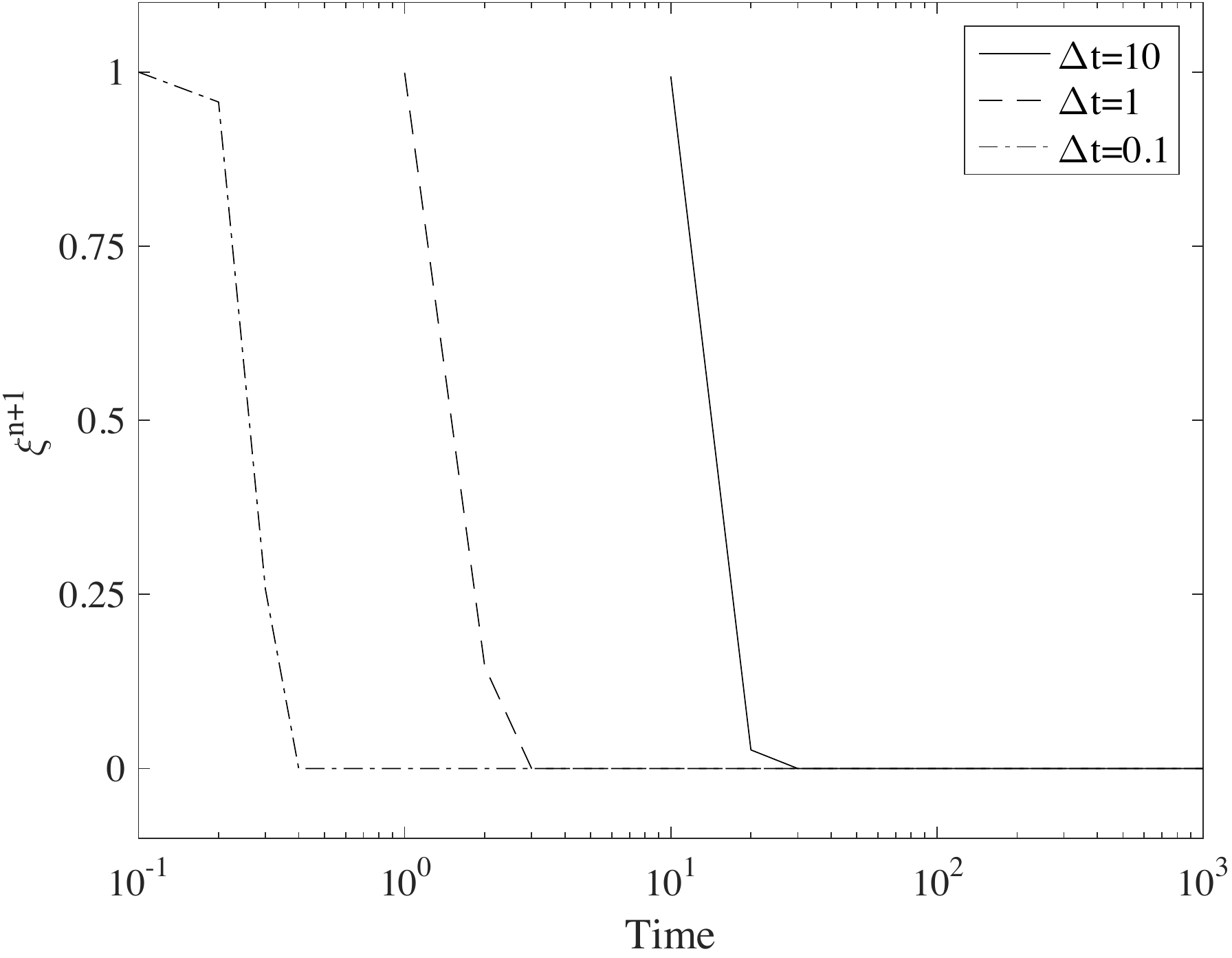}}
\caption{Capillary wave problem (matched density $\rho_1=\rho_2=1$):
  (a) Comparison of capillary amplitude histories obtained with
  a range of smaller ${\Delta}t$, and also comparison with results
  obtained by the semi-implicit method from \cite{DongS2012},
  and (b) corresponding time histories of $\xi^{n+1}$.
  (c) Comparison of capillary amplitude histories
  obtained with a range of large time step sizes ${\Delta}t$, 
  and (d) corresponding time histories of $\xi^{n+1}$ in (d).
}
\label{fig:capwavevarydt}
\end{figure}

To simulate this problem, we
consider a rectangular computational
domain $\Omega=\{(x,y)| \;0\leq x \leq 1, -1 \leq y \leq 1   \}.$
The upper and bottom sides of the domain are assumed to be
solid  walls of neutral wettability.
On the left and right sides, all the variables are assumed to be periodic at $x=0$ and $x=1.$ The equilibrium position of the fluid interface is assumed to coincide with $y=0.$ The initial perturbed profile of the fluid interface is given by $y=H_0 \cos(k_w x)$, where $H_0=0.01$ is the initial amplitude, $\lambda_w=1$ is the wavelength of the perturbation profiles, and $k_w=2\pi/\lambda_w$ is the wave number. Note that the initial capillary amplitude $H_0$ is small compared with the dimension of the domain in the vertical direction.
Therefore, the effect of the walls at the domain top/bottom will be small.

The computational domain is partitioned with $240$  quadrilateral elements,
with $10$ and $12$ elements respectively in $x$ and $y$ directions. The elements are uniform in the $x$ direction, and are non-uniform and clustered around the regions $-0.015\leq y \leq 0.015$ in the $y$ direction. 
In the simulations, the external body force $\bs f$ in equation  \eqref{eq:NphaseEq1} is set to $\bs f=\rho \bs g_r,$ where $\bs g_r$ is the gravitational acceleration, and
the source term in equation \eqref{eq:NphaseEq3} is set to $g=0.$ On the upper and bottom wall, the boundary condition \eqref{eq:bc1} with $\bs w=\bs 0$ is imposed for the velocity, and the boundary conditions \eqref{eq:bc2} and \eqref{eq:bc3} with $d_a=d_b=0$ are imposed for the phase field functions.
The initial velocity is set to zero, and the initial phase field function is prescribed as follows:
\begin{equation}
\phi(\bs x,0)=\tanh\frac{y-H_0\cos(k_w x)}{\sqrt{2}\eta}.
\end{equation}
We list in Table \ref{table:capillary} the values for the physical and numerical parameters involved in this problem.

Let us first focus on the case with a matched density for the two fluids,
i.e., $\rho_1=\rho_2=1$, and study the effects of several parameters on
the simulation results with the data
from Figures \ref{fig:capwavevarycoeff} and \ref{fig:capwavevarydt}.
Figure \ref{fig:capwavevarycoeff}(a) shows a spatial resolution test.
Here we compare the time histories of the capillary wave amplitude obtained with several element orders ranging from 6 to 14 in the simulations. These results are obtained with a fixed mobility $m=5\times 10^{-4},$ interfacial thickness $\eta=0.005,$ ${\Delta}t=10^{-5}$ and $C_0=0.$ The history curves corresponding to different element orders overlap with one another (with elements orders $8$ and above),
suggesting independence of the results with respect to the grid resolution.
In the forthcoming simulations, we will fix element order equal to $10$.

\begin{figure}[tbp]
 \subfigure[$(\rho_1,\rho_2)=(1,10)$]{ \includegraphics[scale=.29]{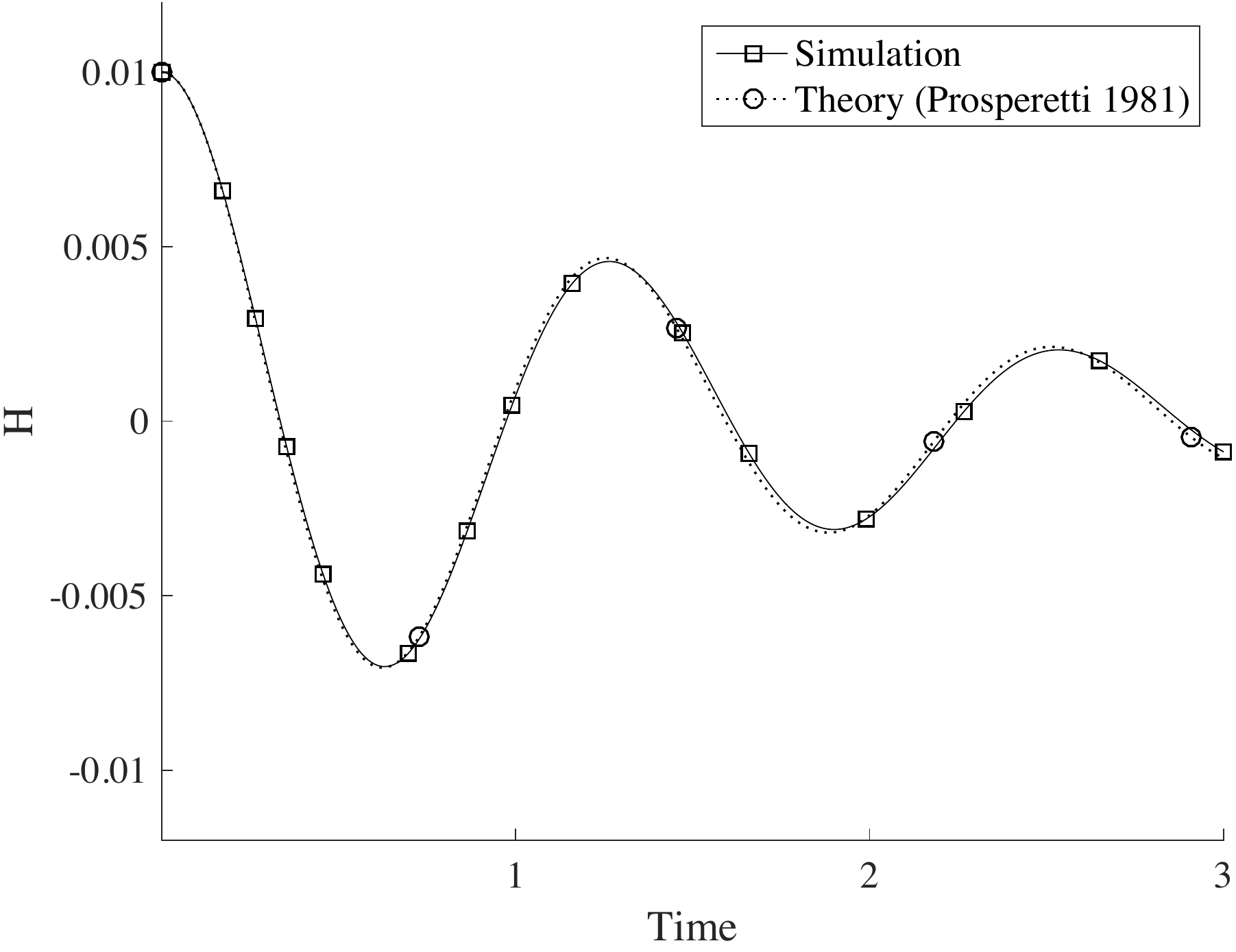}} 
\subfigure[$(\rho_1,\rho_2)=(1,100)$]{ \includegraphics[scale=.29]{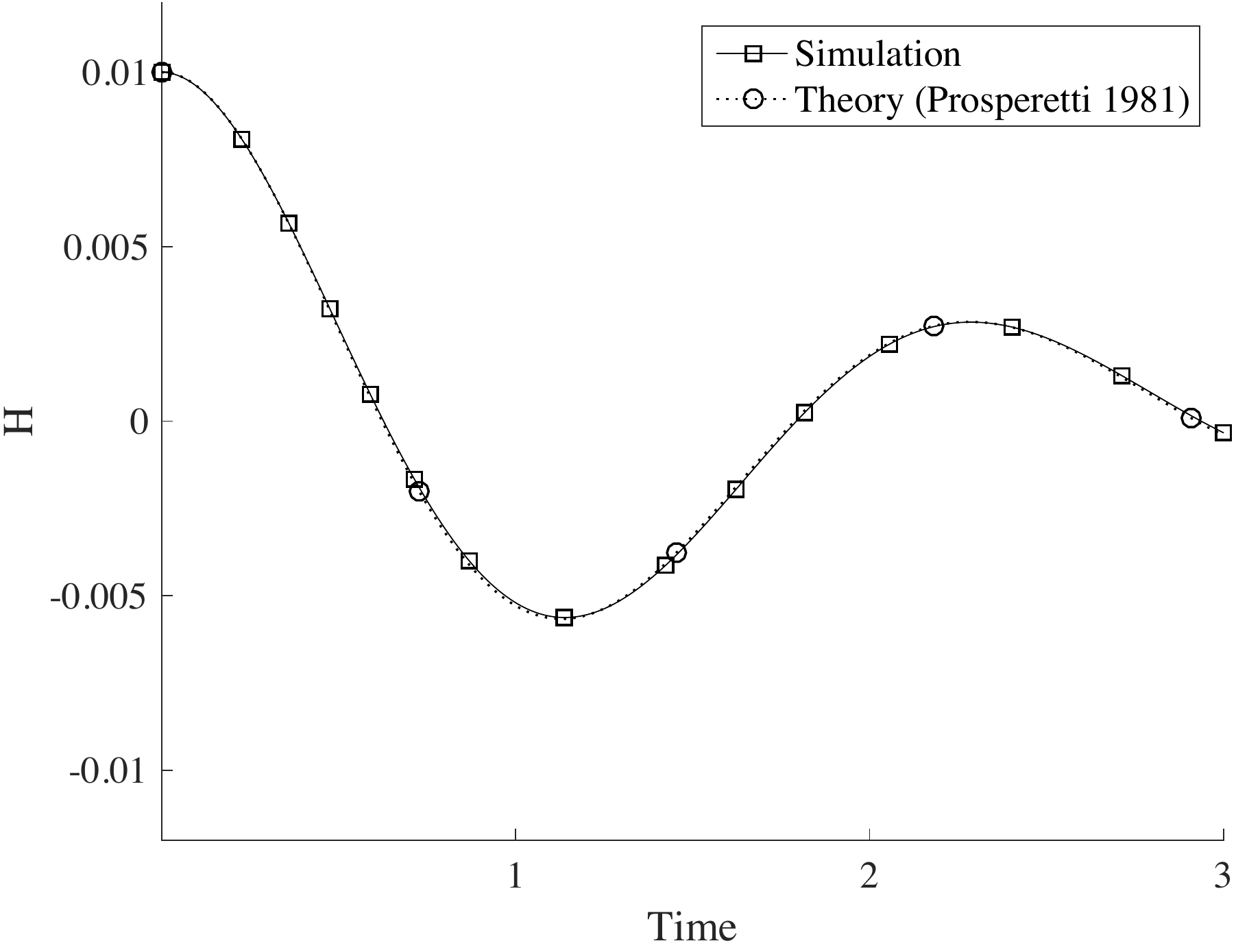}} 
\subfigure[$(\rho_1,\rho_2)=(1,1000)$]{ \includegraphics[scale=.29]{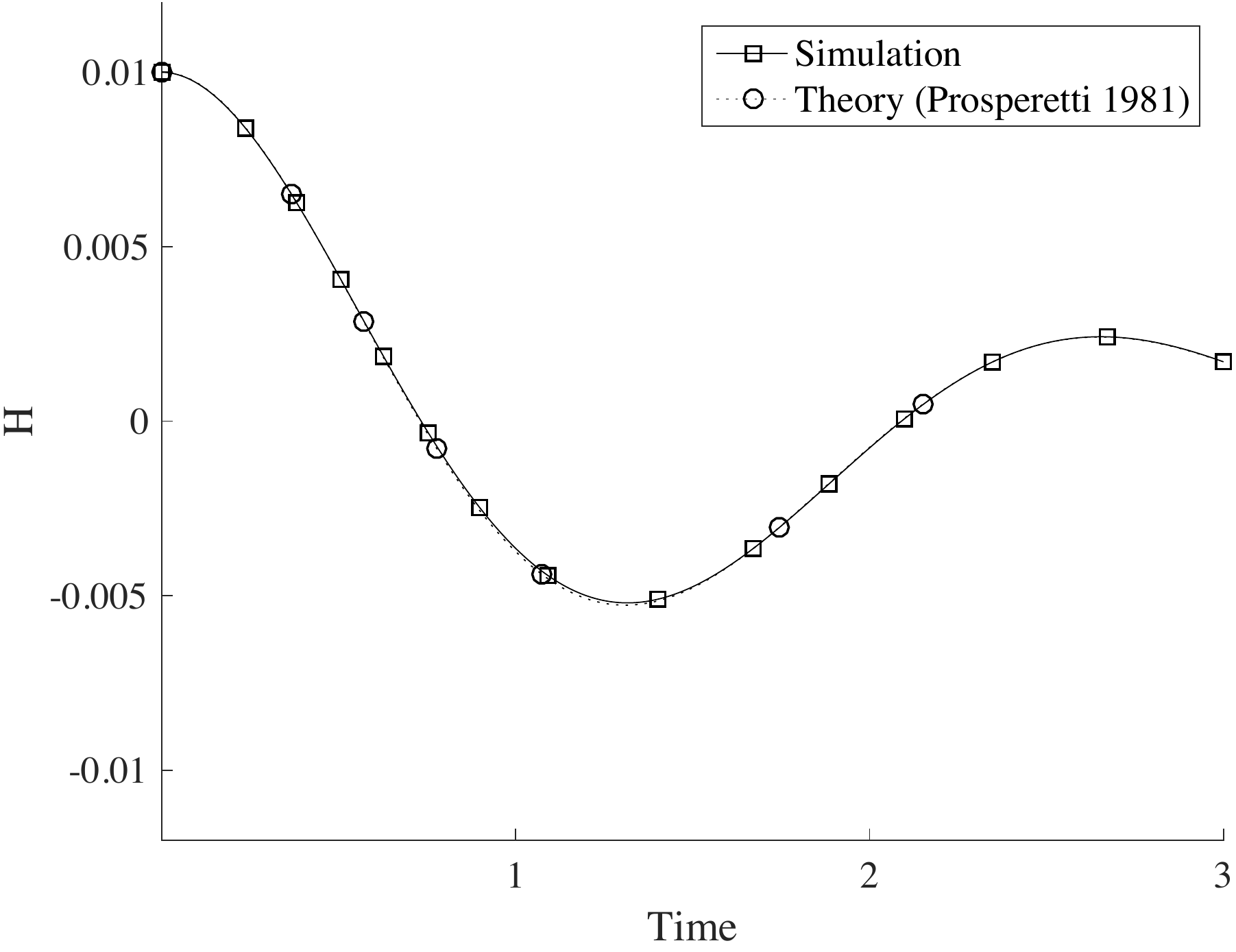}} 
\caption{
  Capillary wave problem (different density ratios):
  Comparison of capillary amplitude histories with Prosperetti's exact solution for density ratios (a) $\frac{\rho_2}{\rho_1}=10$ (b) $\frac{\rho_2}{\rho_1}=100$ (c) $\frac{\rho_2}{\rho_1}=1000$.
  The simulation results are obtained with
  $\Delta t=10^{-5}$ for (a)-(b)
  and $\Delta t=2.5e-6$ for (c). 
}
\label{fig:caprho}
\end{figure}

Figure \ref{fig:capwavevarycoeff}(b) shows the effect of the interfacial thickness
scale $\eta$ on the simulation results.
In this figure we compare time histories of the capillary amplitude obtained with the interfacial thickness scale parameter $\eta$ ranging from $0.01$ to $0.003.$ These results correspond to a fixed mobility $5\times 10^{-4},$ ${\Delta}t=10^{-5}$ and $C_0=0.$  Some influence on the amplitude and the phase of the history curves can be observed
as $\eta$ decreases from $0.01$ to $0.007.$ From $\eta=0.007$ and below, on the other hand, the history curves essentially overlap with one another and little difference can be discerned among them, suggesting a convergence of the results with respect to $\eta.$

Figure \ref{fig:capwavevarycoeff}(c) demonstrates the effect of the parameter $C_0$ on the simulation results with a fixed $m=5\times 10^{-4}$ and $\eta=0.005$.
No difference can be discerned among the numerical results obtained with
different $C_0$ values.
In the following simulations, we will employ $C_0=0.$

In Figure \ref{fig:capwavevarycoeff}(d), we compare the time history of the capillary amplitude obtained using $m=5\times 10^{-4}$ and $\eta=0.005$ with the exact solution by \cite{Prosperetti1981}. The difference between the numerical result and the theoretic solution is negligible, showing that our method produces physically accurate numerical results.

Next, we demonstrate the robustness of our method with various time step sizes,
ranging from very small to very large values.
In Figure \ref{fig:capwavevarydt}(a),
we show time histories of the capillary amplitude obtained with time
step size in a range of smaller values,
$\Delta t=10^{-2}\sim 5\times 10^{-6}$.
The results correspond to a fixed mobility $m=5\times 10^{-4}$ and
interfacial thickness $\eta=0.005$ in the simulations.
For comparison, this figure also includes the results obtained
using the semi-implicit method from~\cite{DongS2012} with
time step sizes $\Delta t=0.001$ and $0.002$.
All the simulations using the current method are stable,
and the time history curves collapse to a single one when
the time step size is reduced to $\Delta t=10^{-4}$ and below,
suggesting convergence with respect to $\Delta t$.
In contrast, the simulation using the semi-implicit scheme of
\cite{DongS2012} blows up with $\Delta t=0.002$.
With $\Delta t=0.001$, the semi-implicit simulation is stable. But
the resultant history curve contains considerably larger errors in both
amplitude and phase,
with respect to the theoretical solution, than the result obtained
using the current method with the same time step size.
In Figure \ref{fig:capwavevarydt}(b), we plot the corresponding time histories of $\xi^{n+1}=\frac{R^{n+1}}{\sqrt{E^{n+1}}}$ computed from equation \eqref{eq:nonlinear}.
Note that $\xi^{n+1}$ should physically take the unit value, because
$R(t) = \sqrt{E(t)}$ by definition when deriving the dynamic equation
about $R(t)$. A substantial deviation of $\xi^{n+1}$ from the unit value
indicates that $R(t)$ is no longer an accurate approximation of $\sqrt{E(t)}$,
and therefore the simulation will contain significant errors. So the value
of $\xi^{n+1}$ is a good indicator of the accuracy of the simulation results.
We observe that when ${\Delta}t$ is small (less than $10^{-3}$),
$\xi^{n+1}$ is identical to $1.$ But when ${\Delta}t=10^{-2}$, $\xi^{n+1}$ decreases sharply
to $0.2$ and then remains at that level over time. This
suggests that the simulation is no longer accurate and the result
contains significant errors with this $\Delta t$.

Thanks to its energy-stable nature, our algorithm can produce stable
simulation results even with very large time step sizes.
This is demonstrated by Figures \ref{fig:capwavevarydt}(c) and (d)
for a range of large time step sizes.
Here we increase the time step sizes to $\Delta t=0.1,1,10$
and depict the time histories of the capillary amplitude
in Figure \ref{fig:capwavevarydt} (c) for
a much longer simulation (up to $t=1000$).
The long time histories demonstrate that the computations with these large ${\Delta}t$ values are indeed stable using the current algorithm. On the other hand, because these time step sizes are very large,
we do not expect that these results will be accurate.
It is observed that the characteristics of these history curves are quite different from
those obtained with smaller time step sizes (comparing with
e.g.~Figure \ref{fig:capwavevarydt}(a)).
Figure \ref{fig:capwavevarydt}(d) shows the corresponding time histories
of $\xi^{n+1}$ with this range of large $\Delta t$ values.
The behaviors of $\xi^{n+1}$ are quite different from those
obtained with small ${\Delta}t$ (see Figure \ref{fig:capwavevarydt}(b)).
The general characteristics seem to be that, when $\Delta t$ increases to
a fairly large value the computed
$\xi^{n+1}$ would decrease over time and its
history curve would level off at a certain value for a given ${\Delta}t$.
With a larger ${\Delta}t$, the $\xi^{n+1}$ history tends to level off at a smaller value.
With very large ${\Delta}t$ values, $\xi^{n+1}$ will essentially tend to zero.
Note that these simulations are performed using the method with
the approximation~\eqref{eq:vort_approx} when solving
equation \eqref{eq:weakpold}. The results with
the large $\Delta t$ values demonstrate that the approximation
\eqref{eq:vort_approx} on the boundary does not weaken
the stability of our method.

Let us next investigate the effect of the density ratio
on the dynamics of the fluid interface.
In these tests we fix $\rho_1=1$ and $\mu_1=0.01$ for the first fluid and vary the density $\rho_2$ and the dynamic viscosity $\mu_2$ of the second fluid systematically
while the relation $\nu=\frac{\mu_1}{\rho_1}=\frac{\mu_2}{\rho_2}$ is maintained as required by the theoretic solution in \cite{Prosperetti1981}.
In Figure \ref{fig:caprho}, we show the time histories of the  capillary amplitudes
corresponding to density ratios ${\rho_2}/{\rho_1}=10,100,1000$,
and compare them with the theoretical solutions from \cite{Prosperetti1981}.
The  simulation results are obtained with a fixed interfacial thickness $\eta=0.005$,
mobility $m=10^{-5}$, and $\Delta t=10^{-5}$
for Figures \ref{fig:caprho}(a)-(b)
and $\Delta t=2.5\times 10^{-6}$
for Figure \ref{fig:caprho}(c).
We observe that the density contrasts have a dramatic effect on the motions of
the interfaces. Increase in the density ratio  increases
the oscillation period and the attenuation of the amplitude becomes more pronounced.
It can also be observed that
the history curves from the simulations agree well with those of the
exact solutions for all the density ratios,
indicating that our method has captured the dynamics of the
fluid interfaces correctly.

The capillary wave problem and in particular the comparisons
with Prosperetti's exact solution for this problem demonstrate
that our algorithm produces
physically accurate results for a wide range of density ratios (up to density ratio $1000$ tested here), and that
it is stable for large time step sizes in long-time simulations.

\subsection{Rising Air Bubble in Water}

\begin{table}[tbp]
\centering 
\begin{tabular}{l c| l c} 
\hline 
Parameter & Value & Parameter & Value \\  
\hline 
 $(x_0,y_0)$         &$(0.08,-0.3)$            &   $R_0$           &  $0.1$             \\
  $\sigma$        &   10         &    $|\bs g_r|$ (gravity)          &     1.0          \\
 $ \rho_1$         & 1.0           &   $ \rho_2$           &  $100$ or $1000$             \\
 $ \mu_1$          & 0.01           & $\mu_2$             & $ 0.5$              \\
$\rho_0$ & ${\rm min}( \rho_1,\rho_2) $   &$\nu_m$ & $2\times \frac{{\rm max}(\mu_1, \mu_2)}{{\rm min}(\rho_1, \rho_2)}$  \\
$C_0$ & $0$ & $S$ & $\sqrt{\frac{4\gamma_0\lambda}{m\Delta t}}$\\
$m$   &$10^{-7}$            &    $\eta$         & $0.01$              \\
$\lambda$& $\frac{3}{2\sqrt{2}}\sigma \eta$   &
$\Delta t$    & $5\times 10^{-5}$ (or varied)   \\ 
$J$ (temporal order)   &  2& Number of elements  & 146  \\
Element order & $10$ \\
\hline 
\end{tabular}
\caption{Simulation parameter values for the rising bubble problem.}
\label{table:risingbubble} 
\end{table}

In this section, we test the proposed method
using a two-phase flow problem in an irregular domain, involving
realistically large density ratios and viscosity ratios
encountered in the physical world.
The density ratio and viscosity ratio of the two fluids correspond to
those of the air and water in the majority of cases considered here.
But for some cases
the density ratio of the fluids will also be artificially varied,
 and for those cases we will still
refer to these fluids as air and water.

\begin{figure}[htbp]
 \subfigure[$t=0.00005$]{ \includegraphics[scale=.23]{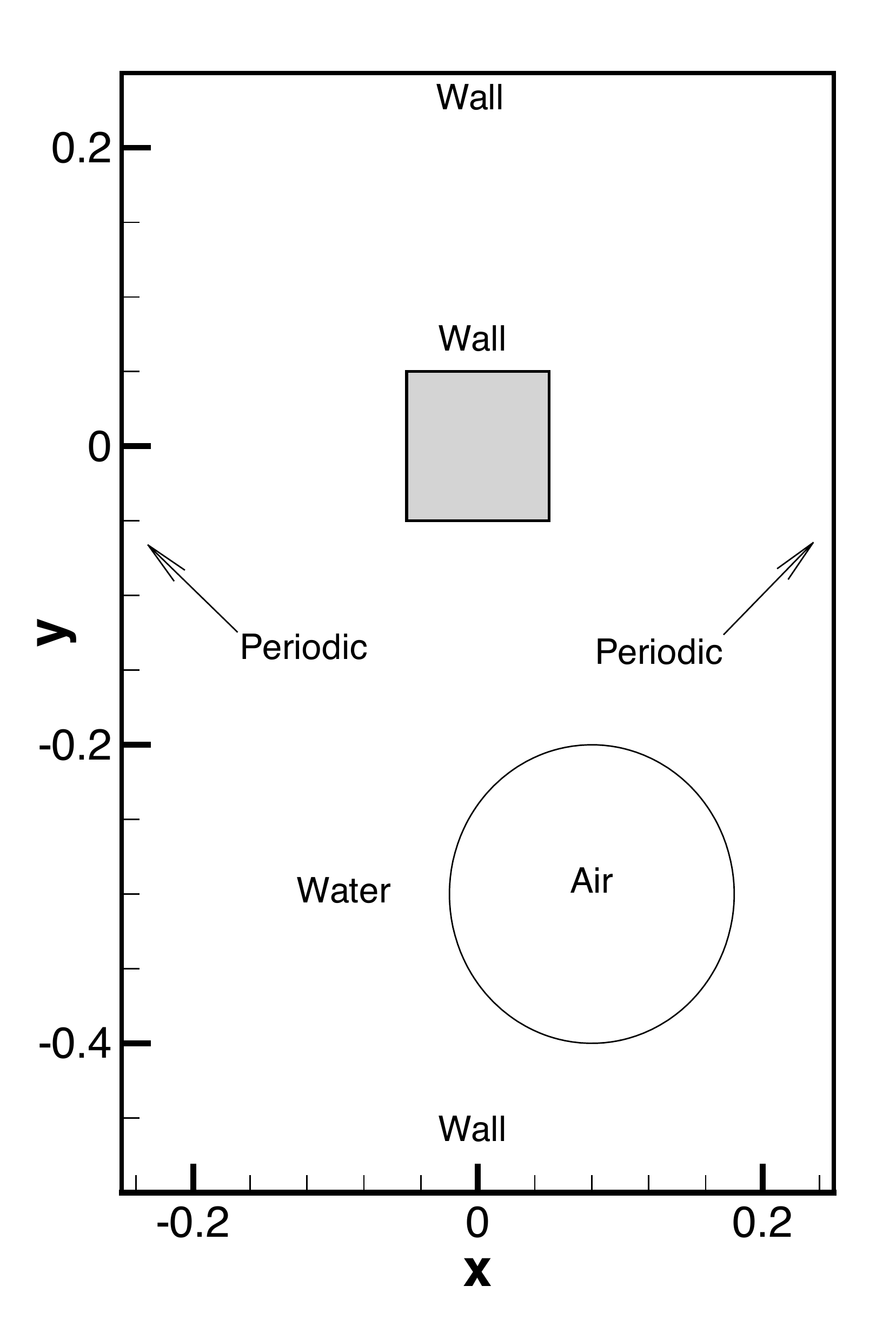}} 
\subfigure[$t=1.0$]{ \includegraphics[scale=.23]{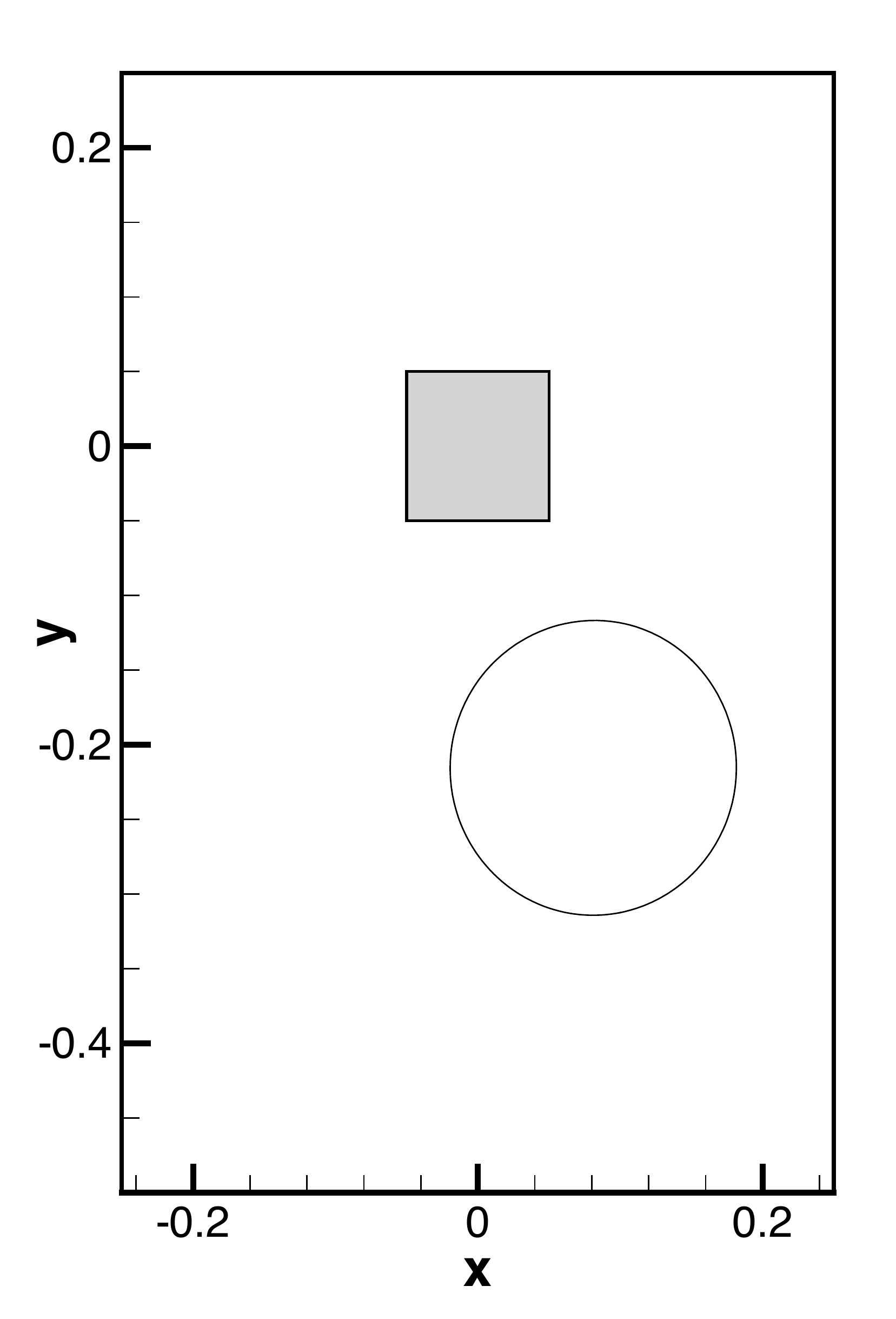}}
\subfigure[$t=1.6$]{ \includegraphics[scale=.23]{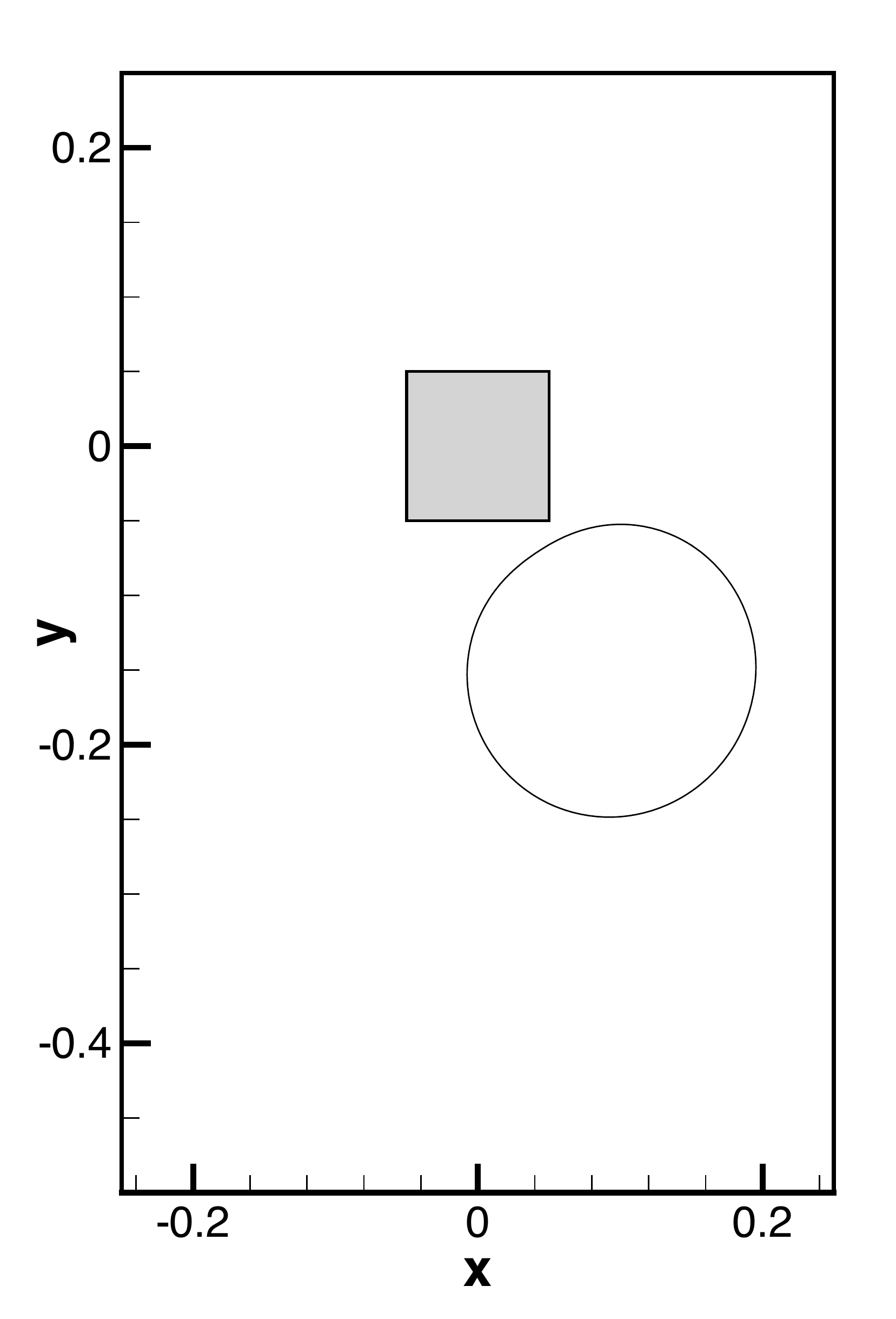}} 
\subfigure[$t=1.7$]{ \includegraphics[scale=.23]{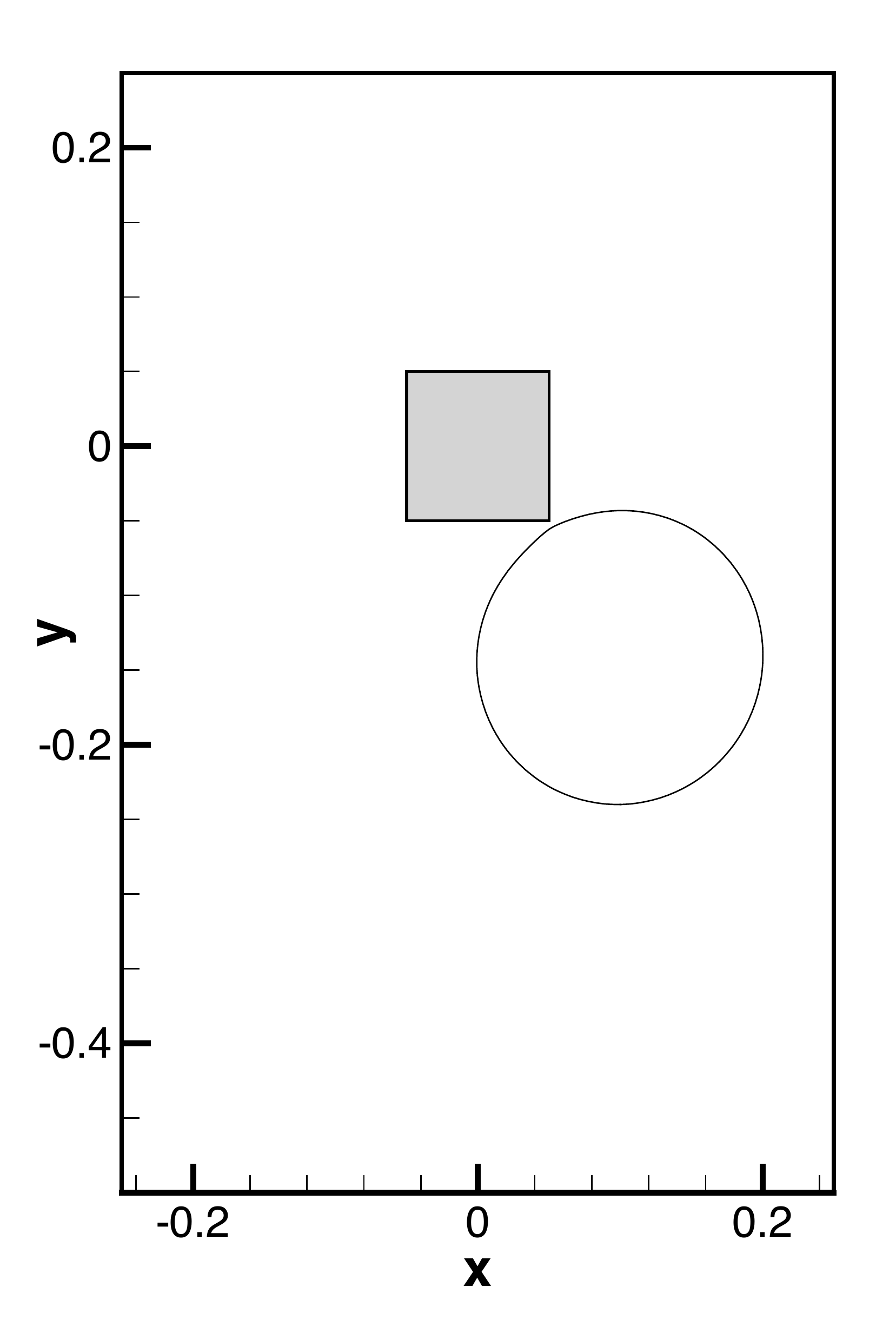}}\\
 \subfigure[$t=1.75$]{ \includegraphics[scale=.23]{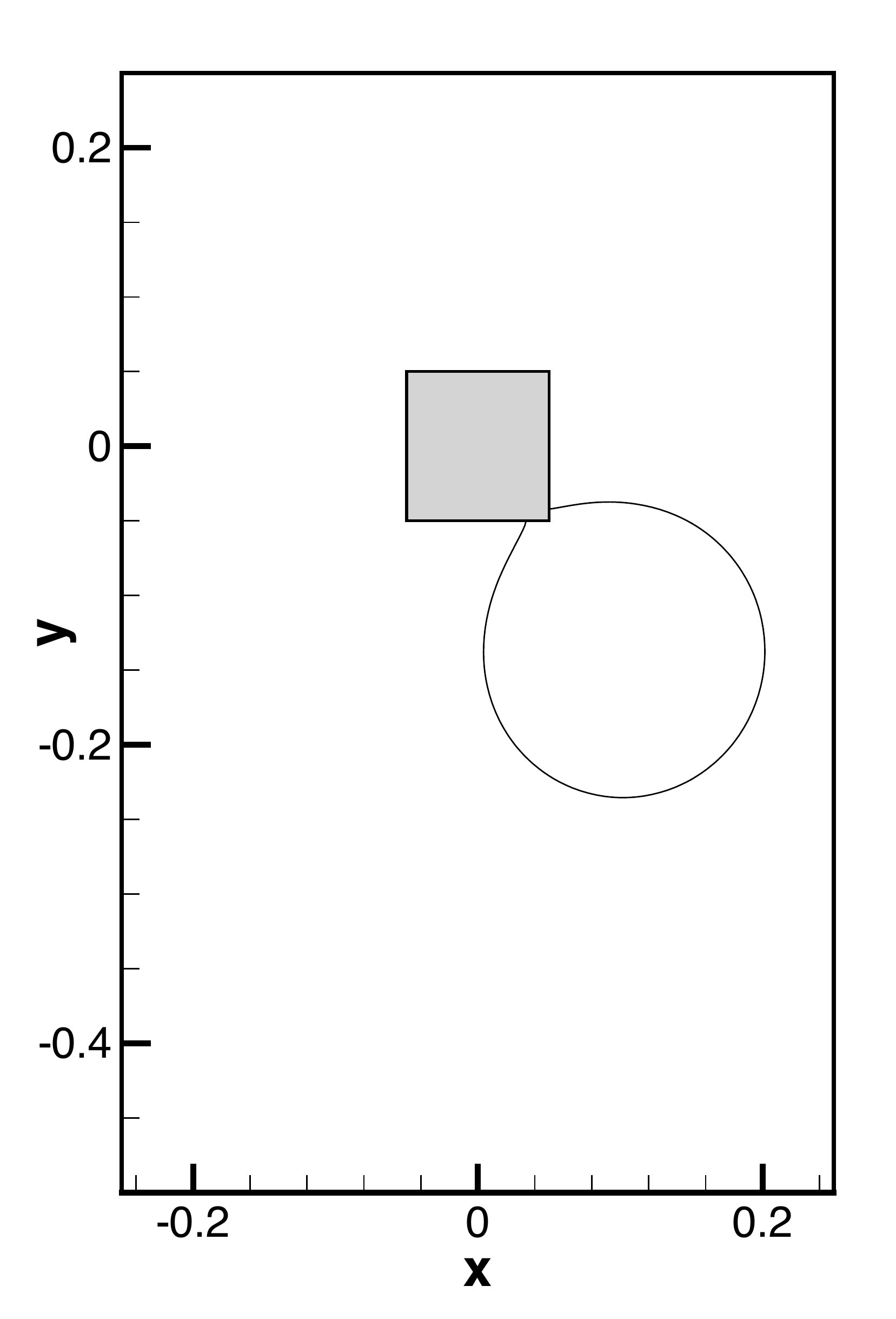}} 
\subfigure[$t=1.9$]{ \includegraphics[scale=.23]{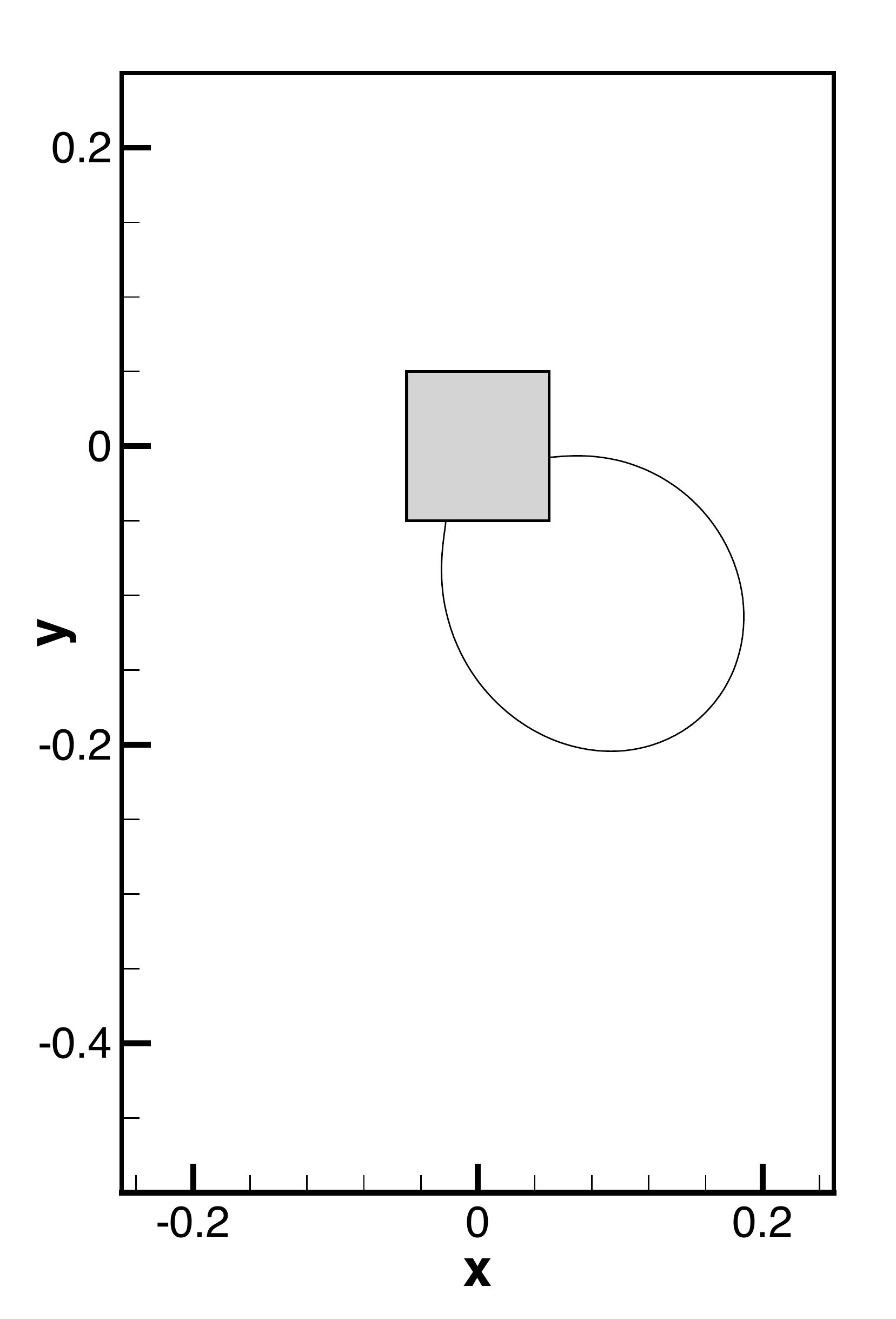}}
\subfigure[$t=2.1$]{ \includegraphics[scale=.23]{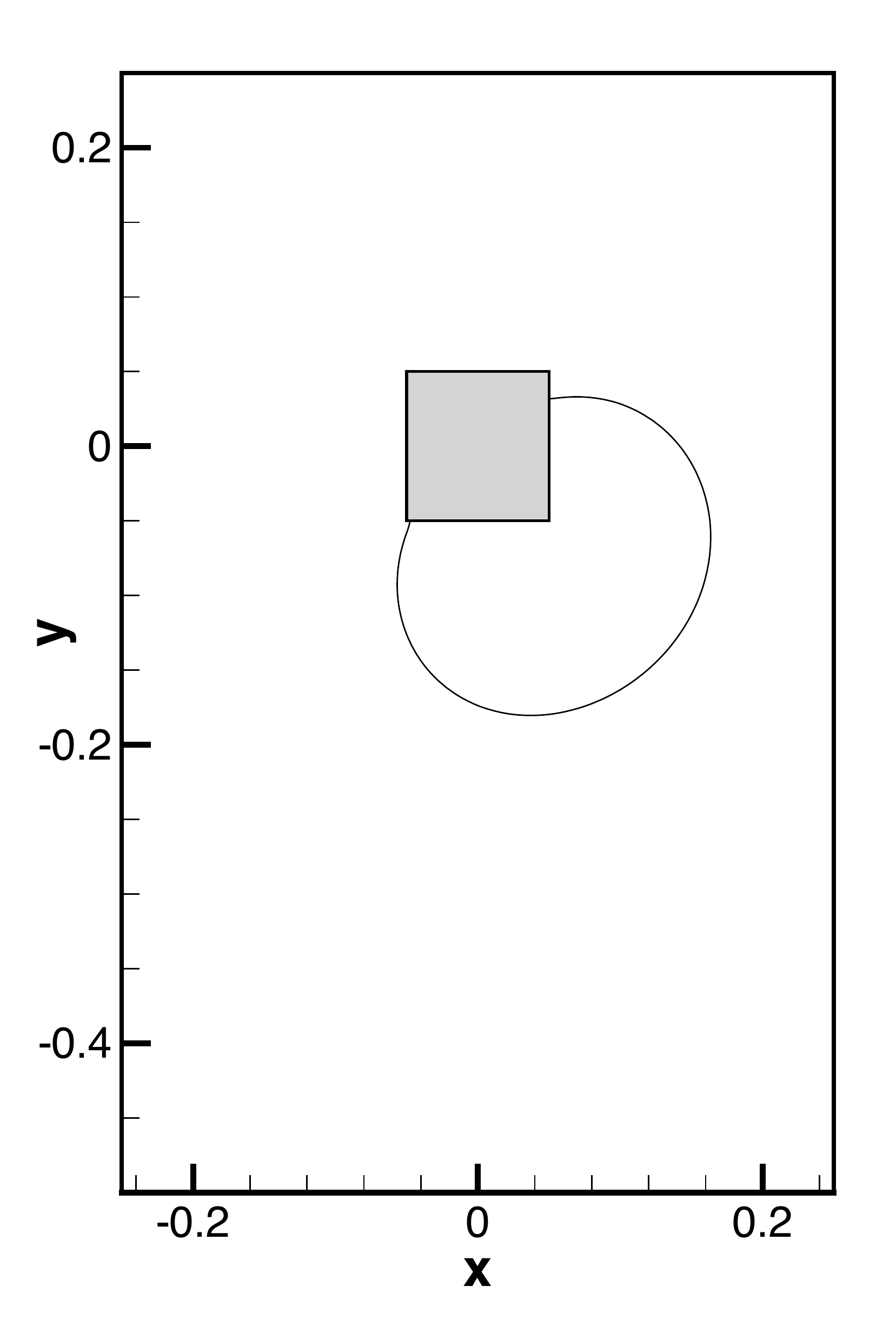}} 
\subfigure[$t=2.25$]{ \includegraphics[scale=.23]{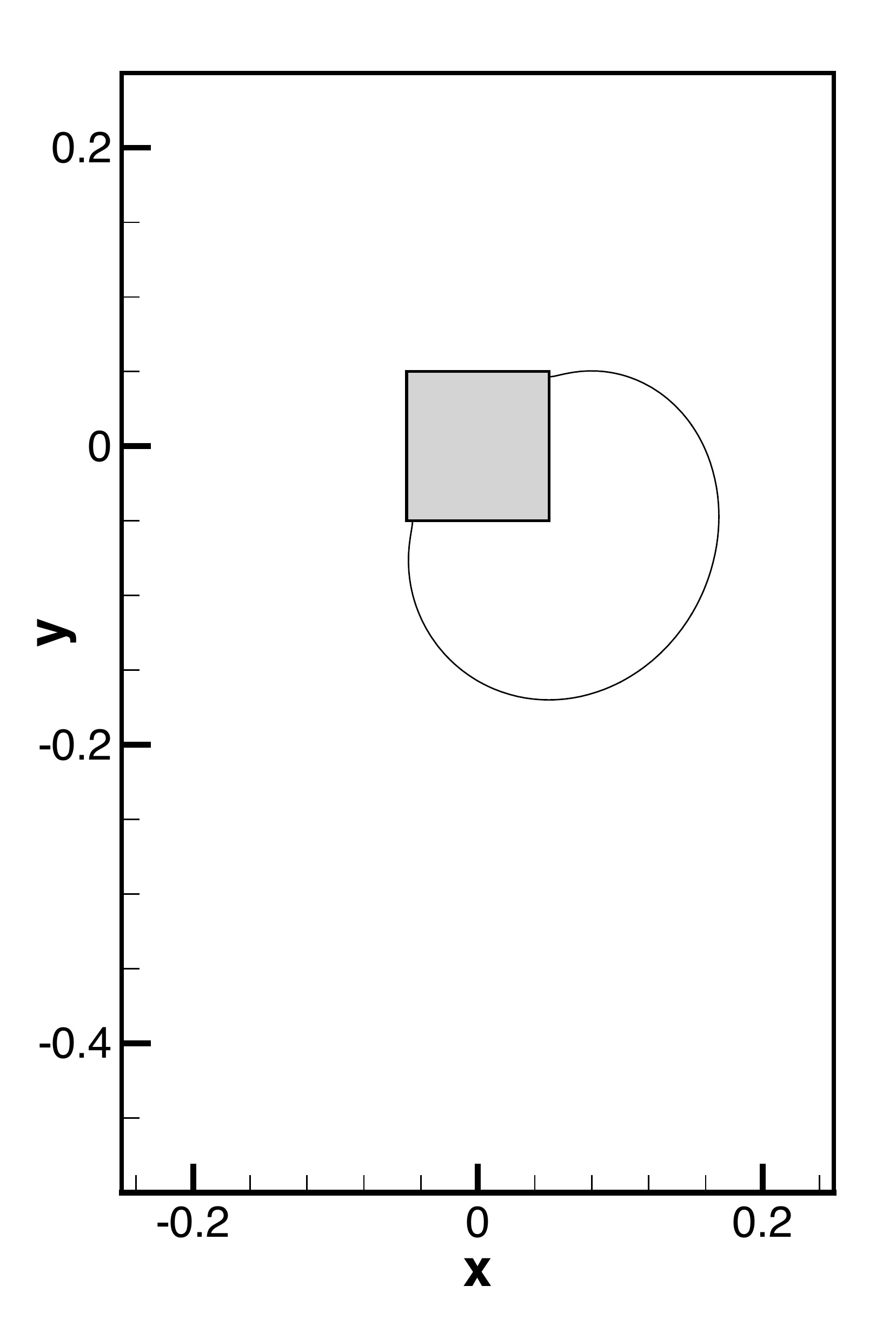}}\\
 \subfigure[$t=2.85$]{ \includegraphics[scale=.23]{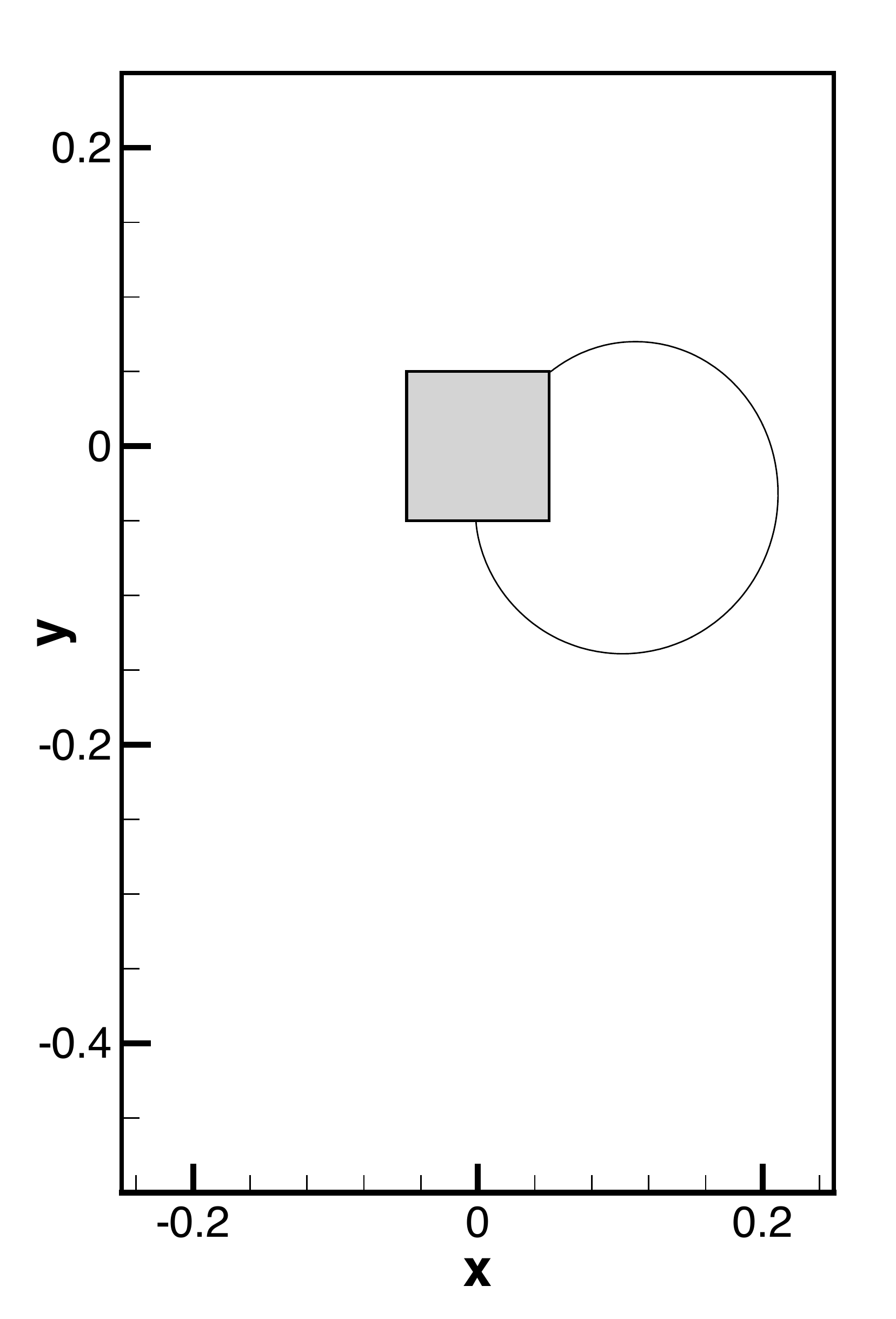}} 
\subfigure[$t=3.25$]{ \includegraphics[scale=.23]{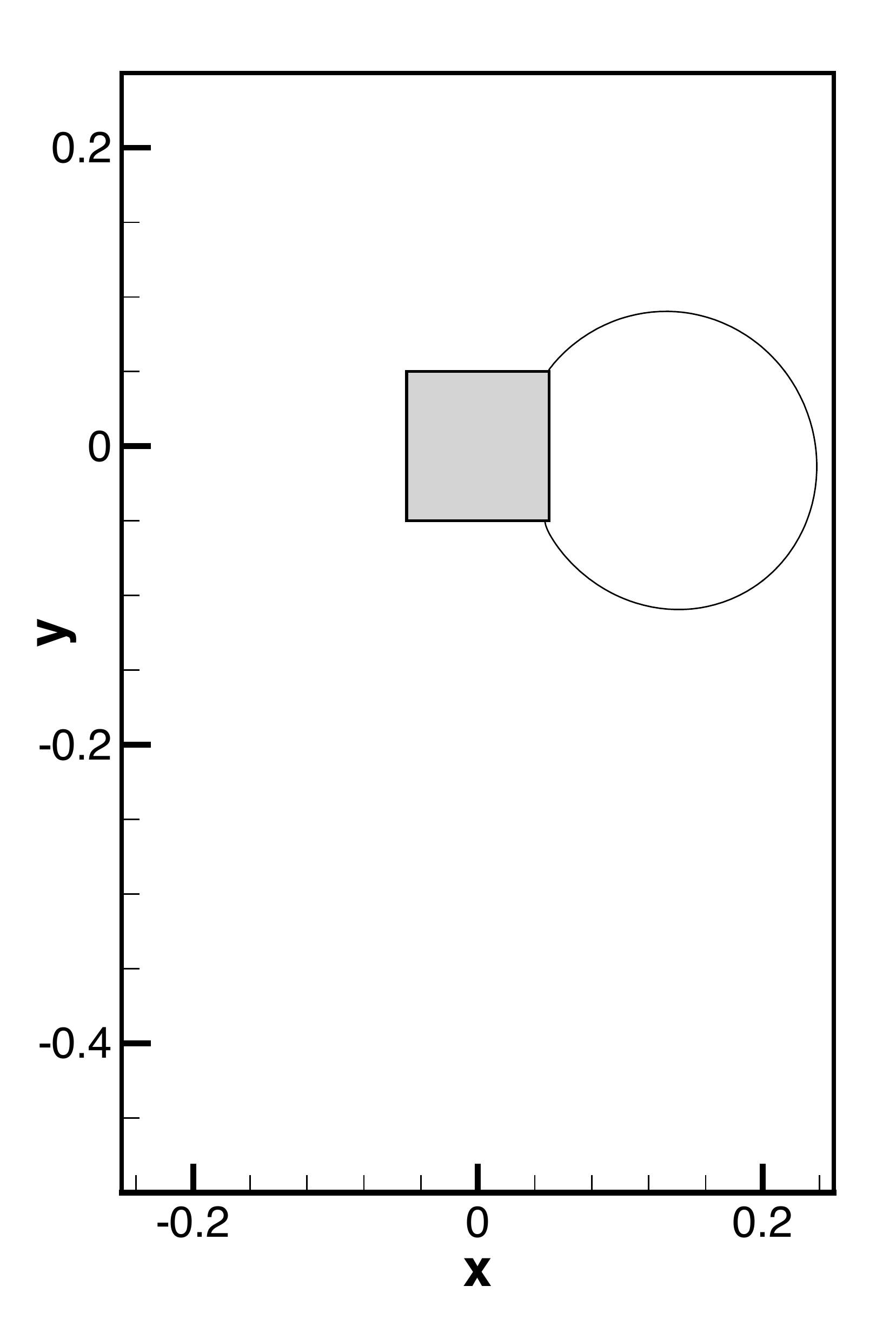}}
\subfigure[$t=4.2$]{ \includegraphics[scale=.23]{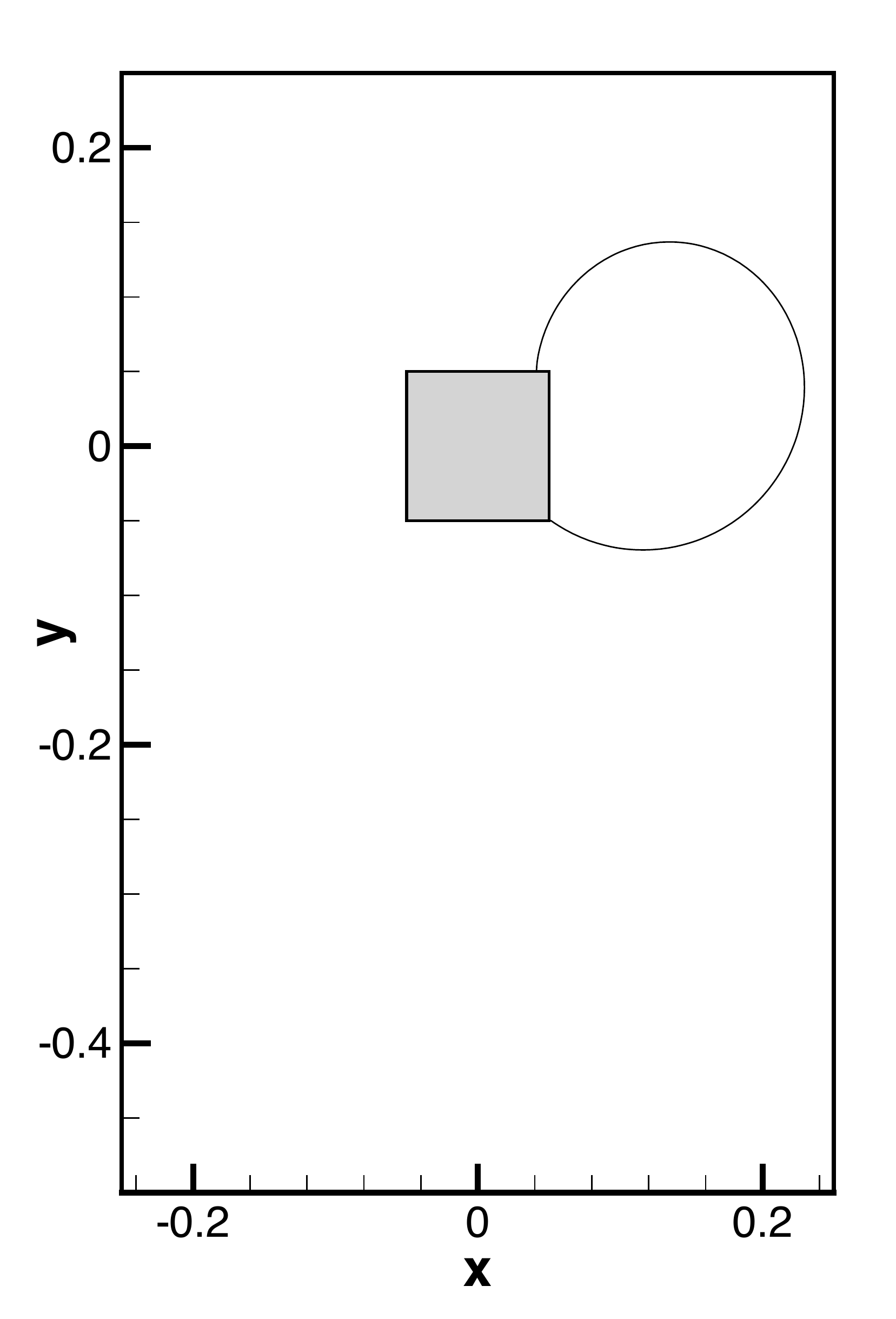}} 
\subfigure[$t=6.0$]{ \includegraphics[scale=.23]{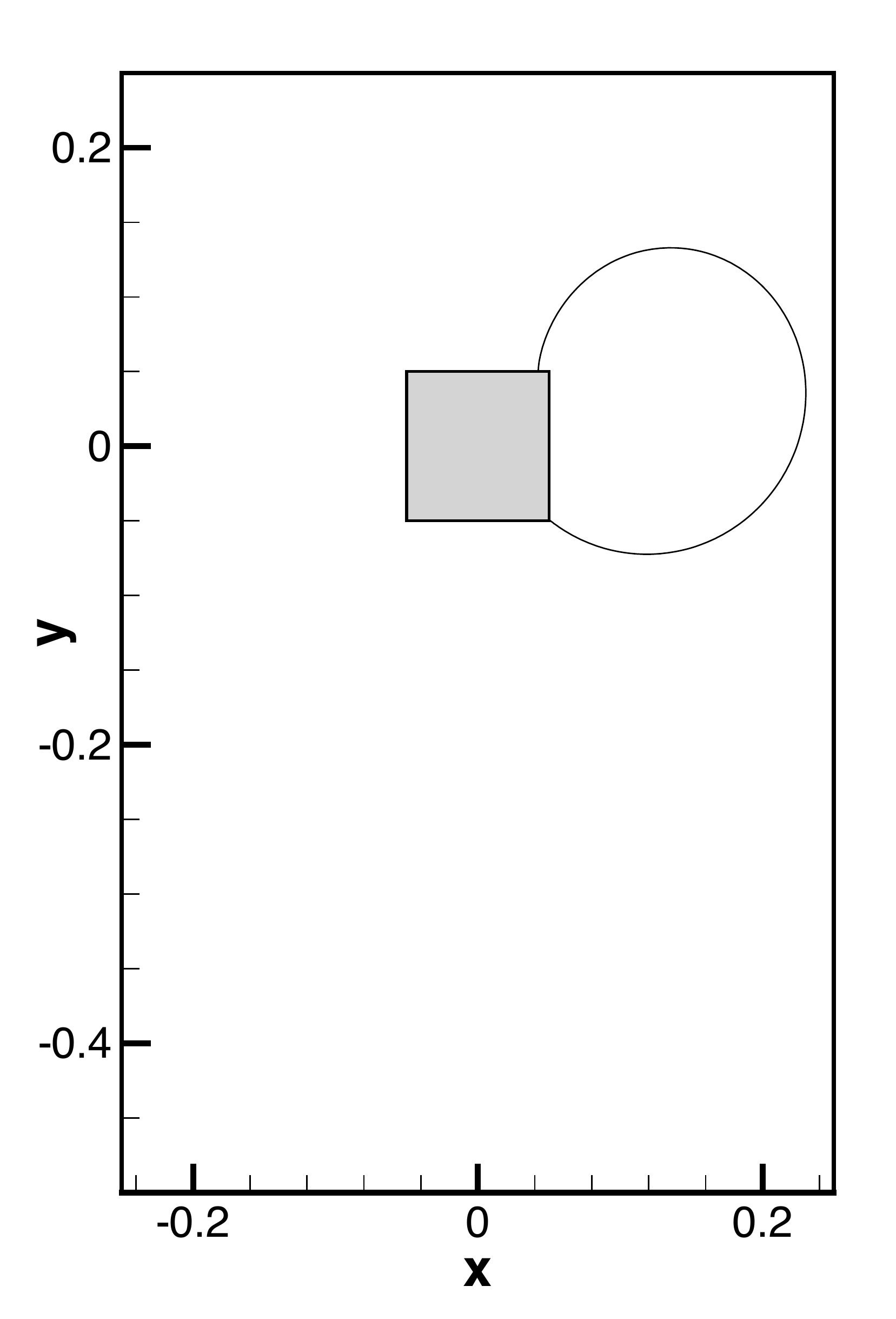}}\\
\caption{Time sequence of snapshots of an air bubble rising in water with
  a smaller density contrast $(\rho_1,\rho_2)=(1,100)$.
  Fluid interface is visualized by the contour levels $\phi(\bs x,t)=0$.
}
\label{fig:risingbubblecase1}
\end{figure}

To be more specific, we consider a rectangular domain,
defined by $\Omega=\big \{(x,y)| -0.25 \leq x\leq 0.25, \; -0.5 \leq y \leq 0.25   \big \}$. The top and bottom of the domain are solid walls,
and in the horizontal direction it is periodic.
A square region inside this domain, centered at
$(0,0)$ with a dimension $l=0.1$ on each side, is occupied by
a solid object; see Figure \ref{fig:risingbubblecase1}(a) for a sketch
of the configuration.
The flow (and computational) domain consists of the space between
the interior solid-square object and the top/bottom walls on the outside.
This domain is filled with water, which traps an air bubble inside;
see Figure \ref{fig:risingbubblecase1}(a). The gravity is assumed to
point downward.
At $t=0,$ the air bubble is circular
and centered at $(x_0,y_0)=(0.08,-0.3)$ with a
radius $R_0=0.1,$ and it is at rest. 
Then the system is released, and the air bubble rises through
the water and interacts with the walls of the interior object
and the outside domain wall. We would like to simulate this
dynamic process using the current method.

The computational domain is partitioned with $146$  quadrilateral elements,
with $10$ and $15$ uniform elements respectively in the
$x$ and $y$ directions (with $4$ elements excluded
 from the region of the solid square). 
 In the simulations, the external body force $\bs f$ in equation  \eqref{eq:NphaseEq1} is set to $\bs f=\rho \bs g_r,$ where $\bs g_r$ is the gravitational acceleration
and $\rho$ is the mixture density.
The source term in equation \eqref{eq:NphaseEq3} is set to $g=0.$
We further assume that all the wall surfaces have a neutral wettability.
In other words, if the air-water interface intersects
the outside domain walls or the walls of the interior solid square,
the contact angle on the wall would be $90^o.$
On the outside top/bottom walls and the surfaces of
the interior solid square, the boundary
condition \eqref{eq:bc1} with $\bs w=\bs 0$ is imposed for the velocity,
and the boundary conditions \eqref{eq:bc2} and \eqref{eq:bc3} with $d_a=d_b=0$ are imposed for the phase field function.
On the two sides of the domain in the horizontal direction ($x=\pm 0.25$)
periodic boundary conditions are imposed for all flow variables.
The initial velocity is set to zero, and the initial phase field function is prescribed as follows:
\begin{equation}
\phi(\bs x,0)=-\tanh\frac{ \sqrt{(x-x_0)^2+(y-y_0)^2-R_0}  }{\sqrt{2}\eta}.
\end{equation}
In the simulations, we treat air as the first fluid and
water as the second fluid. We employ a viscosity
ratio $\frac{\mu_2}{\mu_1}=50$ and a density ratio $\frac{\rho_2}{\rho_1}=1000$
for the air and water. Another artificial
density ratio $\frac{\rho_2}{\rho_1}=100$
is also considered in the tests.
The normalized physical and numerical parameters involved in
this problem are listed in Table~\ref{table:risingbubble}.

We first consider the case with an artificially smaller density contrast
for air and water,
$(\rho_1,\rho_2)=(1,100).$
Figure \ref{fig:risingbubblecase1} is a sequence of snapshots in
time of the fluid interfaces for this case,
visualized by the contour level $\phi(\bs x,t)=0.$
These results are computed with a time step size
$\Delta t=5\times 10^{-5}$.
From $t=0$ to about $t=1.6$ (Figure \ref{fig:risingbubblecase1}(a)-(c)), the air bubble rises through the water inside the container,
and gradually approaches the bottom
right corner of the interior square object. A slight deformation of the bubble
can be noticed. The air bubble collides onto the square object at
approximately $t=1.7$ (Figure \ref{fig:risingbubblecase1}(d)), and is drawn
rapidly towards the cylinder due to the strong effect of surface
tension (Figure \ref{fig:risingbubblecase1}(e)-(g)).
From $t=2.25$ to $t=6$ (Figure \ref{fig:risingbubblecase1}(h)-(l)),
the air bubble slides on the bottom surface of the interior square object, and moves
toward the right.
It eventually settles down at an equilibrium position, still attached to
the right and top sides of the solid square.

\begin{figure}[htbp]
\subfigure[$t=0.00005$]{ \includegraphics[scale=.18]{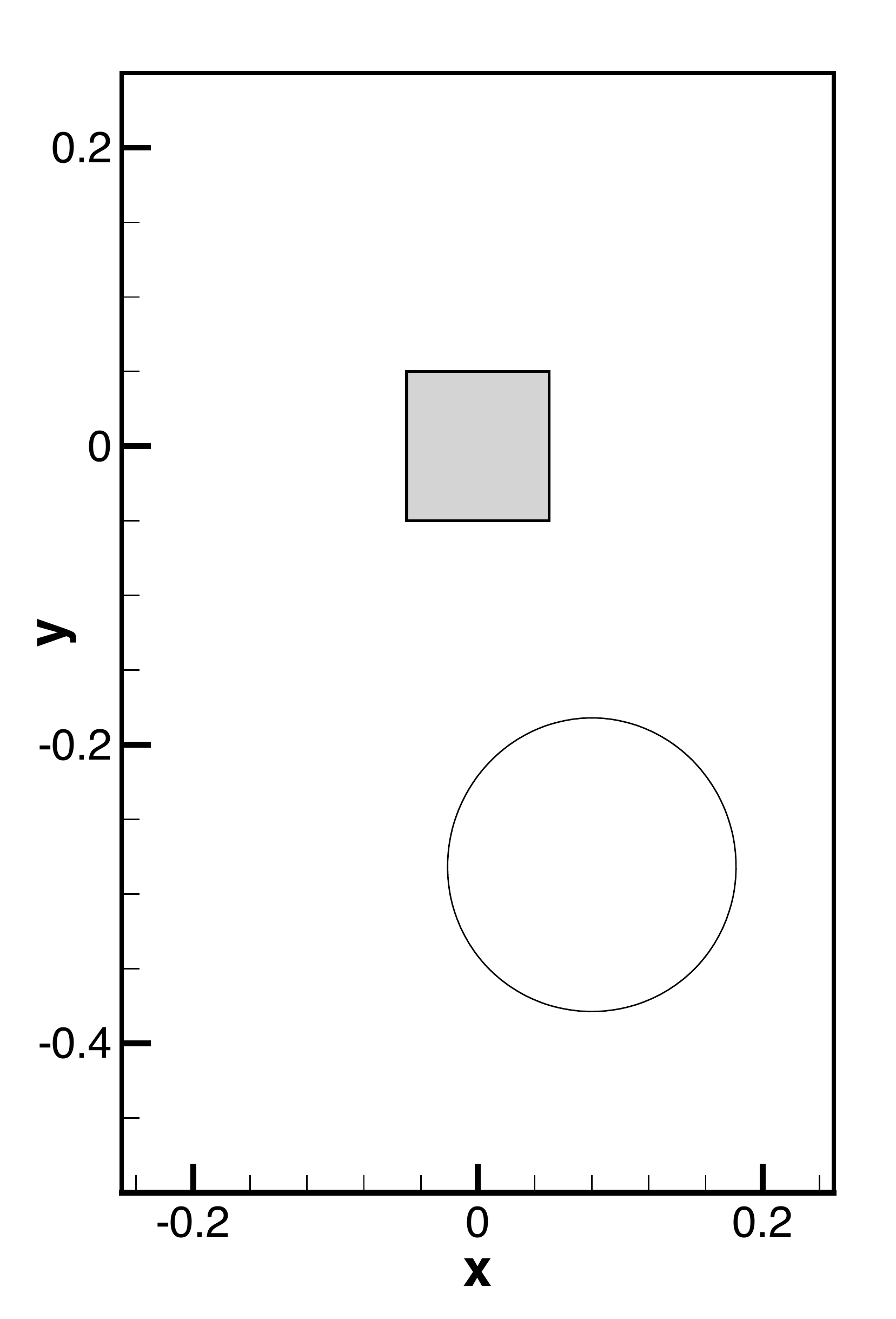}} 
\subfigure[$t=0.5$]{ \includegraphics[scale=.18]{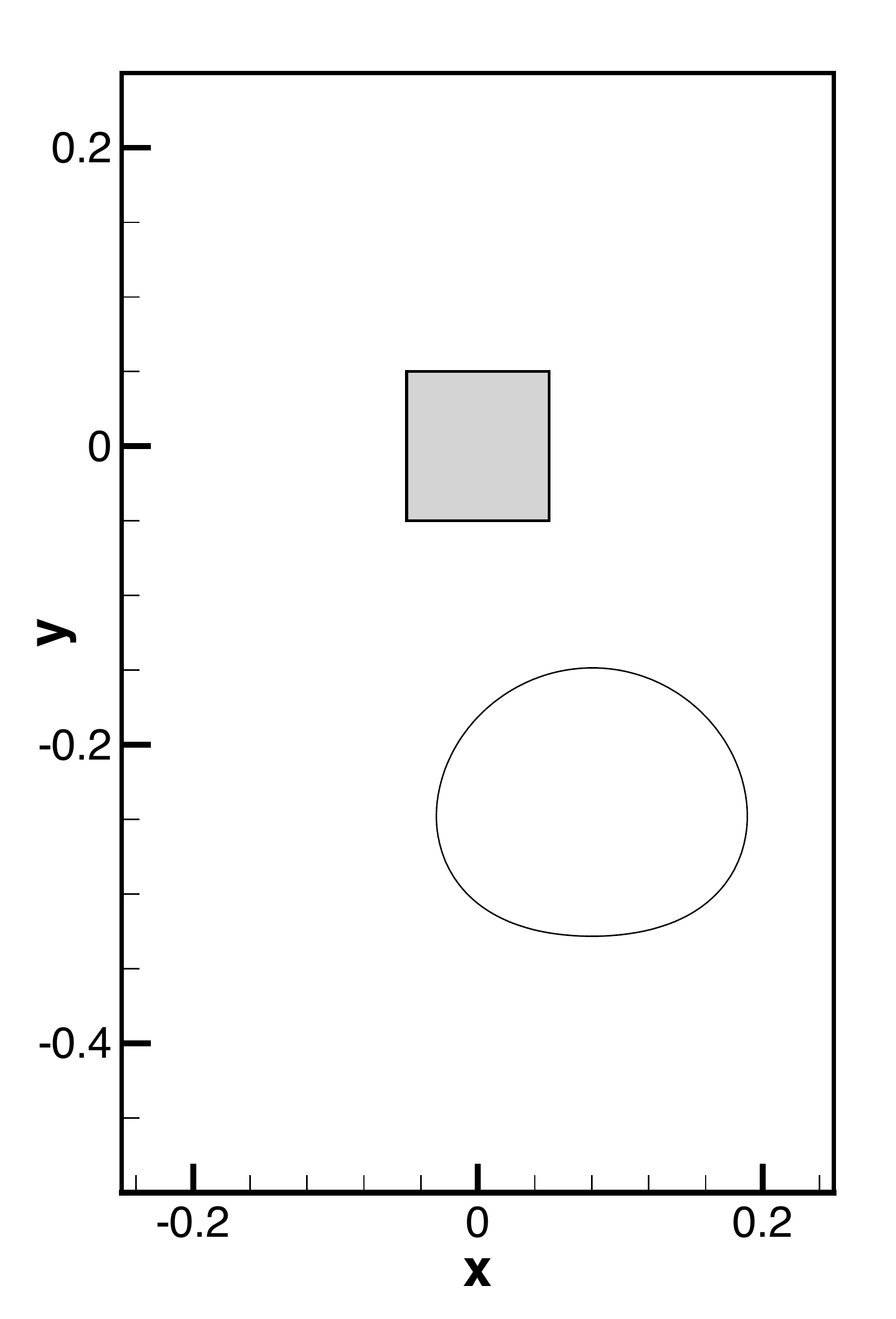}}
\subfigure[$t=0.85$]{ \includegraphics[scale=.18]{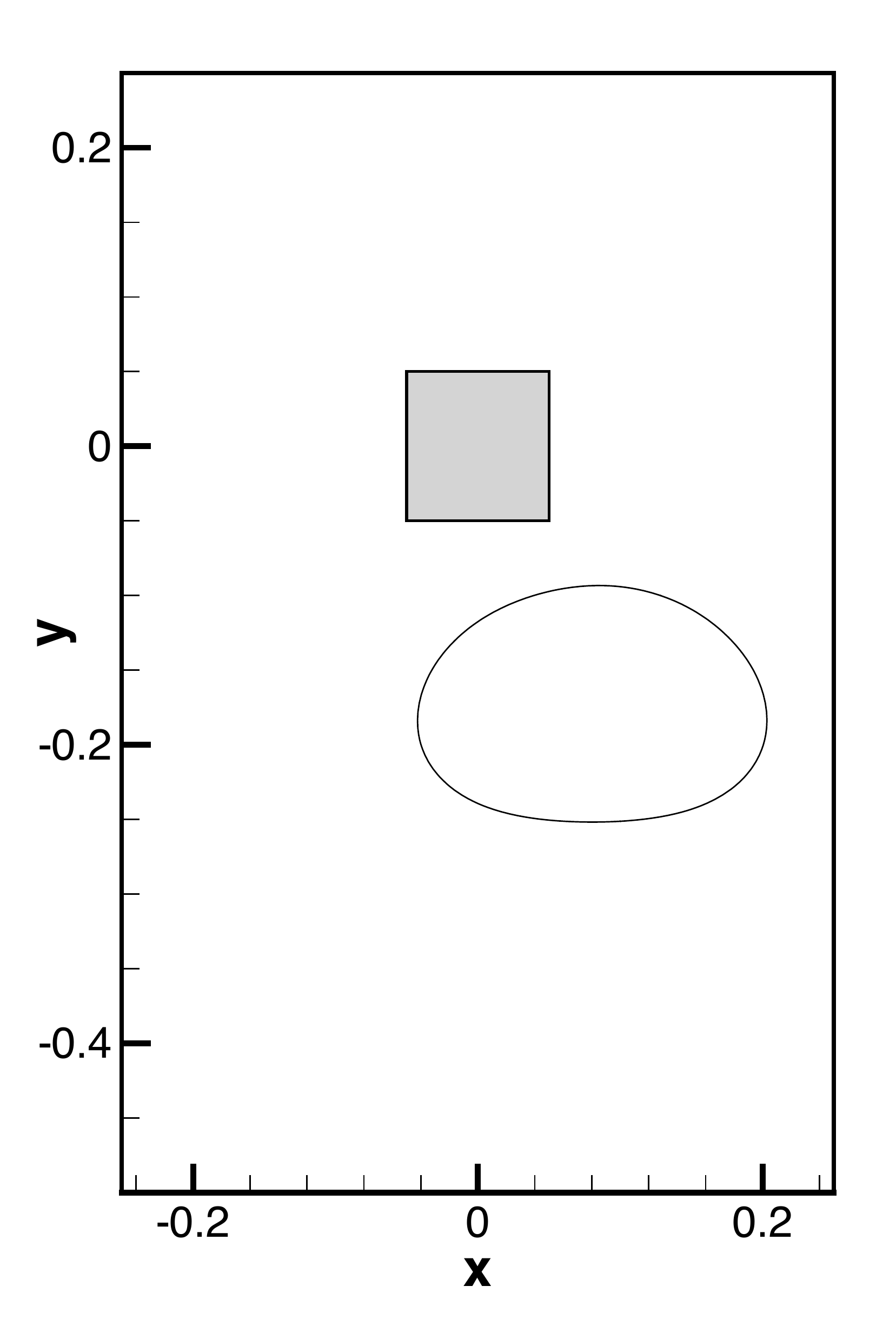}}
\subfigure[$t=1.0$]{ \includegraphics[scale=.18]{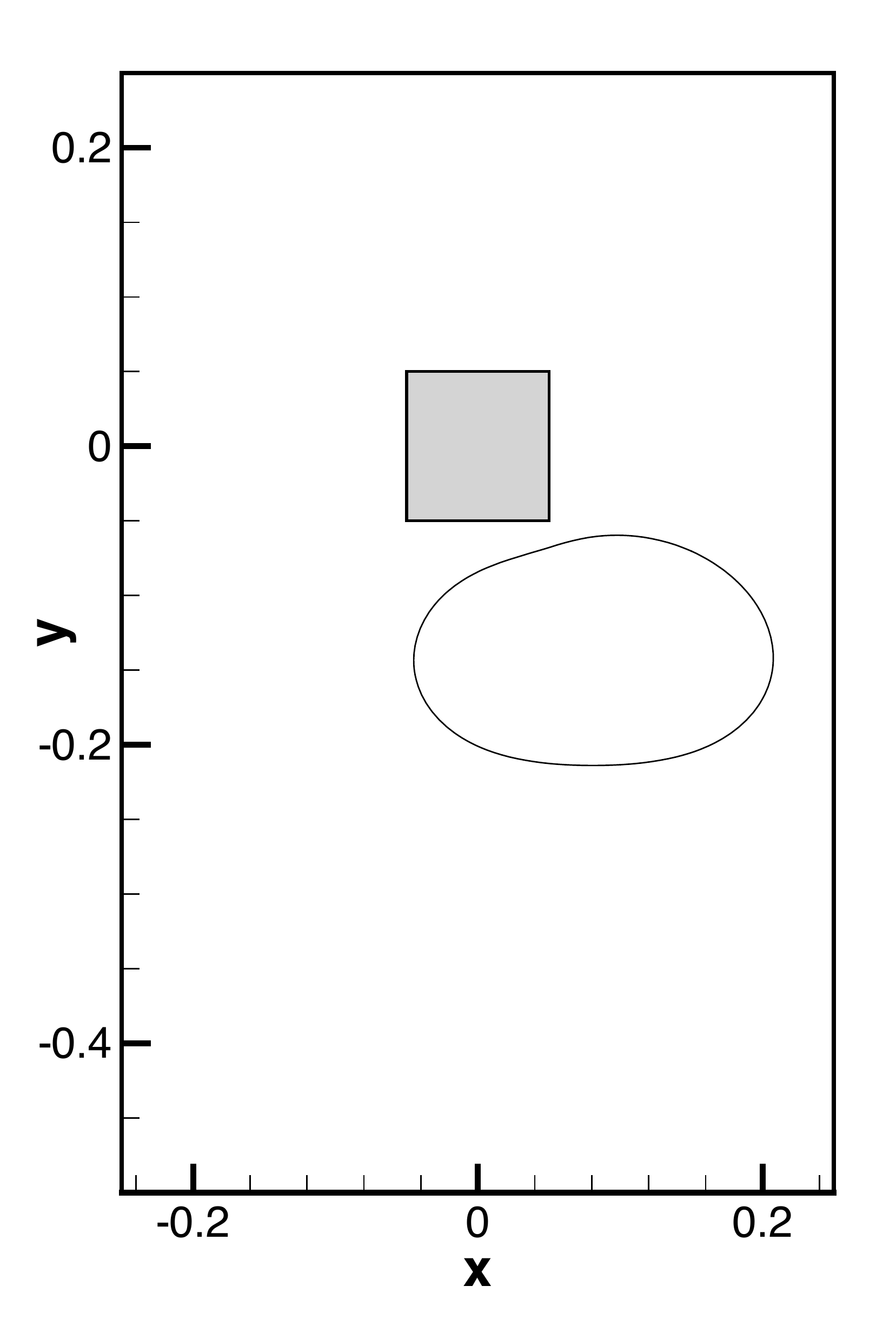}}
\subfigure[$t=1.1$]{ \includegraphics[scale=.18]{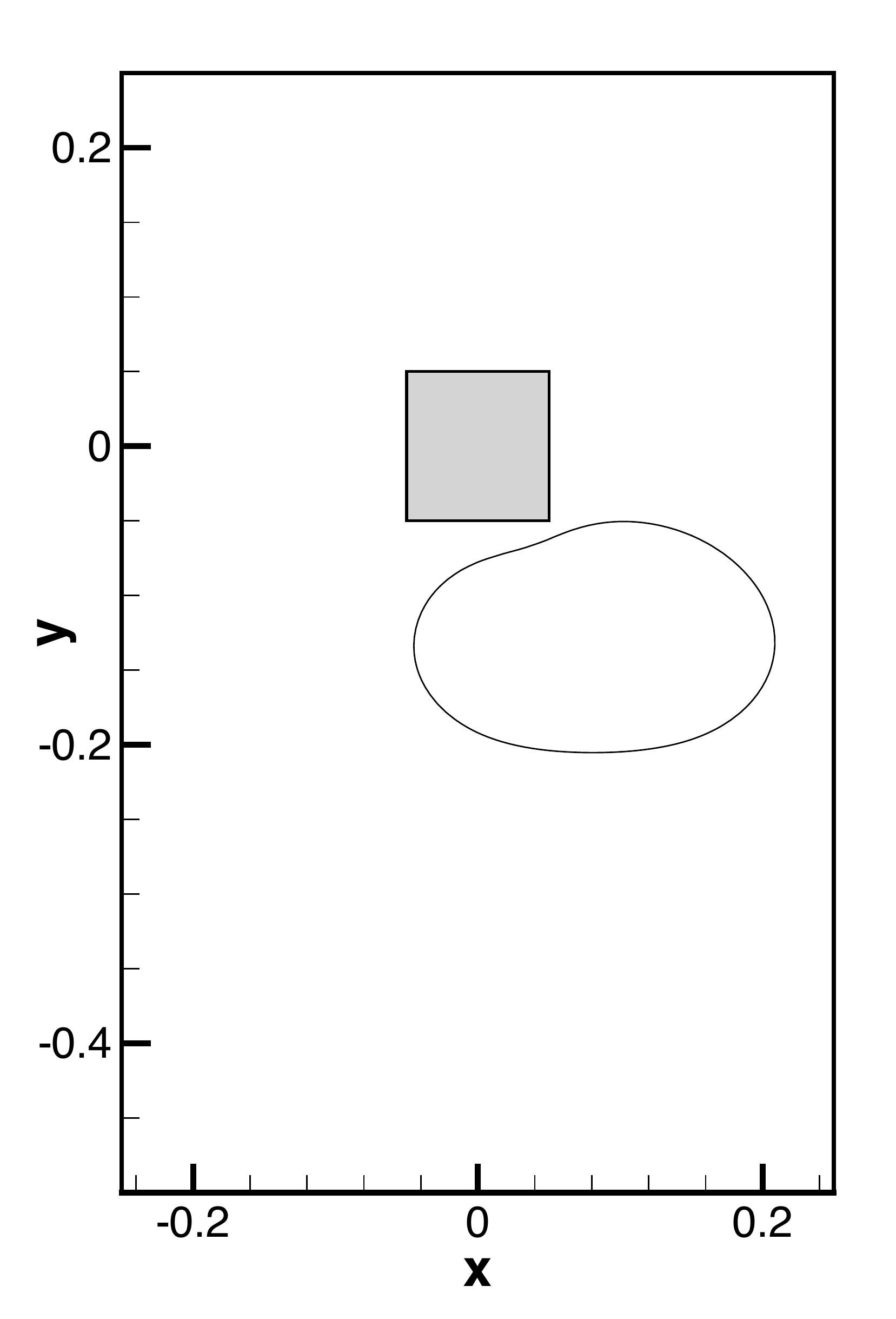}}\\
 \subfigure[$t=1.2$]{ \includegraphics[scale=.18]{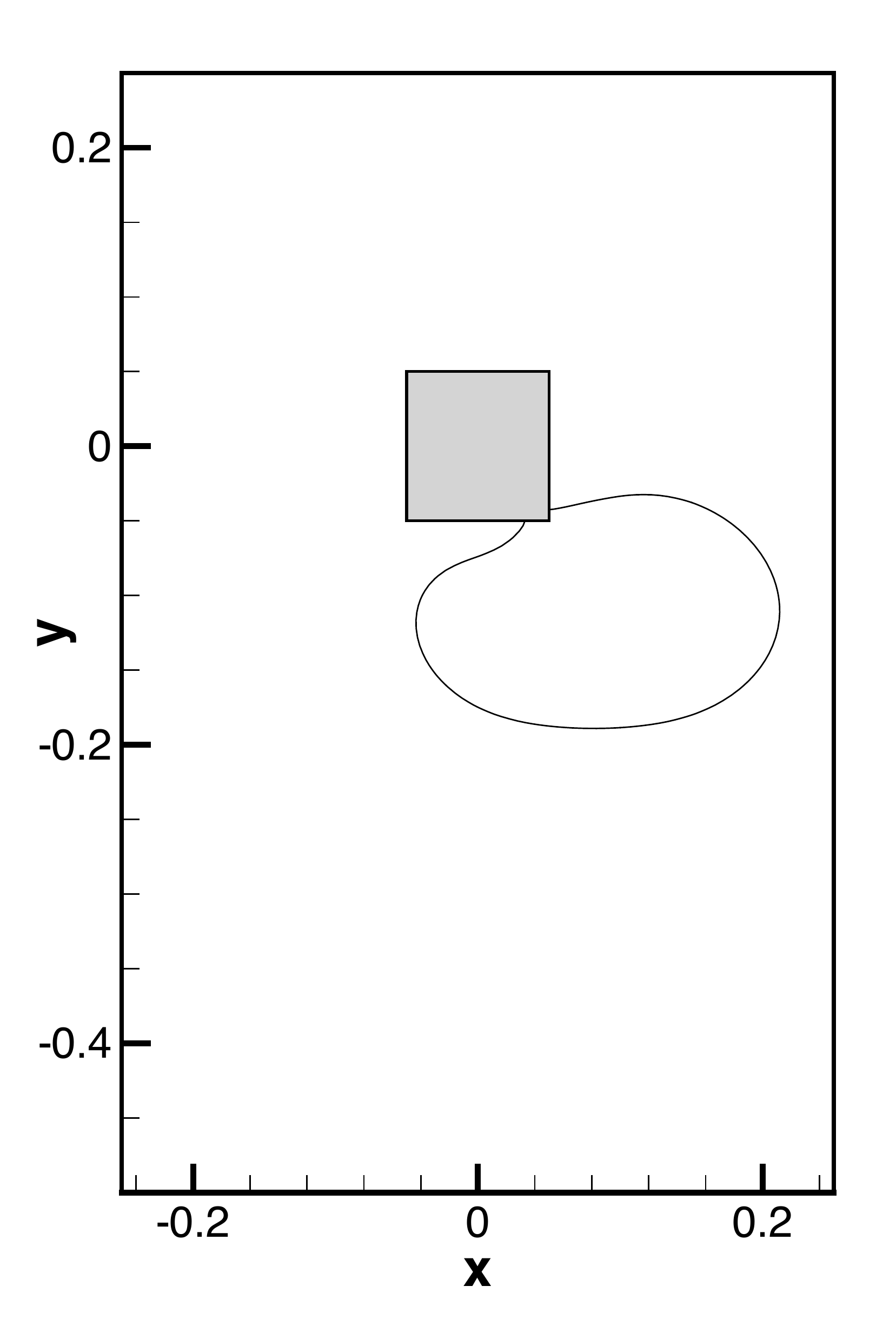}} 
\subfigure[$t=1.3$]{ \includegraphics[scale=.18]{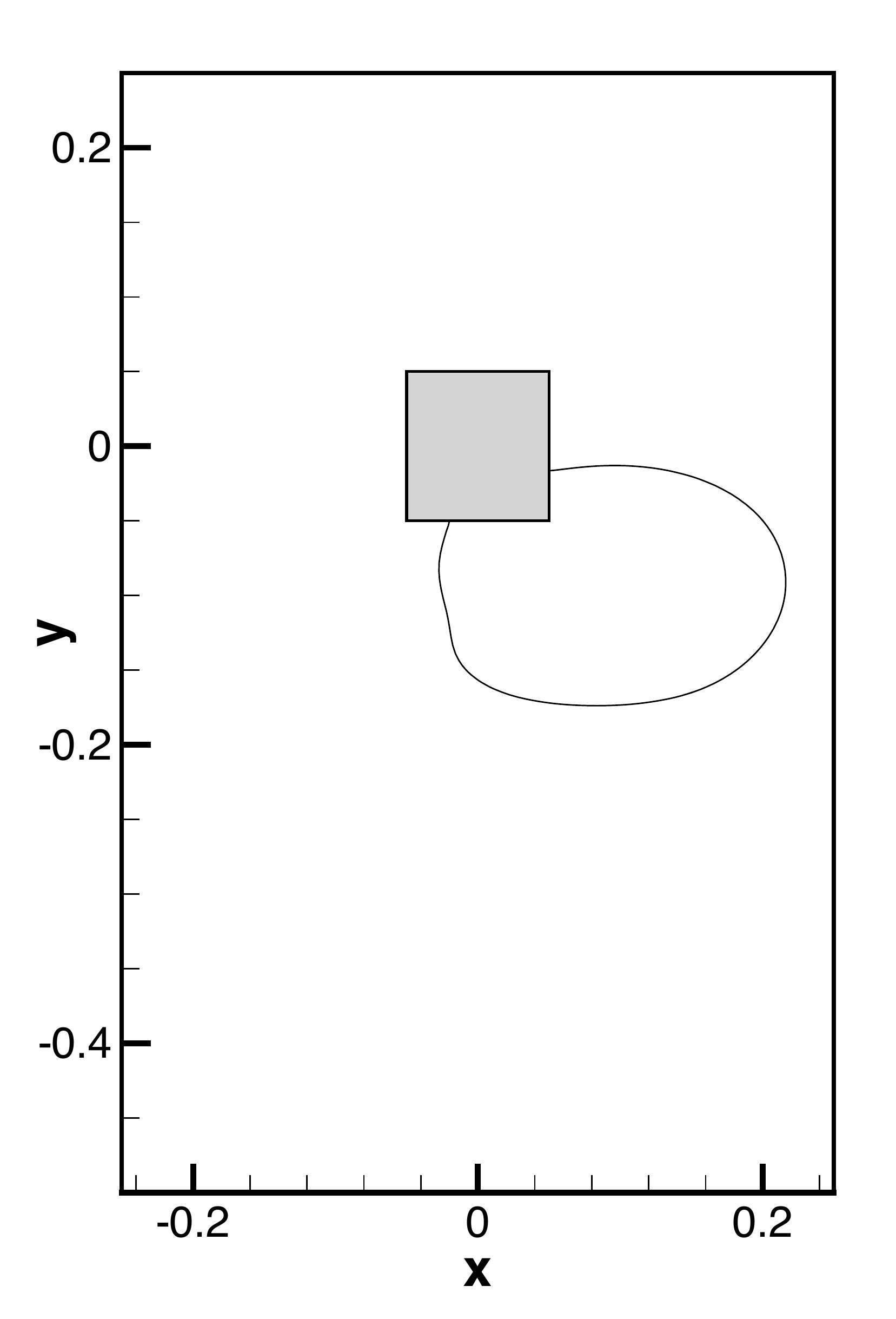}} 
\subfigure[$t=1.55$]{ \includegraphics[scale=.18]{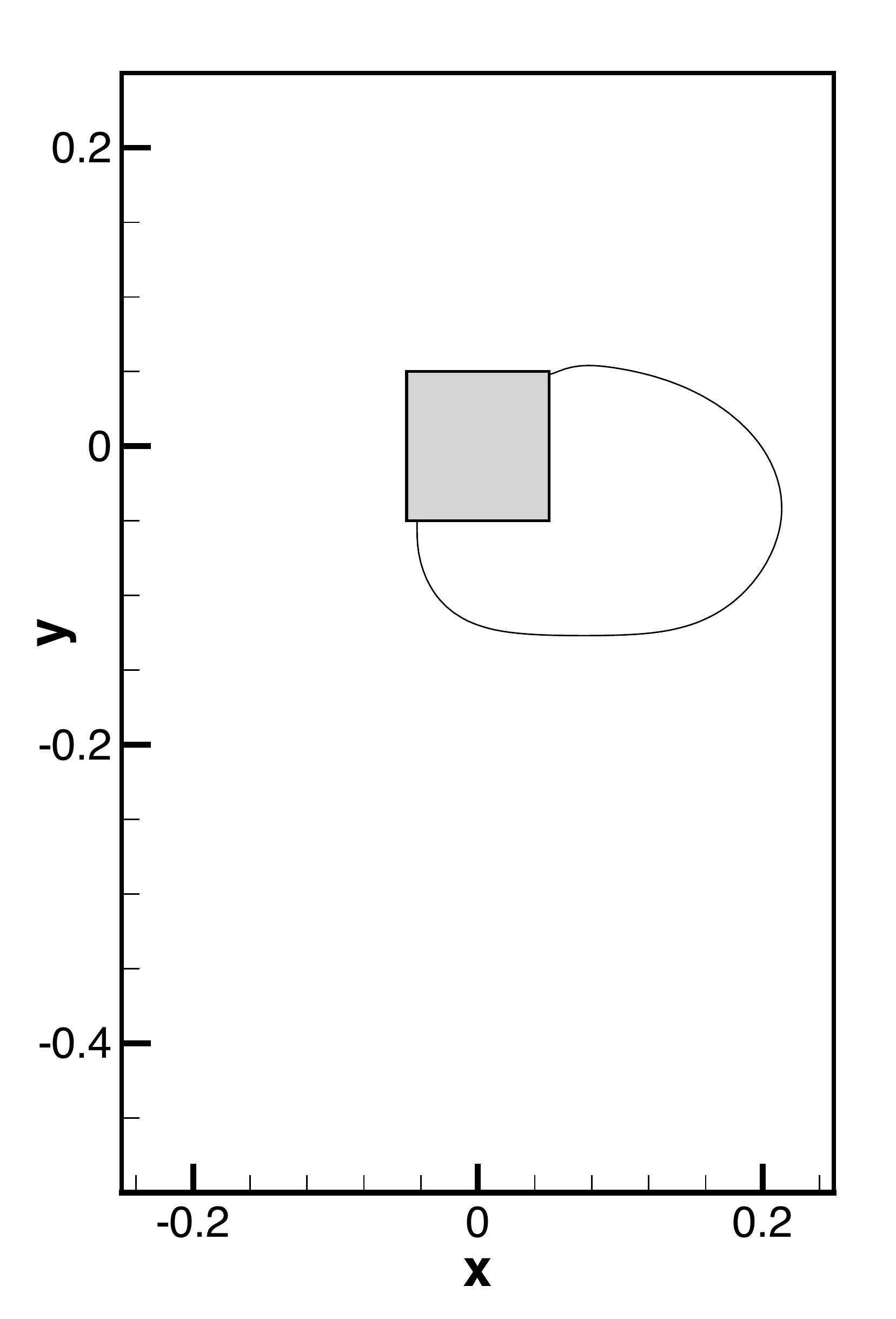}}
\subfigure[$t=1.8$]{ \includegraphics[scale=.18]{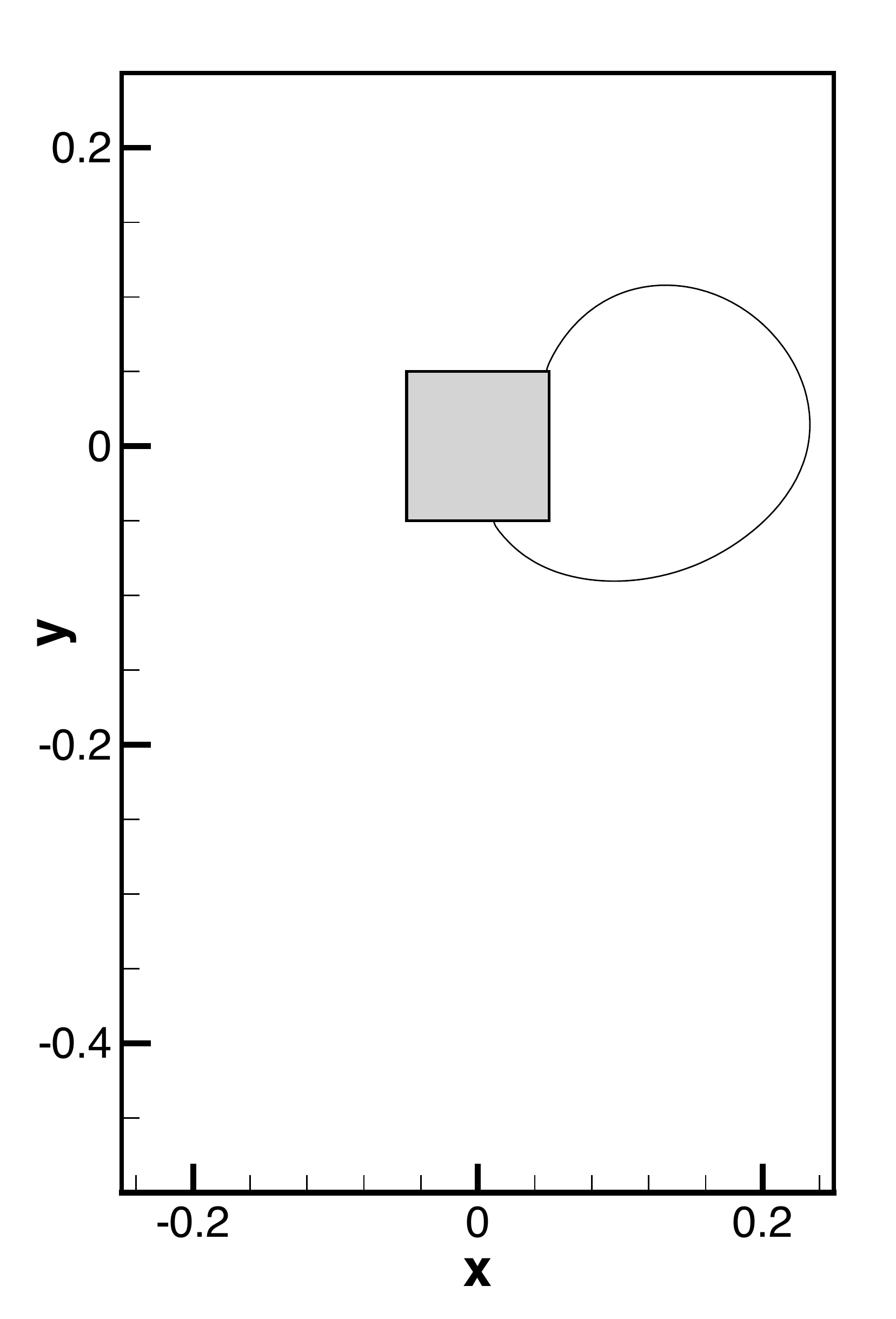}}
\subfigure[$t=1.9$]{ \includegraphics[scale=.18]{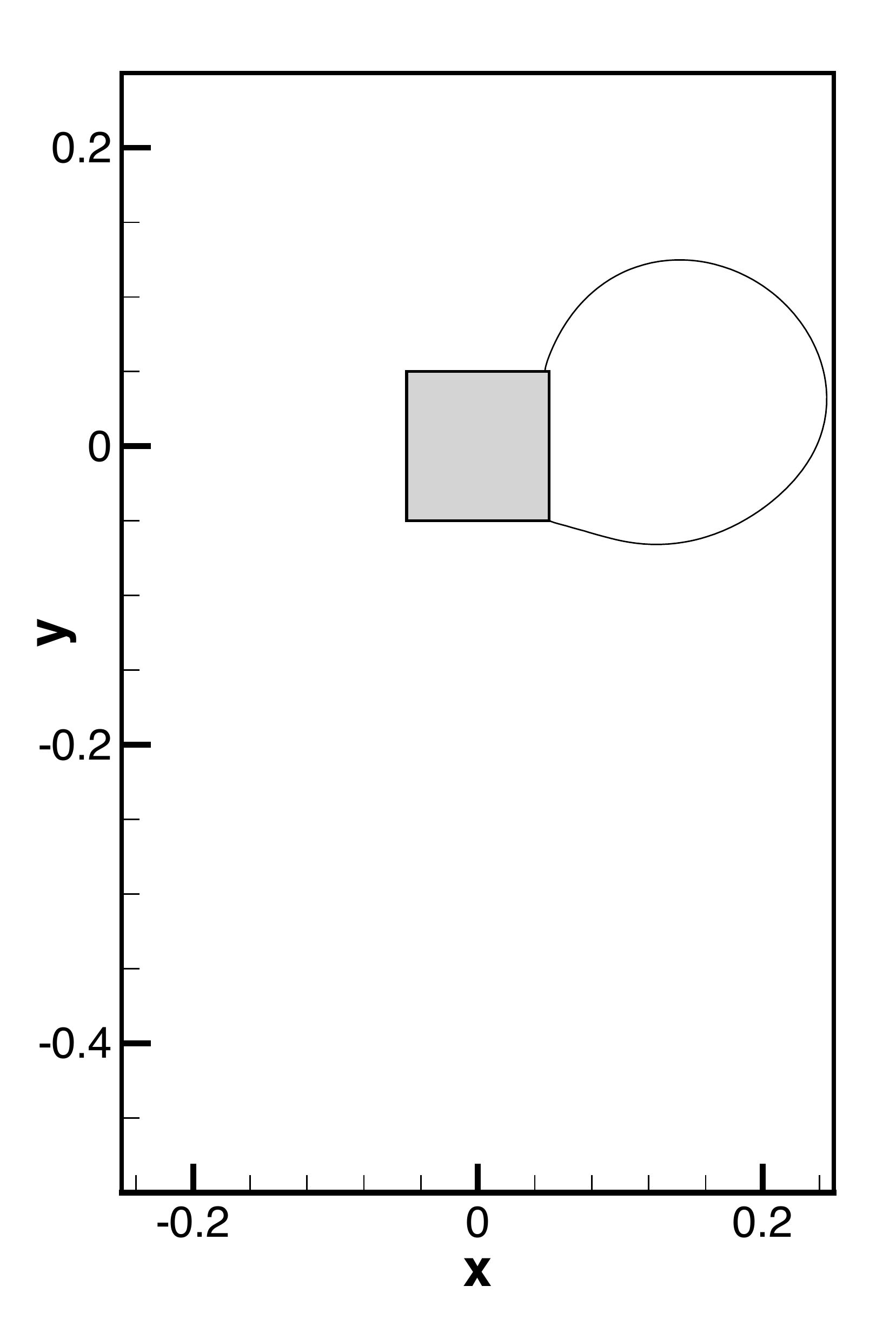}}\\
 \subfigure[$t=2.7$]{ \includegraphics[scale=.18]{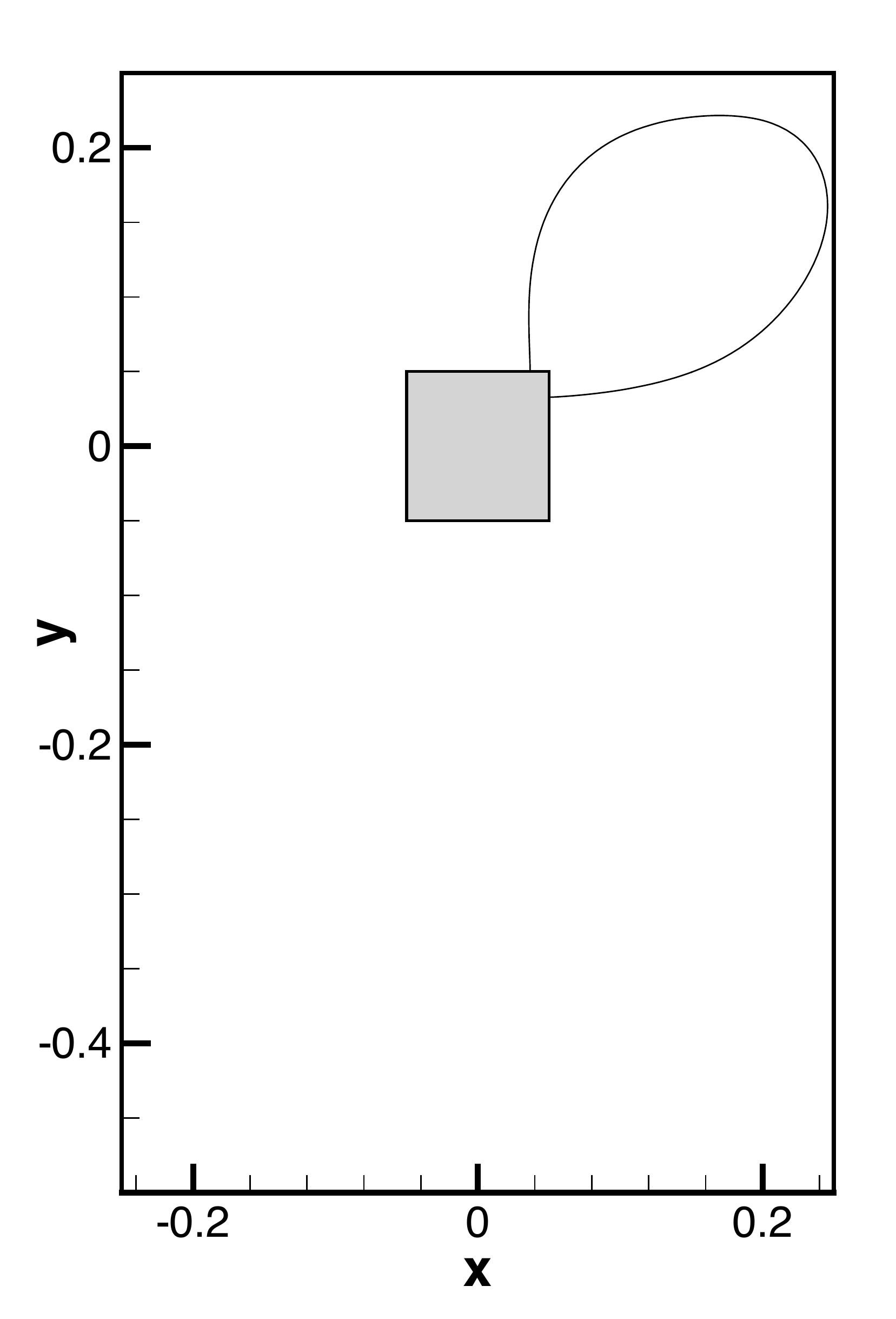}} 
\subfigure[$t=3.25$]{ \includegraphics[scale=.18]{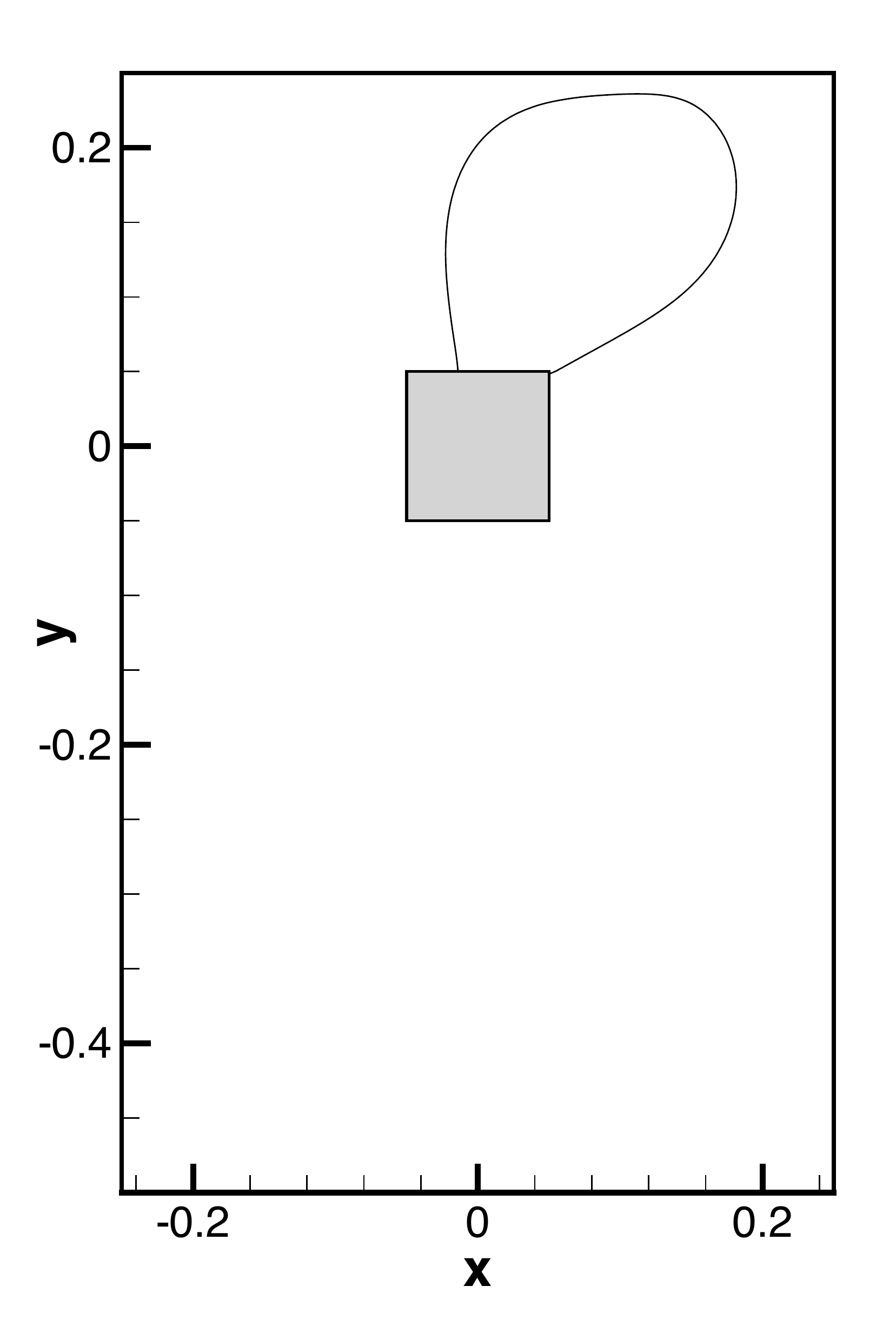}} 
\subfigure[$t=3.55$]{ \includegraphics[scale=.18]{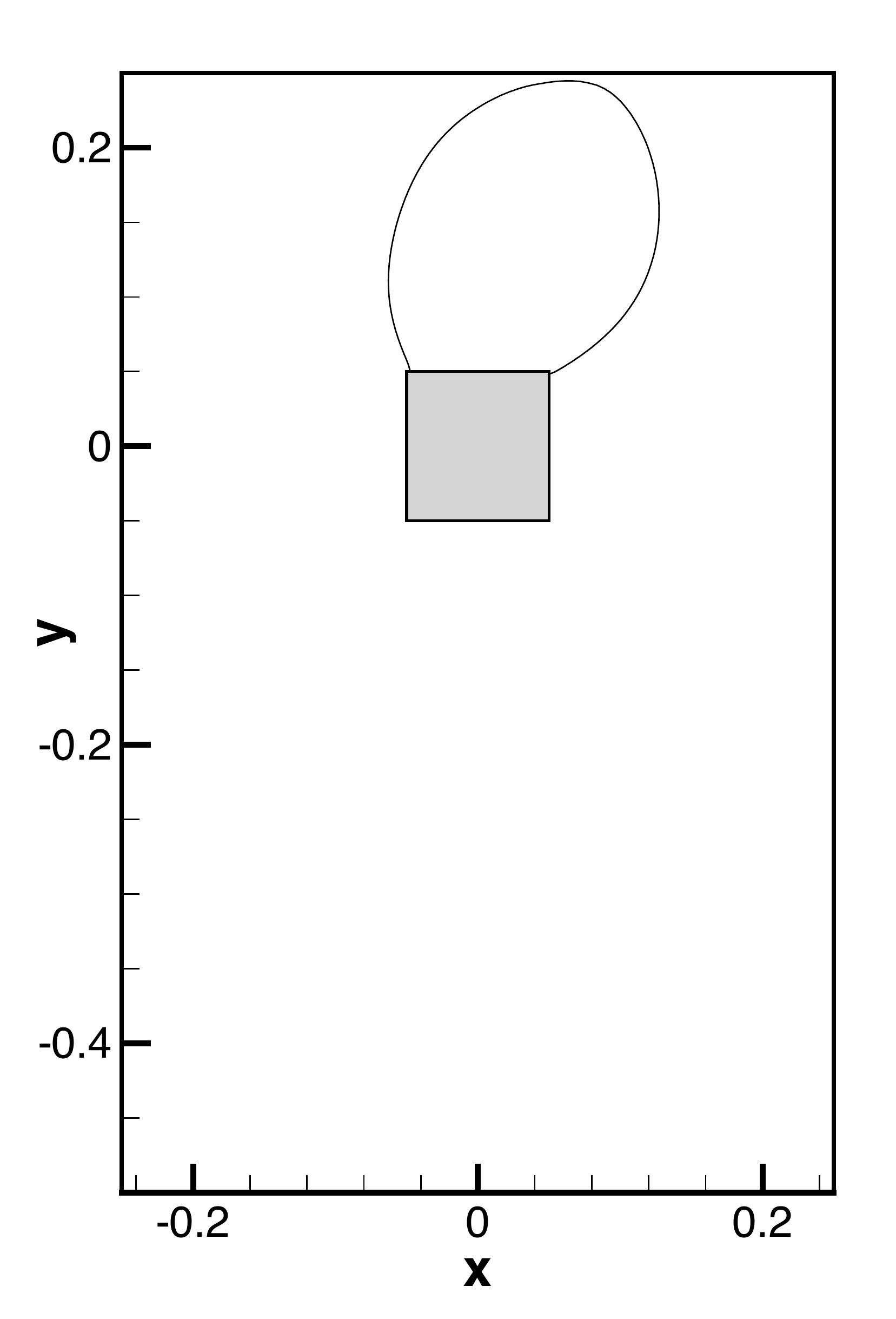}}
\subfigure[$t=3.6$]{ \includegraphics[scale=.18]{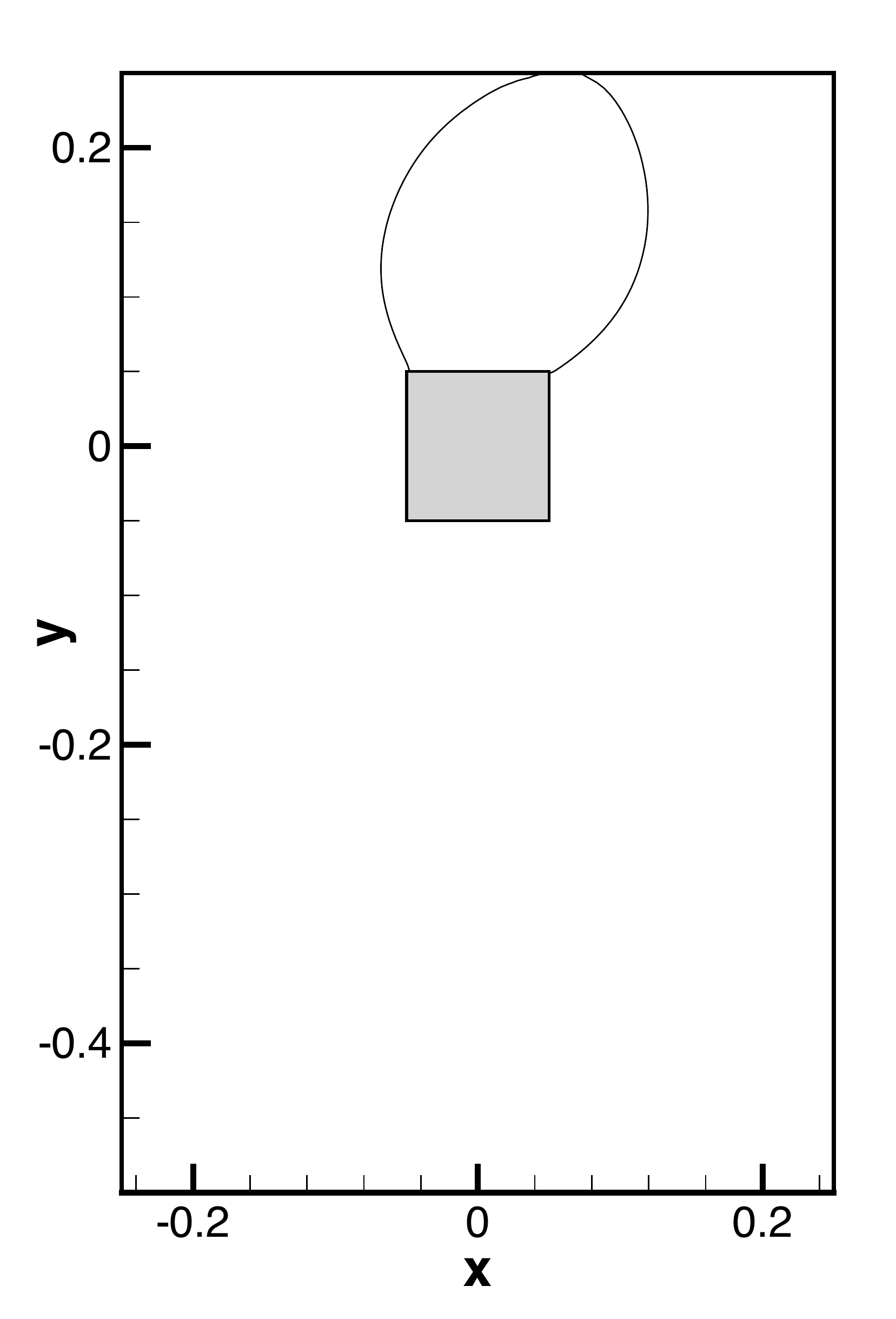}}
\subfigure[$t=3.65$]{ \includegraphics[scale=.18]{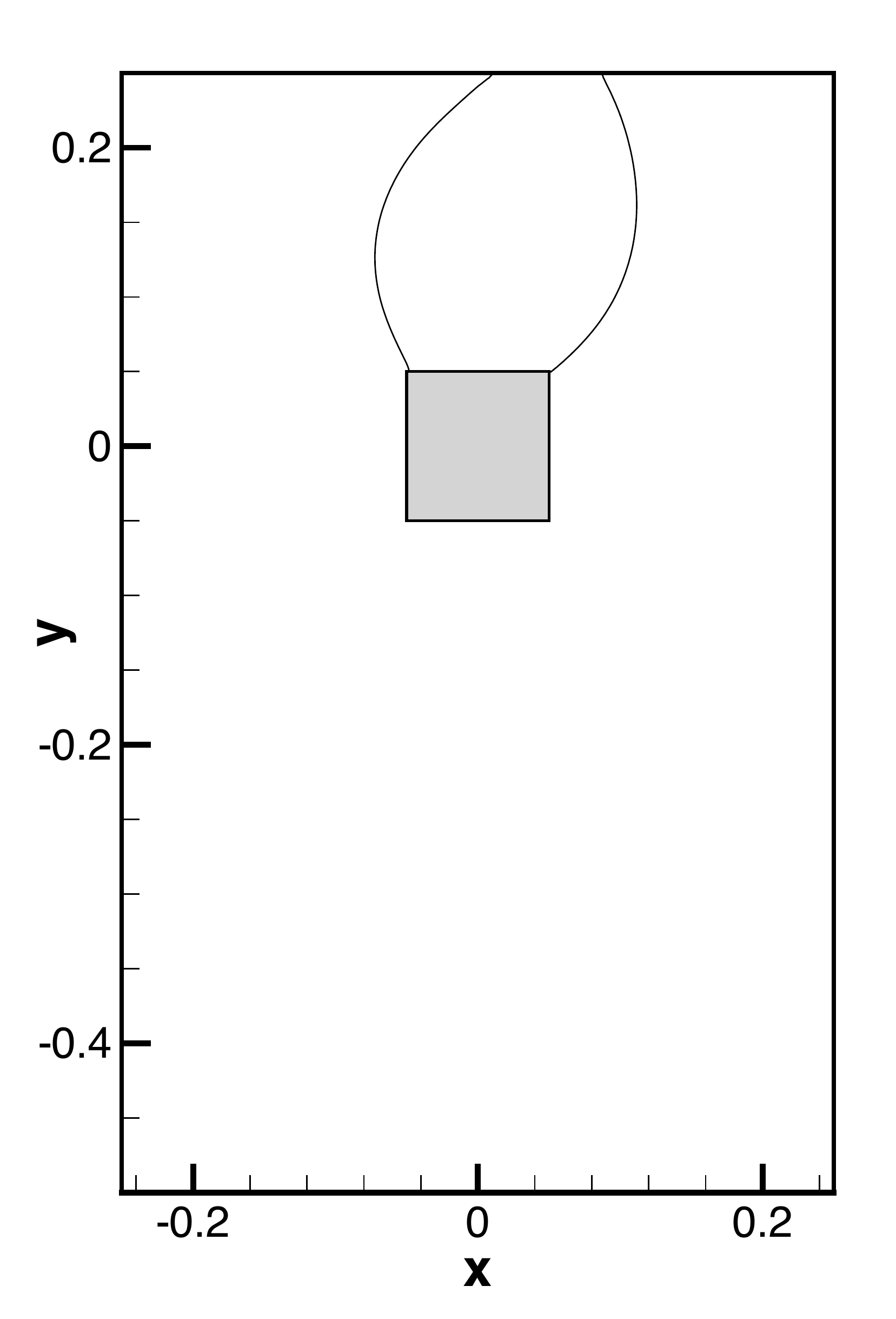}}\\
 \subfigure[$t=3.95$]{ \includegraphics[scale=.18]{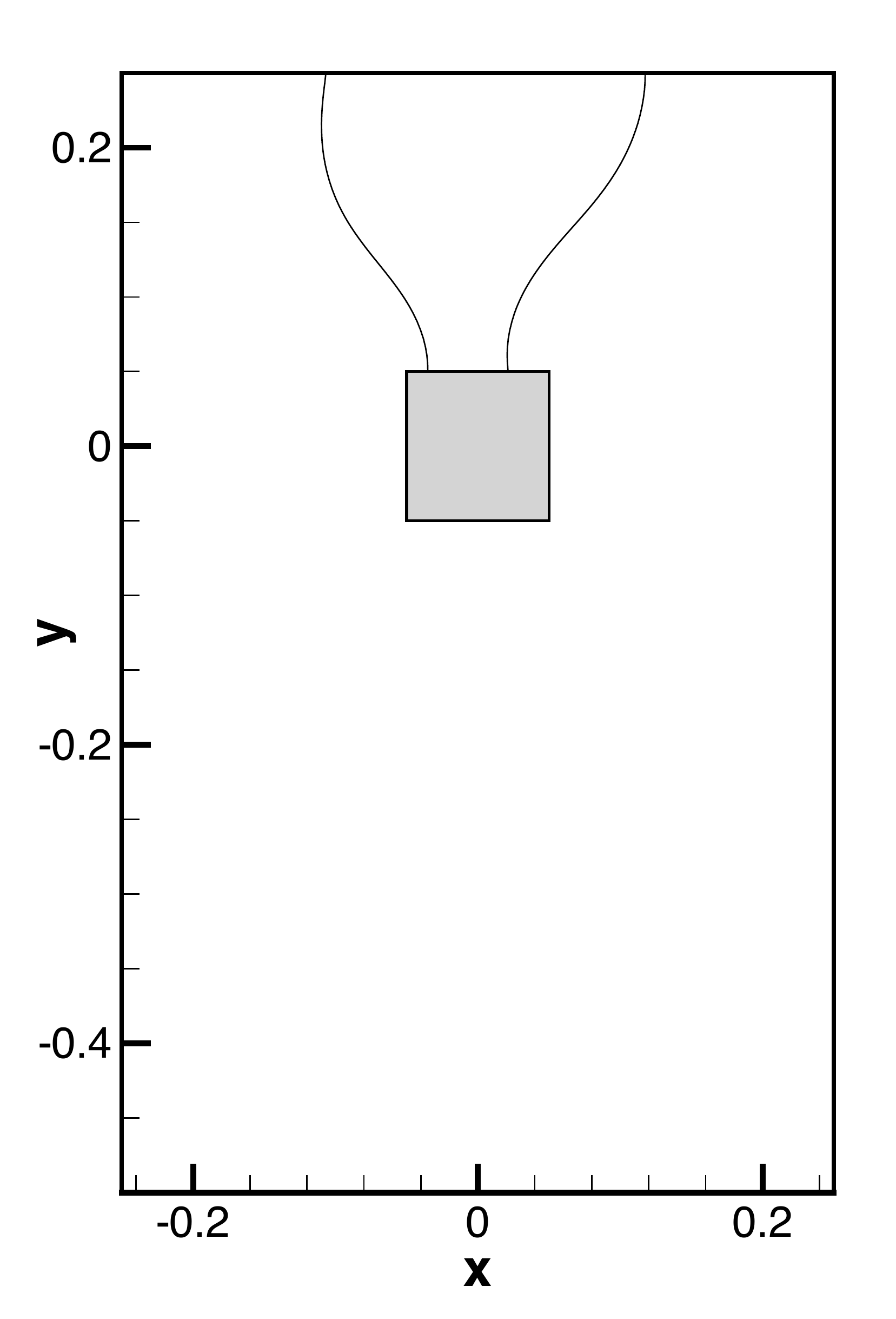}} 
\subfigure[$t=4.4$]{ \includegraphics[scale=.18]{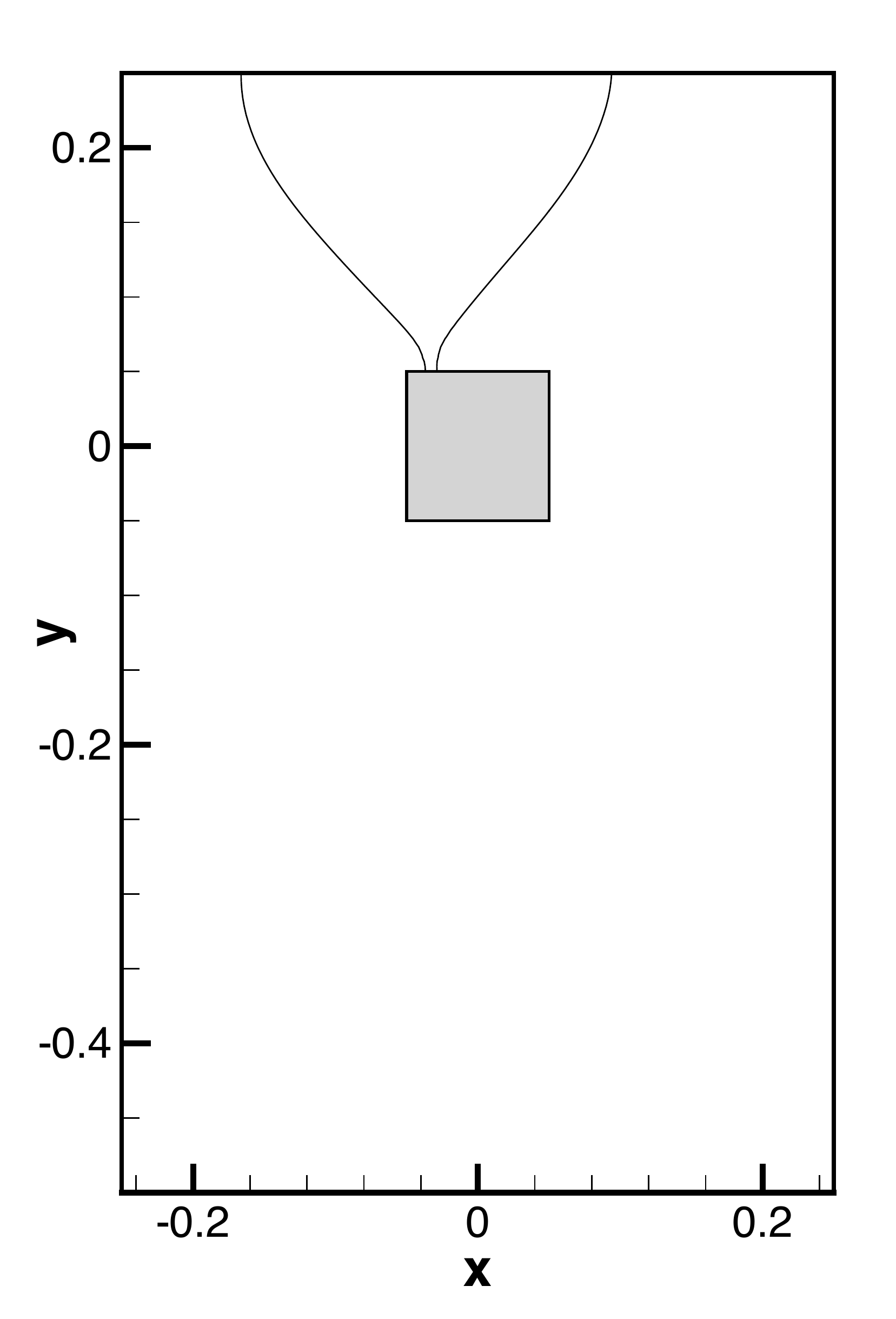}} 
\subfigure[$t=4.45$]{ \includegraphics[scale=.18]{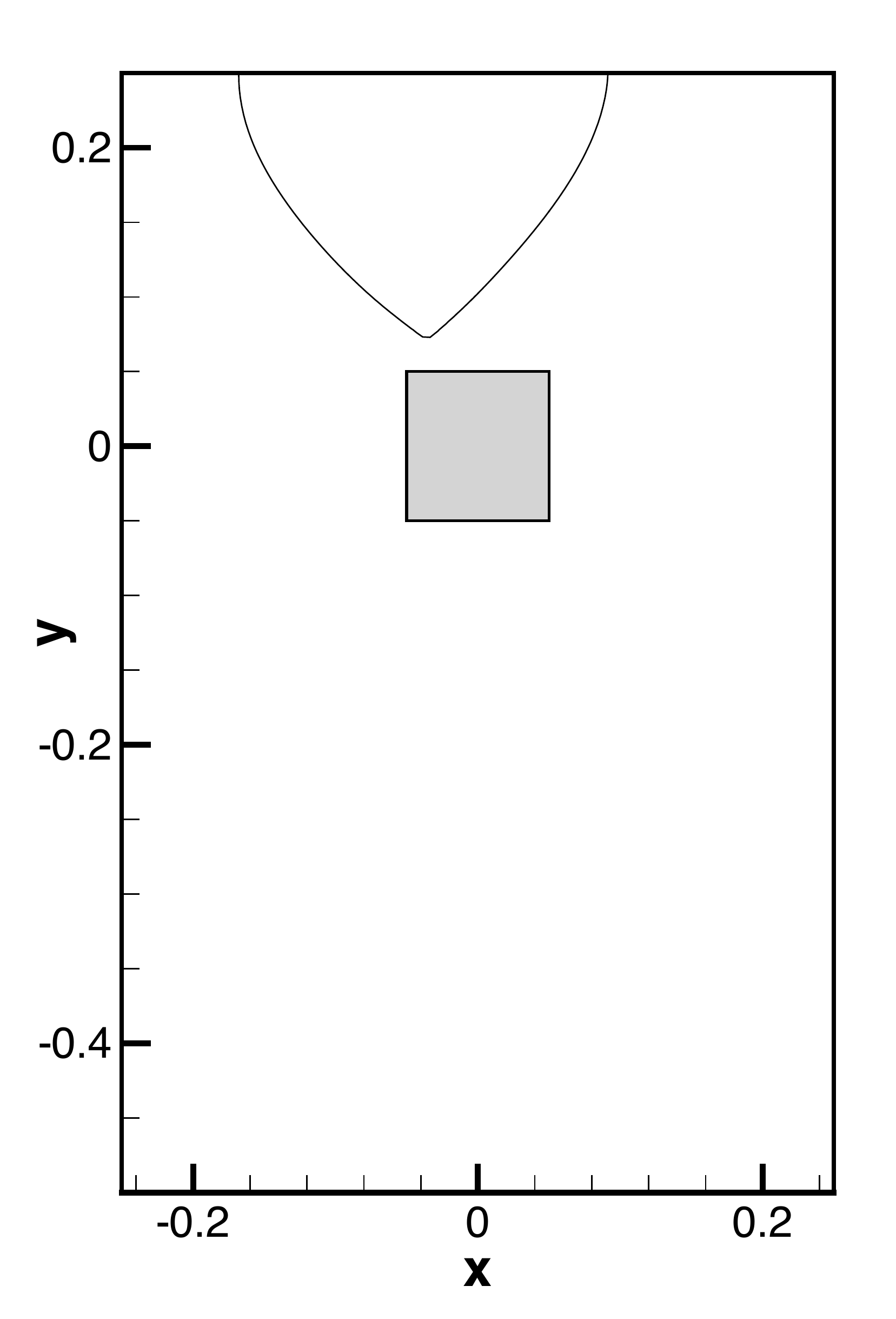}}
\subfigure[$t=5.35$]{ \includegraphics[scale=.18]{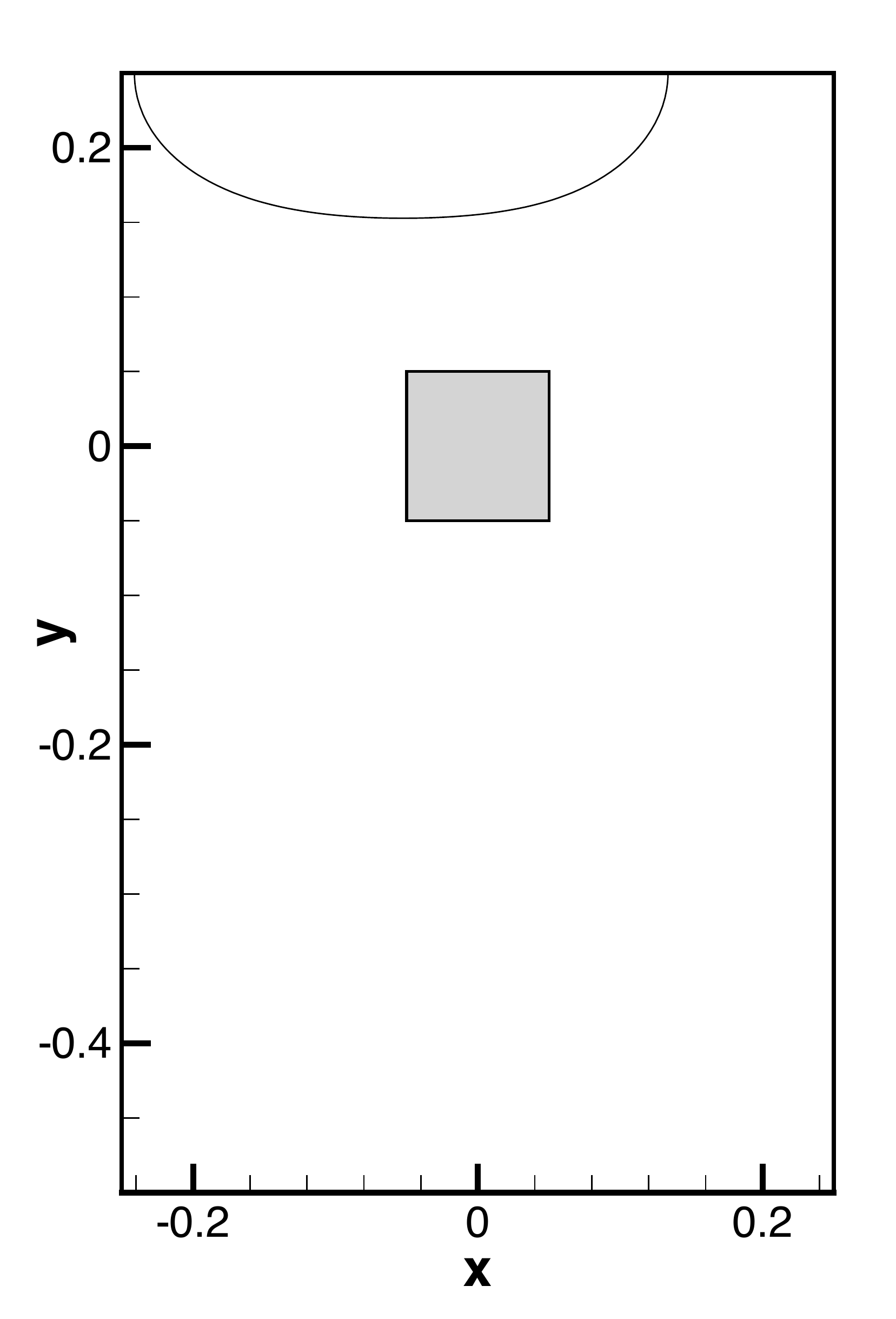}}
\subfigure[$t=6.0$]{ \includegraphics[scale=.18]{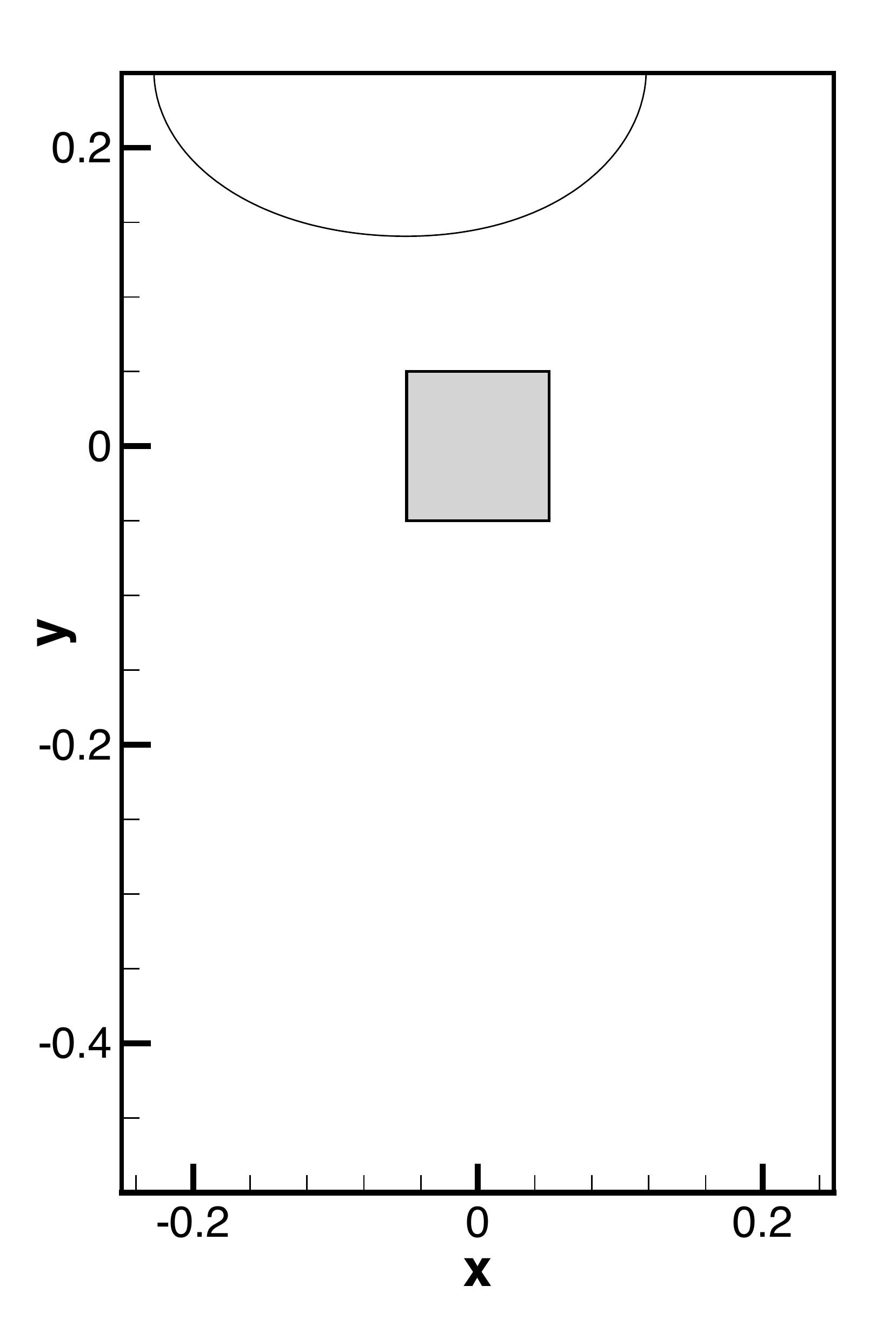}}
\caption{
  Time sequence of snapshots of an air bubble rising in water with
  a normal density contrast  $(\rho_1,\rho_2)=(1,1000)$.
  Shown are the contour levels $\phi(\bs x,t)=0$.
} 
\label{fig:risbubblecase2}
\end{figure}

Let us now look into the case with a normal
density contrast for air and water,
$(\rho_1,\rho_2)=(1,1000).$ In Figure \ref{fig:risbubblecase2}, we plot a temporal sequence of snapshots of the air-water interface for this case,
computed with $\Delta t=5\times 10^{-5}$.
The dynamical characteristics observed here are notably  different and much more complicated than those in the previous case with a smaller density contrast.
As the air bubble rises through the water from $t=0$ to $t=0.85$
(Figure \ref{fig:risbubblecase2}(a)-(c)), the bubble experiences
more considerable deformation and one can observe a flat bottom side in the shape of the bubble.
Notice also that the rise of the bubble is more rapid than in the previous case.
From $t=1.0$ to $t=1.1$ (Figure \ref{fig:risbubblecase2}(d)-(e)),
the  air bubble approaches the bottom right corner of
the solid square, and an indentation appears on the shoulder of the bubble.
The air bubble collides onto the square and is attached to its surface
near the bottom right corner from $t=1.2$ to $t=1.55$
(Figure \ref{fig:risbubblecase2}(f)-(h)). Afterwards, the bubble
slides on the surface of the solid square, from the bottom to the right
and then to the top of the interior square (Figure \ref{fig:risbubblecase2}(g)-(m)).
The bubble deformation is very pronounced during this process.
Due to strong buoyancy,
the bubble is stretched upward while still attached to
the top surface of the solid square, and it approaches
the top wall of the domain (Figure \ref{fig:risbubblecase2}(m)).
From $t=3.6$ to $t=4.4$ (Figure \ref{fig:risbubblecase2}(m)-(q)),
the air bubble touches the upper domain wall
and forms contact lines on that wall.
Simultaneously, the area of contact between the bubble and
the top surface of the interior square shrinks, and the bubble
deforms into a funnel (Figure \ref{fig:risbubblecase2}(p)-(q)).
From $t=4.45$ to $t=6.0$ (Figure \ref{fig:risbubblecase2}(r)-(t)),
the air bubble detaches from the interior solid square,
and forms a dome attached to the upper domain wall over time.

\begin{figure}[tbp]
  \centering
    \includegraphics[width=0.45\textwidth,height=0.35\textwidth]{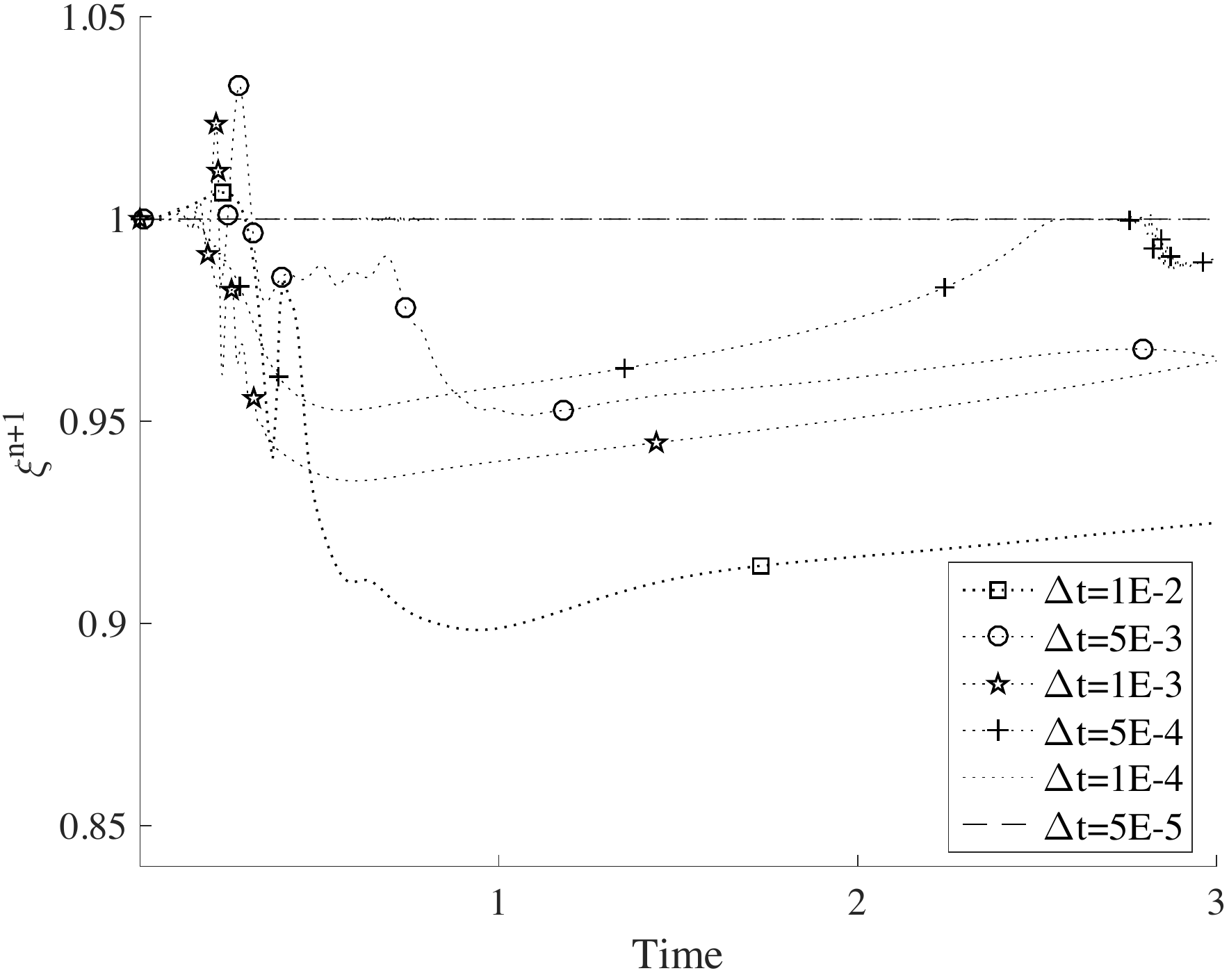}
    \caption{\small  Comparison of time histories of $\xi^{n+1}$ obtained
      with various time step sizes
      ${\Delta}t=10^{-2} \sim 5\times 10^{-5}$ for the rising air
      bubble problem with density contrast $(\rho_1,\rho_2)=(1,1000)$.
      }
    \label{fig:risbubbledt}
\end{figure}

In order to demonstrate the robustness of our method,
we have performed tests using the second case $(\rho_1,\rho_2)=(1,1000)$
for various time step sizes ranging from
${\Delta}t=10^{-2}$ to $5\times 10^{-5}$.
Figure \ref{fig:risbubbledt} is a comparison of the time
histories of $\xi^{n+1}$ for each time step size.
We observe that even with large time step sizes
(e.g.~${\Delta}t=10^{-2}),$ the computation is stable.
We can observe that when ${\Delta}t$ is relatively small (${\Delta}t\leq 10^{-4}$),
the computed $\xi^{n+1}$ is identical to $1$. For larger ${\Delta}t$,
the computed values for $\xi^{n+1}$ initially oscillate
near $1$ (from $t=0.2$ to $t=0.4$), and then decreases and
settles to some levels less than $1$ over a long time.
With a larger $\Delta t$ we generally attain a $\xi^{n+1}$ level
over time that tends to be smaller. 

\begin{remark}
  The computational cost for solving the nonlinear
  equation \eqref{eq:nonlinear} using the Newton's method,
  including the computations for the coefficients involved therein
  for preparation of the Newton solver,
  is small compared with the total cost within each time step.
  For example, with the same computational setting for producing the results
  of Figure \eqref{fig:risbubblecase2} for the rising air bubble problem,
  the  time spent in the Newton solver (including the time spent on computing
  the coefficients in equation \eqref{eq:nonlinear}) per time step
  accounts for about $8\%$ of the total time per time step with our current method.
  The total wall time per time step is $0.3$ seconds on the average,
  in which around $0.024$ seconds is spent in the Newton solution for
  equation \eqref{eq:nonlinear}. 
  Of the time spent in the Newton method, essentially all is spent on
  computing the coefficients involved in the nonlinear equation
  \eqref{eq:nonlinear}, and the actual Newton iteration time is negligible.
\end{remark}

\section{Concluding Remarks}
\label{sec:summary}


In this paper we have developed an energy-stable scheme for
approximating the coupled system of Navier-Stokes/Cahn-Hilliard
equations for incompressible two-phase flows
with different densities and viscosities for the two fluids.
The scheme employs a scalar-valued auxiliary variable
in the formulation, and it satisfies a discrete energy stability
property regardless of the time step sizes.
The scheme allows an efficient solution procedure
and is computationally attractive.
Within a time step, it computes two copies
of the flow variables (velocity, pressure, phase field function),
by solving individually a linear algebraic system involving
a constant and time-independent coefficient matrix for each of
these field functions.
The coefficient matrices involved in these linear systems
only need to be computed once and they can be pre-computed, even
with large density ratios and large viscosity ratios.
Additionally, the algorithm requires the solution of a
nonlinear algebraic equation about a {\em scalar-valued number}
using the Newton's method within each time step.
The computational cost for this nonlinear solver is very low,
accounting for around $8\%$ of the total solver time per time step,
because this nonlinear equation is about a scalar number,
not a field function.
Because of these properties, the presented method is robust
and computationally efficient.

We have tested the proposed method with
extensive numerical experiments for several two-phase
flow problems involving large density ratios and viscosity ratios.
In particular, by comparing with Prosperetti's exact solution
for the capillary wave problem, we have shown that our method
produces physically accurate results for various density ratios
and viscosity ratios. Large time step sizes have been tested with
the proposed method in two-phase simulations. While the simulation results cannot be
expected to be accurate at those time step sizes, the computations
have been shown to be stable nonetheless, verifying the robustness
of the method developed herein.


Some comments concerning the auxiliary variable strategy
are in order. It should be noted that the coupled
Navier-Stokes/Cahn-Hilliard equations are not a gradient-type system.
As such, we find that the auxiliary variable formulation
as described in e.g.~\cite{ShenXY2018} for gradient flows
does not carry over to Navier-Stokes equations or two-phase
flows. Two lessons are learned:
\begin{itemize}

\item
  Attempts to reformulate the viscous terms in these equations using
  the auxiliary variable, which is analogous to the auxiliary-variable
  treatments of the dissipation terms
  in the evolution equation for gradient flows, lead to very
  poor simulation results. A viable strategy for 
  the two-phase governing
  equations is to employ the auxiliary variable to reformulate the
  convection terms in the momentum equation and the phase field equation.  
  The integral form of the dynamic equation for the scalar-valued
  auxiliary energy variable provides a great deal of flexibility
  to be commensurate with the corresponding treatments of various terms
  in the momentum and phase-field equations to ensure a discrete energy
  stability property.
  
\item
  When solving for the scalar-valued number,
  $\xi^{n+1}=\frac{R^{n+1}}{\sqrt{E^{n+1}}}$, it should be noted that
  in the implementation we have employed equation \eqref{eq:Rid3},
  which is based on 
  the transformed dynamic equation
  \eqref{eq:Rid2} instead of
  the original equation \eqref{eq:scheme6eq}. 
  While equations \eqref{eq:scheme6eq} and \eqref{eq:Rid2}
  are mathematically equivalent, we find that in practice 
  the method is more robust if the implementation is based
  on the transformed equation \eqref{eq:Rid2}.
  
\end{itemize}


An important implication of the current method lies in the following.
It demonstrates that the energy-stable schemes for two-phase flows
can also become 
computationally efficient,
eliminating the expensive re-computations
for the coefficient matrices involved in the associated linear
algebraic systems.
In terms of the amount of operations per time step,
the energy-stable schemes can be competitive. For example,
within each time step,
the amount of operations involved in the current method
is approximately twice that of the semi-implicit method of~\cite{DongS2012},
which is only conditionally stable, or slightly larger than that  due
to the extra nonlinear equation about a scalar-valued number.


\section*{Acknowledgement}
This work was partially supported by
NSF (DMS-1318820, DMS-1522537).
Useful discussions with Professor J. Shen (Purdue University)
are gratefully acknowledged.

\bibliographystyle{plain}
\bibliography{engstab,interface.bib,nphase.bib,obc.bib,mypub.bib,nse.bib,sem.bib,contact_line.bib,multiphase.bib}

\end{document}